\documentclass[%
 amsmath,amssymb,
 aps,pre,
 showpacs,
 superscriptaddress,
twocolumn,
]{revtex4-2}
\UseRawInputEncoding
\usepackage{graphicx}
\usepackage{dcolumn}

\usepackage{bm}
\usepackage{xcolor}

\usepackage{hyperref}

\newcommand{\red}[1]{\textcolor{black}{#1}}

\begin{document}
\preprint{APS}

\title{
Time-reversal and $\mathcal{PT}$ symmetry breaking in non-Hermitian field theories
}

\author{Thomas Suchanek}
 \affiliation{Institut f{\"u}r Theoretische Physik, Universit{\"a}t Leipzig, Postfach 100 920, D-04009 Leipzig, Germany}
\author{Klaus Kroy}
\affiliation{Institut f{\"u}r Theoretische Physik, Universit{\"a}t Leipzig, Postfach 100 920, D-04009 Leipzig, Germany}
\author{Sarah A. M. Loos}
 \email{sl2127@cam.ac.uk}
\affiliation{DAMTP, Centre for Mathematical Sciences, 
University of Cambridge, Wilberforce Road, Cambridge CB3 0WA, United Kingdom
}%

\date{\today}

\begin{abstract} 
We study time-reversal symmetry breaking in non-Hermitian fluctuating field theories with conserved dynamics, comprising the mesoscopic descriptions of a wide range of nonequilibrium phenomena. They exhibit continuous parity-time ($\mathcal{PT}$) symmetry breaking phase transitions to dynamical phases. For two concrete transition scenarios, exclusive to non-Hermitian dynamics, namely oscillatory instabilities and critical exceptional points, a low-noise expansion exposes a pre-transitional surge of the mesoscale (informatic) entropy production rate, inside the static phases. 
Its scaling in the susceptibility contrasts conventional critical points (such as second-order phase transitions), where the susceptibility also diverges, but the entropy production generally remains finite. The difference can be attributed to active fluctuations in the wavelengths that become unstable. For critical exceptional points, we identify the coupling of eigenmodes as the entropy-generating mechanism, causing a drastic noise amplification in the Goldstone mode.
\end{abstract}

\maketitle





\section{Introduction}
Field theories are a powerful framework to study the mesoscale behavior of spatially extended many-body systems. 
They provide analytical access to
their salient universal characteristics such as phase transitions, pattern formation, and the associated breaking of global symmetries, without resorting to the underlying molecular statistical mechanics. In thermal equilibrium, this approach is simplified by profound constraints that prestabilze the possible physical outcomes and their classification. Their breakdown far from equilibrium still poses formidable theoretical challenges.
A particularly strong manifestation of nonequilibrium occurs in \textit{non-Hermitian} field theories, which break the mesoscopic reciprocity principle -- a rampart of equilibrium coarse-graining -- already at linear order. 
The non-Hermitian property suggests to regard them as a special subclass of active field theories~\cite{cates2019active}. Mesoscopic models for a wide range of very different physical systems share this property, including examples from active matter~\cite{Menzel13,Alaimo_2016,Peshkov_2012,Wittkowski_2017,Jülicher_2018,Tiribocchi_15_model_H,liebchen2018synthetic,Weber_2019}, biological systems~\cite{hickey2023nonreciprocal,Ramaswamy2000,Kozyreff07_Swift_H,Bhattacharya20,Kohyama19,John2005_membranes}, chemical systems~\cite{ZHENG2015,Pena04_wave_insta,Okuzono2003,tong2002phase}, and, generically, systems with \textit{nonreciprocal} interactions~\cite{You_2020,Saha_2020,frohoff2023nonreciprocal,frohoff2023non}. Non-Hermitian field theories thus provide a strongly unifying framework for a broad class of nonequilibrium systems. 

\begin{figure*}
\includegraphics[width=.99
\textwidth]{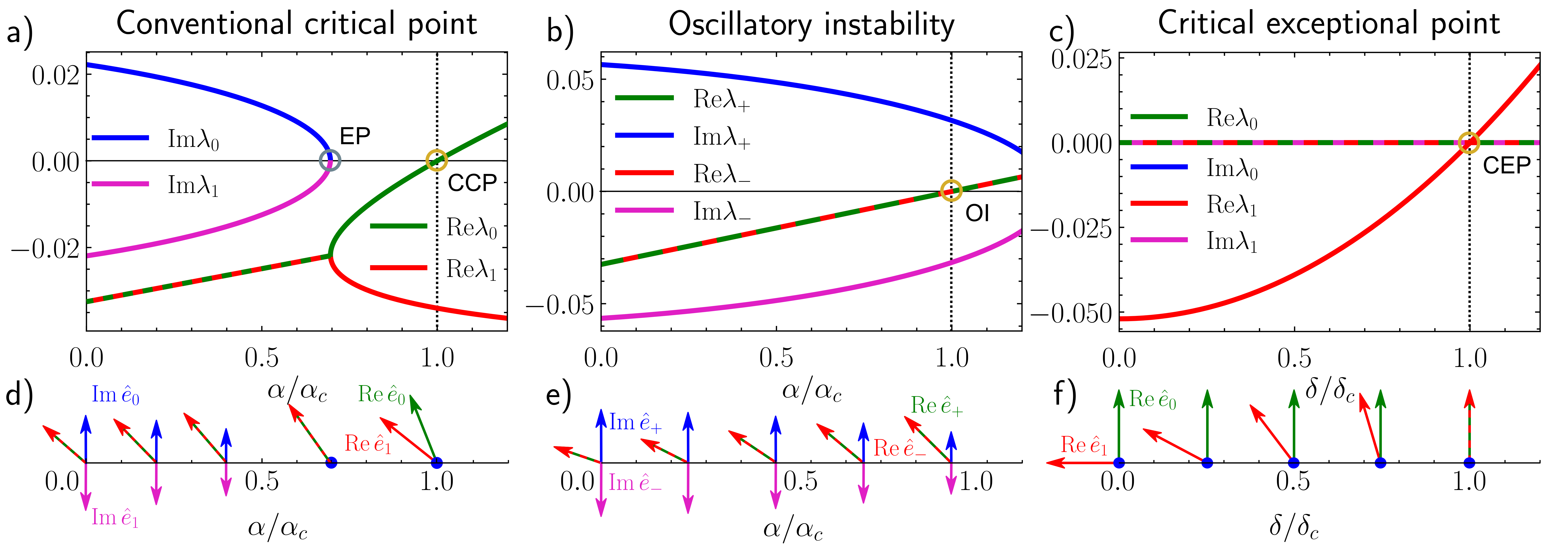}
\caption{\label{fig:pre_1} 
\red{Three types of linear instabilities in non-Hermitian field theories: the first column (a,d) depicts a conventional critical (CCP) point (at $\alpha=\alpha_c$, dashed line); 
the middle column (b,e) depicts an oscillatory instability (OI) (at $\alpha=\alpha_c$, dashed line); the third column (c,f) shows a critical exceptional point (CEP) (at $\delta=\delta_c$, dashed line). Whereas a CCP marks a static-static transition as familiar from equilibrium second-order phase transitions, the OI and CEP are continuous, $\mathcal{PT}$ symmetry-breaking, static-dynamic phase transitions. The top row (a,b,c), depicts the characteristic behavior of the two eigenvalues with the largest real parts,
the second row (d,e,f) depicts the relative orientation of the corresponding eigenvectors of the first Fourier component of the linearized operator $\mathcal{J}$. For the conventional critical point (a,d), the eigenvalue $\lambda_0$ of the unstable eigenmode $\hat{e}_0$ is real and the eigenvectors $\hat{e}_0$ and $\hat{e}_1$ remain linearly independent at the transition. Panels (a,b) also feature a non-critical exceptional point (EP) at $\lambda_0=\lambda_1$, $\alpha\approx 0.7\alpha_c$, preceding the CCP. At the oscillatory instability (b,e), the eigenvectors $\hat{e}_+$ and $\hat{e}_-$ form a complex conjugate pair that becomes unstable, while their eigenvalues $\lambda_-$ and
$\lambda_+$ retain a non-zero imaginary part. At the CEP (c,f), a formerly stable mode ($\mathrm{Re}\lambda_1<0$) becomes unstable, when it merges with Goldstone mode $\hat{e}_0$ with $\lambda_0=0$. Since the formerly linearly independent eigenmodes $\hat{e}_0$ and $\hat{e}_1$ coalesce, the dimensionality of the eigenspace drops by one. The concrete values for the paradigmatic example of a nonreciprocal Cahn-Hilliard model, discussed in the companion article Ref.~\cite{suchanek2023entropy}, serves to illustrate the general situations. For all panels, the model parameters are $\beta=0.05$, $\gamma=0.015$, $\kappa=0.01$, and for (a,d): $\delta=0.03$, (b,e): $\delta=0.06$, (c,f): $\alpha=-0.07$.
}}
\end{figure*}

One of the most striking implications of non-Hermitian dynamics, which has recently gained renewed interest, is the emergence of dynamical phases \red{via {parity-time ($\mathcal{PT}$) symmetry-breaking phase transitions}. 
{In this scenario, a static stationary phase with parity symmetry (e.g., a static, mirror-symmetric pattern), becomes dynamical upon loosing this symmetry.
If the dynamical equations are parity symmetric (meaning that there is no external parity-breaking, e.g., by an external field acting on the system), such dynamical  states are manifestations of spontaneous symmetry breaking. They must then, by symmetry, always occur as pairs with opposite parity. {A paradigmatic example are pairs of traveling waves that run in opposite directions and are equally likely to appear.}
These dynamical system states can be transformed into each other by the action of the $\mathcal{PT}$ operator (combined parity and time inversion).
Such $\mathcal{PT}$-symmetry breaking transitions}} generate a variety of dissipative structures, such as 
wave trains~\cite{MALOMED1984,Cummins93,Dangelmayr_1997,Zhabotinsky95,Pan94,BESTEHORN1989}, traveling domains~\cite{Coullet89,Rabaud90_viscous,Goldstein91PT} or states~\cite{Ophaus21,frohoff2021localized,Stegemerten22}, or emergent chiral patterns~\cite{Fruchart2021,zhang2020reconfigurable,Liao21_chiral}. 
Important instances of {$\mathcal{PT}$-symmetry breaking,} static-dynamic phase transitions, exclusive to non-Hermitian dynamics, 
are \textit{oscillatory instabilities}~\cite{Cross93,cross_greenside_2009}, (which may also lead to standing oscillatory patterns~\cite{Pena04_wave_insta,cross_greenside_2009}), and the recently uncovered \textit{critical exceptional points} (CEP). \red{While all these concepts are defined mathematically in Secs.~\ref{sec:overview-transitions}, \ref{sec:EPRosc}, \ref{sec:irrexp}, and~\ref{sec:con}, we aim to briefly explain here in words their characteristics. Oscillatory instabilities are characterized by an oscillatory component of their dynamics even before the transition, reflected in a non-vanishing imaginary parts of the associated eigenvalues of the linearized dynamical operator [see Fig.~\ref{fig:pre_1}b) for an illustration]. This oscillatory component becomes apparent when the system is perturbed, and, after crossing the transition, may stabilize into stable periodic motion.
Critical exceptional points are secondary transitions occurring in phases of broken continuous symmetry, such that there exists a Goldstone mode. At the CEP, a critical mode coalesces with the Goldstone mode, as illustrated in Fig.~\ref{fig:pre_1}c.} 

Beyond this rich phenomenology, one should expect that the breaking of reciprocity, \textit{at linear order}, also has crucial implications at the level of small perturbations \red{and fluctuations}. In the context of open quantum systems, it was indeed recently shown that non-Hermitian dynamics can entail a drastic mode-selective amplification of thermal noise~\cite{hanai2020critical,Zhang19}. Similar mechanisms have also been observed in the context of shear flows~\cite{Farrell94}, fluctuation-induced pattern formation~\cite{Biancalani17} and optical resonators~\cite{Zhong20}.
This motivated us to take a closer look at the fluctuations in a general class of classical non-Hermitian field theories. Our investigation was mainly guided by the following two questions.

First, in contrast to  (equilibrium or equilibrium-like) static phases, the above-mentioned dynamical phases exhibit a systematic mass flow associated with a spontaneously broken $\mathcal{PT}$ symmetry. One may naturally wonder whether this qualitatively affects also the fluctuations.
In this respect, it is important to distinguish 
between two completely different types of symmetry breaking and to understand their connection: $\mathcal{PT}$ and time-reversal symmetry breaking (TRSB),
the first affecting the systematic dynamics, the second a genuine feature of \red{nonequilibrium} fluctuations. 
In order to characterize $\mathcal{PT}$, we employ a small noise approximation that allows us to resort to bifurcation theory. To characterize TRSB, we bring to bear the concept of informatic entropy production for field theories, established in Ref.~\cite{Nardini2017}.

Secondly, we would like to contribute  to the general debate whether and to what extent entropy production has universal properties in the vicinity of phase transitions, which can help to classify phase transitions far from equilibrium. In recent years, this question has been addressed  from various angles in different models~\red{\cite{seara2021,caballero2020stealth,meibohm2023landau,fiore2021current,barato2010entropy,tome2012entropy,Paoluzzi22,GrandPre21,Bowick22}}, but not yet from the general perspective of non-Hermitian field theories.
Recalling that the very essence \red{and nontrivial features} of equilibrium phase transitions can be pinned down to fluctuations of ``static" observables (measured at a single time) becoming systematic, one may naturally ask, to what extent this picture extends to dynamical observables (such as the entropy production) in nonequilibrium systems.

In the following, we uncover generic characteristics of TRSB fluctuations for different types of phase transitions in a broad class of physical systems that can be addressed by non-Hermitian field theories, independent of their specific molecular details.
Our approach enables us to establish rigorous connections between TRSB and the spectrum and geometry of the eigenmodes of the dynamical operators.
We pay special attention to the fluctuations associated with the formation of dissipative structures, and consider in detail the two most common static-dynamic transition scenarios, namely, oscillatory instabilities and CEPs. 
We cover certain key aspects of this program in the two companion papers~\cite{suchanek2023irreversible,suchanek2023entropy}. In Ref.~\cite{suchanek2023irreversible}, we provide a compact and accessible bird's eye perspective on the most striking results. In Ref.~\cite{suchanek2023entropy}, we specifically analyze TRSB in the nonreciprocal Cahn-Hilliard model, which is a prototypical example for the abstract class of non-Hermitian field theories studied here. The explicit analytical findings and numerical results presented there are in perfect agreement with the general theory developed below, and selectively employed for illustrative purpose in \red{Figs.~\ref{fig:pre_1} and \ref{fig:pre_2}.}

The remainder of this paper is organized as follows. In Sec.~\ref{sec:framework}, we introduce the \red{concept} of non-Hermitian field theories, the types of studied phase transitions, and the employed {measure} for TRSB. In Sec.~\ref{app:epdef}, we provide a detailed characterization of the entropy production in various possible phases and across their transitions, with a focus on the {low} noise \red{regime}. Our most important result is an explicit representation of the informatic entropy production rate $\mathcal{S}$ for fields in terms of 
the eigensystem of the associated linearized dynamical operator (Sec.~\ref{sec:eprsing}). From this expression, we can infer the relevant characteristics of $\mathcal{S}$, and show that the entropy production surges within static phases close to transitions to a dynamical phase. 
In Sec.~\ref{sec:nonHermitian-PT-CEP}, we shed light onto the underlying dissipative mechanism around CEPs. 
To improve readability, some of the technical aspects were relegated to an Appendix.


\section{Framework}\label{sec:framework}
\subsection{Class of studied field theories}
 We consider nonequilibrium field theories  for ${N}$ scalar field components $\phi_i(\boldsymbol{r},t)$, $i=1,\dots,{N}$, where $t$ denotes time and the spatial coordinates $\boldsymbol{r}$ reside on 
a bounded $d$-dimensional domain \red{of volume} $V$ with periodic boundary conditions. We assume conserved stochastic dynamics of the form
\begin{align}\label{equ:moda}
    &\dot{\phi}_i=-\nabla\cdot (\boldsymbol{J}^\mathrm{d}_i+\sqrt{2\epsilon}\boldsymbol{\Lambda}_i),
    ~~~\boldsymbol{J}^\mathrm{d}_i=\ -\nabla\mu_i \, ,
\end{align}
where we used the compact notation $\dot \phi \equiv \partial_t \phi$. The space-time Gaussian white noise $\boldsymbol{\Lambda}$ satisfies 
\begin{align}
\big\langle{\boldsymbol{\Lambda}}_{in}(\boldsymbol{r},t){\boldsymbol{\Lambda}}_{jm}(\boldsymbol{r}',t')\big\rangle=\delta_{ij}\delta_{nm}\delta(\boldsymbol{r}-\boldsymbol{r}')\delta(t-t'),
\end{align}
with $n,m=1,\dots,{N}$, and $k,l=1,\dots,d$, and $\epsilon$ denoting the noise intensity. 

We explicitly allow the Jacobian $ \mathcal{J}\equiv- {\delta \nabla\boldsymbol{J}^\mathrm{d}}/{\delta \phi}$ of the dynamical operator, 
 represented in an appropriate basis (see Sec.~\ref{sec:overview-transitions}) to be \textit{non-Hermitian}, which can only occur in the presence of a nonequilibrium deterministic current $\boldsymbol{J}_i^\mathrm{d}$, i.e., if the chemical potential ${\mu}_i$ cannot be represented as gradient of a scalar potential.
We aim to study systems that are intrinsically out of equilibrium, which should be contrasted to boundary-driven systems that are subject to a form of global energy injection, associated with an explicit external symmetry breaking. A common example for the latter involves a current $\boldsymbol{J}^\mathrm{d}=f{\phi}$ due to a constant external force $f$ acting uniformly on ${\phi}$. 
To exclude such more ``conventional" setups from our discussion, we assume that Eq.~\eqref{equ:moda} is symmetric with respect to a parity inversion, $\mathcal{P}: \boldsymbol{r}\mapsto-\boldsymbol{r}$.

We focus on the regime of low noise intensity $\epsilon$, where we can build on concepts of bifurcation theory, familiar from deterministic nonlinear dynamics.

\subsection{Types of studied continuous phase transitions}\label{sec:overview-transitions}

We consider transition scenarios of the dynamics~\eqref{equ:moda} that are caused by a linear instability of its dynamical operator 
  $  F\equiv-\nabla\cdot\boldsymbol{J}^\mathrm{d}.$ From a nonlinear-dynamics perspective, these transitions are bifurcation points of $F$~\cite{Hohenberg_2015,Bose_2019}.
 By $\phi^*$ we denote fixed points of $F$, which are at the same time solutions of Eq.~\eqref{equ:moda} for $\epsilon=0$.
 Based on the properties of the spectrum of the \textit{linerarized dynamical operator}\red{, i.e.,} the Jacobian  
 \begin{align}\label{def:Jacobian}
 \mathcal{J}\equiv\frac{\delta F}{\delta \phi^*}, 
 \end{align} 
we can distinguish two types of bifurcations. Primary bifurcations arise from solutions that have the full symmetry of Eq.~\eqref{equ:moda}. According to the Goldstone theorem~\cite{Goldstone62}, all real parts of the eigenvalues of $\mathcal{J}$ are negative before the transition. A bifurcation occurs where the real part of one or more eigenvalues vanishes. Secondary bifurcations emanate from  solutions that already spontaneously break a native continuous symmetry of Eq.~\eqref{equ:moda}. Here, the Goldstone theorem ensures the existence of at least one eigenvector (eigenmode) with eigenvalue zero, which is called a \textit{Goldstone} mode, while the real parts of all other eigenvalues remain negative away from the transition.
 Based on the signs of the real parts of their eigenvalues, we refer to eigenmodes as stable or unstable. When applying the above notions, we assume that $F$ and $\mathcal{J}$ have representations in a countable Fourier basis. Since parity inversion corresponds to complex conjugation in Fourier space, the presumed $\mathcal{P}$ symmetry of Eq.~\eqref{equ:moda} then implies that
 $\mathcal{J}$ has a real Fourier representation~\footnote{This implication does not hold for hydrodynamic theories of flocking, where the dynamical operator of the polarization field commonly includes advection terms~\cite{Marchetti2013}, and therefore spatial derivatives of uneven order, yet is still parity invariant}. 

We address three transition scenarios, which are briefly introduced here and schematically \red{illustrated in Fig.~\ref{fig:pre_1}.} 
As a useful reference point, first we consider \textit{conventional critical transitions} that are characterized by the vanishing of a {single} real eigenvalue \red{of the Jacobian in Eq.~\eqref{def:Jacobian}}, while its eigenvectors form a complete set, i.e., an eigenbasis \red{[Fig.~\ref{fig:pre_1} (a,d)].} This scenario is familiar from second-order phase transitions in equilibrium systems. 
The same type of instability can however also be encountered in non-Hermitian field theories. Next, we consider two transition scenarios that are exclusive to non-Hermitian systems, namely, oscillatory instabilities and critical exceptional points (CEPs).
Both of these instabilities may lead into dynamical phases and are then associated with a breaking of $\mathcal{PT}$ symmetry (with $\mathcal{PT}:\mathbf{r},t\mapsto -\mathbf{r},-t$)~\cite{Dangelmayr_1997,Fruchart2021,suchanek2023irreversible}. 
The characteristic feature of an \textit{oscillatory instability}, \red{illustrated in Fig.~\ref{fig:pre_1} (b),} is that the real parts of the eigenvalues of the pertinent modes vanish,  while the imaginary parts remain nonzero (as in the well-known case of the Hopf bifurcation). Here, we consider only primary instabilities of this type.
In contrast, CEPs are \textit{exceptional points}, meaning points where two (or more) modes merge, that also are, at the same time, \textit{critical points}. 
Thus, at CEPs, the directions of two eigenmodes become aligned, while the real parts of their eigenvalues vanish, \red{as illustrated in Fig.~\ref{fig:pre_1} (c,f)}. Consequently, \red{the eigenvectors form a basis} in the vicinity of the CEP, but not at the CEP itself. 
 A particularly interesting $\mathcal{PT}$ symmetry-breaking scenario can then arise when the CEP is a secondary bifurcation point, and a formerly stable mode co-aligns with a \textit{Goldstone} mode~\cite{Dangelmayr87,Dangelmayr_1997,Fruchart2021}. In Sec.~\ref{sec:nonHermitian-PT-CEP}, we give a more rigorous mathematical description of CEPs and discuss the characteristic associated mode dynamics.
 As long as only one pair of eigenvectors converges in the CEP, not only the real parts but also the imaginary parts of the corresponding eigenvalues vanish at the critical point, as implied by the complex conjugate root theorem~\footnote{Assuming that two eigenvectors with complex eigenvalues $\lambda_0$ and $\lambda_1$ merge at a certain fixed point $\phi^*$, the complex conjugate root theorem~\cite{jeffrey2005complex} implies that the complex conjugates of these are also eigenvectors with eigenvalues $\bar{\lambda}_0$ and $\bar{\lambda}_1$, which accordingly also merge.}.

\subsection{Irreversibility measure}
To quantify the time-reversal symmetry breaking (TRSB) of the fluctuating field dynamics, we need an irreversibility measure for the field dynamics. We follow Refs.~\cite{Nardini2017,Li_2021,Cates2021} and employ what can be regarded as a generalization of the standard definition for the entropy production rate from stochastic thermodynamics~\cite{Seifert2005}, and
will henceforth be referred to as the (mesoscopic informatic) entropy production rate $\mathcal{S}$. 

First, one generalizes the notion of entropy production associated with a ``trajectory'' $\{\phi(t)_{t\in [0,T]}\}$ to the space of field configurations as 
\begin{align}\label{def:Pathprobratio-S}
    {s}[{\phi},0,T]\equiv\log \frac{\mathbb{P}\left[\{\phi(t)_{t\in [0,T]}\}\right]}{\mathbb{P}\left[\{\phi^R(t)_{t\in [0,T]}\}\right]}\,,
\end{align}
where $\mathbb{P}\left[\{\phi(t)_{t\in [0,T]}\}\right]$ and $\mathbb{P}\left[\{\phi^R(t)_{t\in [0,T]}\}\right]$ are the path probabilities of the trajectory and of it's time-reversed realization, respectively. 
	To be well-defined, the probabilities are to be interpreted in terms of the Onsager--Machlup formalism~\cite{Onsager53}.
We consider fields that represent conserved position-like degrees of freedom, and therefore treat them as even under time-reversal. Throughout, we employ It\^{o} calculus for all stochastic dynamical equations.
The average entropy production rate is then defined as
	\begin{equation}\label{def:S}
		{\mathcal{S}}(t)\equiv \lim\limits_{h\rightarrow 0}\frac{\langle s[\phi,t,t+h]\rangle}{h}\,,
	\end{equation}
 where $\langle\cdot\rangle$ denotes the noise average. 
 A direct connection to the overall {thermodynamic} entropy production and total heat dissipation in \red{any underlying particle} system is, by construction, elusive, since 
 we here consider a coarse-grained formulation of the dynamics that does not include information about all underlying microscopic degrees of freedom; see \cite{Cates2021} for an in-depth discussion. Nevertheless, ${\mathcal{S}}$ can serve as a useful {measure} for the strength of the average TRSB \red{at the mesoscopic level}. We are specifically interested in the asymptotics, $t\rightarrow\infty$.
Particularly informative is the behavior of the limit
\begin{align}\label{def:S*}
\mathcal{S}^*:=\lim_{\epsilon\to 0}\mathcal{S}, 
\end{align}
which reveals the TRSB at leading order in $\epsilon$. 
\vspace*{0.2cm}\\
\section{Irreversibile fluctuations in non-Hermitian field theories}\label{app:epdef}

In this section, we prove general statements for the TRSB of field theories of \red{the type of Eq.~}\eqref{equ:moda}, with a special focus on the behavior at phase transitions. 
First, we derive an explicit representation for the informatic entropy production rate $\mathcal{S}$ for fields in terms of the statistics of the deterministic current (Sec.~\ref{sec:EPRnew}). Using this expression, we can deduce the general scaling of $\mathcal{S}$ with the noise intensity in a generic dynamical phase (see Sec.~\ref{sec:eprtr}).
To tackle the behavior around the phase transitions, we derive in Sec.~\ref{sec:eprsing} an expression for $\mathcal{S}$ in terms of the eigenvalues and eigenvectors of the associated linearized dynamical operator. From that, we can infer the general properties of $\mathcal{S}$ in the vicinity of the different phase transition scenarios. \red{Thereby, we focus on the regime of small noise intensities $\epsilon \ll 1$ and derive asymptotic expressions for the steady-state entropy production rate $\mathcal{S}$ in dynamical phases and at and near transitions from static to dynamical phases. The limit $\epsilon\rightarrow 0$ is especially informative. 
Most importantly, it turns out that the leading order contribution in $\epsilon$ exhibits universal behavior, unveiling the characteristic properties of the entropy production at and near the transition. In contrast, field theories with strong noise are dominated by their nonlinearities and exhibit non-universal behavior that cannot generally 
be treated by analytical means.
The limit moreover reflects the leading TRSB contribution that dominates the whole regime of small noise intensities. 
Lastly, only in this regime, the bifurcation-analytic viewpoint, and the concepts of linear stability and critical phenomena are applicable on the coarse-grained field level.}

\begin{figure*}
\includegraphics[width=.99
\textwidth]{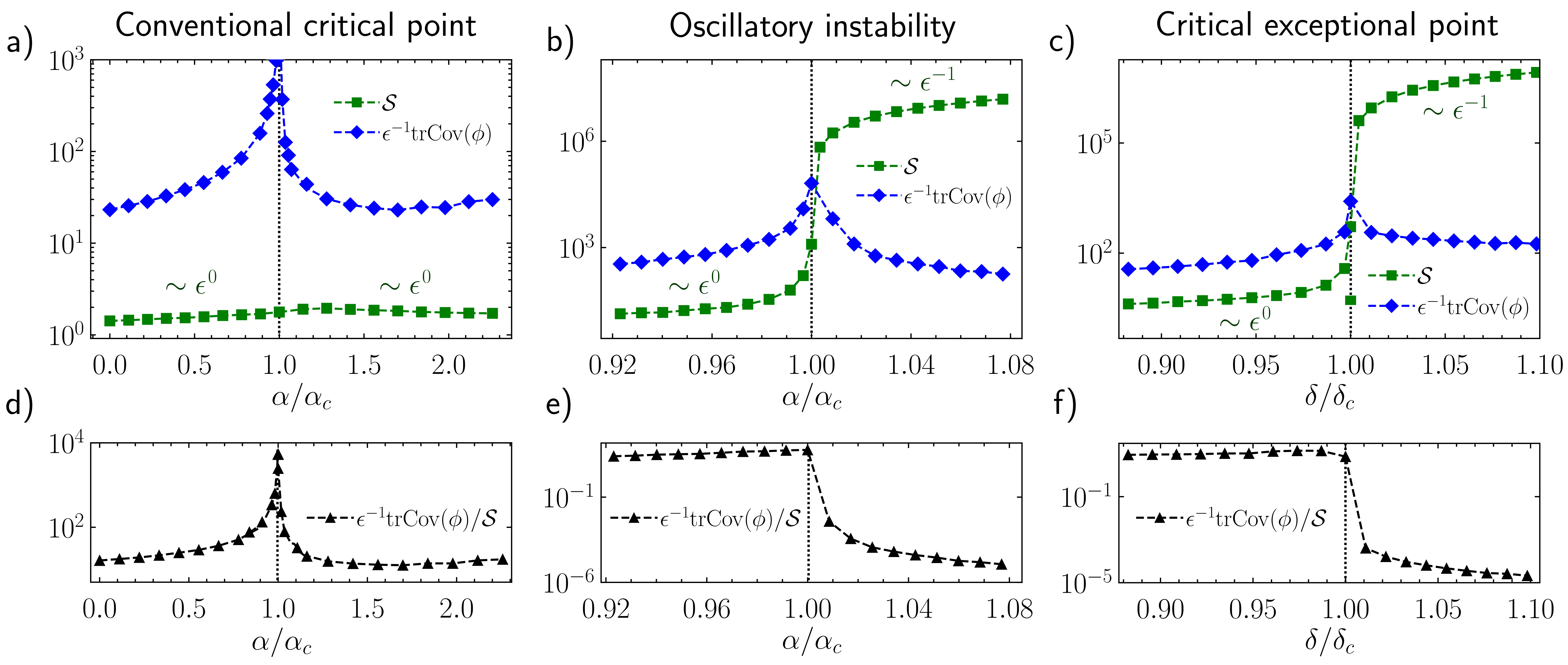}
\caption{\label{fig:pre_2} 
\red{Strength and time-reversal symmetry breaking (TRSB) of fluctuations in the vicinity of a conventional critical point (a,d), an oscillatory instability (OI) (b,e) and a critical exceptional point (CEP) (c,f); demonstrated using the same examplary model and the sampe parameters as in Fig.~\ref{fig:pre_1}. 
The entropy production rate $\mathcal{S}$, defined in Eq.~\eqref{def:S}, quantifies the TRSB, while $\epsilon^{-1}\mathrm{tr}\mathrm{Cov}(\phi)$, defined in Eq.~\eqref{eq:covphi}, quantifies the typical strength of the fluctuations. In the limit of vanishing noise intensity, $ \lim_{\epsilon\rightarrow 0}\epsilon^{-1} \mathrm{tr}\mathrm{Cov}(\phi)=\chi$ approaches the statistic susceptibility defined in Eq.~\eqref{eq:sus}. To approximate it, a sufficiently small noise intensity of $\epsilon=10^{-10}$ was chosen here. While $\epsilon^{-1}\mathrm{tr}\mathrm{Cov}(\phi)$ peaks at all three types of transitions, $\mathcal{S}$ displays regular behavior at the conventional critical point (a), as we analytically show on general grounds in Sec.~\ref{sec:typical}. In contrast, for the OI and the CEP (b,c), Eq.~\eqref{S_dynamicalPhase_SM} predicts $\mathcal{S}\sim \epsilon^{-1}$ inside the dynamical phase ($\alpha>\alpha_c$, or $\delta>\delta_c$, respectively), such that it increases unboundedly for $\epsilon \to 0$. In the static phase, both transitions are heralded by a surging entropy production, (for $\alpha<\alpha_c$, or $\delta<\delta_c$, respectively). Specifically, Eqs.~\eqref{equ:resosc} and \eqref{expfinal} predict $\mathcal{S}^*\sim\chi=\lim_{\epsilon\to 0}\epsilon^{-1}\mathrm{tr}\mathrm{Cov}(\phi)$. The ratio of entropy production and fluctuation strength shown in (e,f) confirms that both quantities grow at the same rate. }
}
\end{figure*}
	\subsection{TRSB and statistics of deterministic current }\label{sec:EPRnew}
\noindent

	We now derive a general expression that relates $\mathcal{S}$ with the statistical properties of the deterministic current,
starting from a recently derived expression for 
the entropy production
\begin{align}\label{equ:inth}
	s[{\phi},t,t+T]=-\frac{1}{\epsilon}\int\limits_V\!\mathrm{d}\boldsymbol{r}\int\limits_t^{t+T}\!\!\mathrm{d}t\sum\limits_i\dot\phi_i(\boldsymbol{r},t)\mu_i(\boldsymbol{r},t)\,
\end{align}
in field theories of the type \red{of Eq.}~\eqref{equ:moda}~\cite{Nardini2017,Gradenigo2012}.
 Applying It\^o's formula for functionals, as introduced in Ref.~\cite{Itofunc}, to this integral  
 and using integration by parts, we find that the
stochastic entropy production along a field trajectory can be expressed via the following It\^o stochastic integral
 \begin{widetext}
	\begin{align}
		\nonumber
		s[{\phi},t,t+T]=~ &\frac{1}{\epsilon}\int\limits_V\mathrm{d}\boldsymbol{r}\int\limits_t^{t+T}\mathrm{d}t\,\sum\limits_i\big[{\boldsymbol{J}}_i^\mathrm{d}(\boldsymbol{r},t)+\sqrt{2\epsilon}{\boldsymbol{\Lambda}}_i(\boldsymbol{r},t)\big]\cdot {\boldsymbol{J}}_i^{\mathrm{d}}(\boldsymbol{r},t)\\
		&
		+\int\limits_V\mathrm{d}\boldsymbol{r}
		\int\limits_V\mathrm{d}\boldsymbol{r}'
		\int\limits_t^{t+T}\mathrm{d}t\,\sum\limits_{ij}\big[\nabla_{\boldsymbol{r}}\cdot{\boldsymbol{\Lambda}}_i({\boldsymbol{r}},t),\nabla_{{\boldsymbol{r}}'}\cdot{\boldsymbol{\Lambda}}_j({\boldsymbol{r}}',t)\big]\frac{\delta}{\delta\phi_i({\boldsymbol{r}})}\mu_j({\boldsymbol{r}}',t)\,,
	\end{align}
 \end{widetext}
 where $[\cdot,\cdot]$ in the second line denotes the total variation~\cite{gardiner2009}.
	Next, we take the noise average of this equation and simplify it using the following identities
	\begin{align}
		[\nabla\cdot{\boldsymbol{\Lambda}}_i({\boldsymbol{r}},t),\nabla\cdot{\boldsymbol{\Lambda}}_j({\boldsymbol{r}}',t)]&=-\delta_{ij}\nabla^2_{\boldsymbol{r}}\delta({\boldsymbol{r}}-{\boldsymbol{r}}')\,,
		\\
		\frac{\delta}{\delta\phi_i({\boldsymbol{r}})}\mu_j({\boldsymbol{r}}')&=\frac{\partial\mu_j}{\partial\phi_i}({\boldsymbol{r}}')\delta({\boldsymbol{r}}-{\boldsymbol{r}}')\,,
		\\
		\frac{\delta}{\delta\phi_i({\boldsymbol{r}})}\nabla_{{\boldsymbol{r}}'}\cdot{\boldsymbol{J}}_j^{\mathrm{d}}({\boldsymbol{r}}')&=\frac{\delta}{\delta\phi_i({\boldsymbol{r}})}\nabla^2_{{\boldsymbol{r}}'}\mu_j({\boldsymbol{r}}') \nonumber \\
  &=\frac{\partial\mu_j}{\partial\phi_i}({\boldsymbol{r}}')\nabla^2_{{\boldsymbol{r}}'}\delta({\boldsymbol{r}}-{\boldsymbol{r}}')\, .
	\end{align}
	This gives rise to the following expression
	\begin{align}\label{eq:EPRgeneral}
		{\mathcal{S}}(t)=&\int_V\mathrm{d}\mathbf{{\boldsymbol{r}}}
		\frac{\sum_i\langle \vert {{\boldsymbol{J}}}_i^\mathrm{d}({\boldsymbol{r}},t)\vert^2 \rangle}{\epsilon}\nonumber \\
		&+\int_V\mathrm{d}\mathbf{r}\sum_i\left\langle\frac{\delta}{\delta{\phi_i(\boldsymbol{r},t)}}\nabla\cdot {\boldsymbol{J}}_i^\mathrm{d}({\boldsymbol{r}},t) \right\rangle
		\,,
	\end{align}
	which is understood to be regularized by a UV cutoff to suppress unphysical field degrees of freedom~\cite{suchanek2023irreversible}.
 \red{
Otherwise, the entropy production in Eq.~\eqref{eq:EPRgeneral} would contain an infinite number of contributions from the small wavelength regime, where field theories loose their predictive power.
Thus, introducing the UV-cutoff is a possibility to obtain a physically pertinent measure of irreversibility, respecting the limited resolution of the model.}

We note that a common type of models within the general class \red{defined in Eq.}~\eqref{equ:moda} are those where the chemical potential can be decomposed into an equilibrium part, derived from a free energy functional $\mathcal{F}$, and a \red{generic nonequilibrium part} $\mu^\mathrm{a}$:

 \begin{align}\label{equ:minimal}
     \mu_i=\frac{\delta\mathcal{F}}{\delta\phi_i}+ \mu^\mathrm{a}_i \, .
 \end{align}
 \red{In the context of active field theories, $\mu^\mathrm{a}$ is denoted ``active part.''}
 It has been proposed in Refs.~\cite{Wittkowski_2014,cates2019active} that this construction of the chemical potential 
 yields a method for generating mesoscopic minimal models for various forms of nonequilibrium. 
 Substituting this form of the chemical potential in Eq.~\eqref{equ:inth}, we find that the entropy production along a trajectory also decomposes into 
 {an ``active part''}
 that takes the same form as in Eq.~\eqref{equ:inth}, with $\mu$ replaced by $\mu^\mathrm{a}$, and an additional boundary term $\mathcal{F}(t+T)-\mathcal{F}(t)$ which does not contribute in the steady state. 
 Hence, along similar lines as above, we find that the general expression for the steady-state entropy production rate then reads
 \begin{align}\label{eprminimal}
\mathcal{S}=&\int_V\!\mathrm{d}\boldsymbol{r}
\frac{\sum_i\left\langle \boldsymbol{J}^\mathrm{d}_i\nabla{\mu}^\mathrm{a}_i\right\rangle}{\epsilon}-
\int_V\!\mathrm{d}\boldsymbol{r}\sum_i\left\langle\frac{\delta}{\delta{\phi_i}}\nabla^2\mu_i^\mathrm{a} \right\rangle
    \,.
\end{align}
Throughout this paper, we will however retain the more general form of non-Hermitian field theories given in Eq.~\eqref{equ:moda}. The representation in terms of the total deterministic current in Eq.~\eqref{equ:moda} allows a general connection with bifurcation theory to be established\red{, as we demonstrate in the following sections.}

	\subsection{Entropy production in dynamical phases}\label{sec:eprtr}
	First, we consider the steady-state entropy production associated with a 
	dynamical (i.e., time-dependent) state. Making use of the standard procedure of the small noise expansion \cite{gardiner2009}, we express \red{the steady-state dynamical solution} as 
    \begin{align}\label{snexpansion}
        \phi=\phi^*+\sqrt{\epsilon}\Delta\phi+\mathcal{O}(\epsilon)\,,
    \end{align}
    where ${\phi}^*(t)\equiv{\phi}(t)\big\rvert_{\epsilon=0}$ corresponds to the \red{deterministic steady-state solution of the zero-noise case and $\Delta\phi$ is the linear response to the Gaussian noise.} We similarly expand the deterministic current up to its \red{leading $\mathcal{O}(\sqrt{\epsilon})$ fluctuations due to the field fluctuations $\Delta\phi$} as
	\begin{align}\label{eq:detcur}
		\boldsymbol{J}^{\mathrm{d}}_i\big[{\phi}(\boldsymbol{r},t)\big] =&~\boldsymbol{J}_i^{\mathrm{d}}\big[{\phi}^*(\boldsymbol{r},t)\big]\nonumber\\
  &+\sqrt{\epsilon}\sum\limits_j\int\limits_V\mathrm{d}\boldsymbol{r}'\frac{\delta\boldsymbol{J}_i^{\mathrm{d}}(\boldsymbol{r},t)}{\delta{\phi_j}^*(\boldsymbol{r}',t)}{\Delta\phi_j}(\boldsymbol{r}',t)\,.
	\end{align}
	Within the scheme of the small noise expansion, the deterministic solution as well as the statistics of $\Delta\phi$ are independent of the noise intensity $\epsilon$. Further, $\Delta\phi$ is Gaussian with zero mean~\cite{gardiner2009,Borthne2020} so that  $\langle f(\phi^*){\Delta\phi}\rangle=0$. Hence, by combining Eqs.~\eqref{eq:detcur} and \eqref{eq:EPRgeneral}, we find the asymptotic expression
	\begin{align}
		\mathcal{S}&=\epsilon^{-1}\sum\limits_i
		\int\limits_V\mathrm{d}\boldsymbol{r} \big\vert\boldsymbol{J}_i^{\mathrm{d}}(\phi^*(\boldsymbol{r},t))\big\vert^2
		+\mathcal{O}(\epsilon^0)\,,
\label{S_dynamicalPhase_SM}
	\end{align}
which shows that the dominant contribution to  ${\mathcal{S}}$ in the dynamical phase scales like $\mathcal{S}\sim \epsilon^{-1}$.  It thus grows unboundedly in the zero-noise limit, \red{thereby denouncing the dissipative character of the dynamical state.}
 Our analysis of a paradigmatic nonreciprocal model in the two companion papers~\cite{suchanek2023irreversible,suchanek2023entropy}, recovers this scaling. It is further consistent with the findings from Ref.~\cite{Borthne2020} for a polar flocking model, which reported $\mathcal{S}\sim \epsilon^{-1}$ in the state of collective motion.
  
	We now consider, as two typical representatives of dynamical phases, traveling waves and oscillating patterns.
In the first case $\phi_i^*(\boldsymbol{r},t)\equiv\varphi_i(\boldsymbol{r}-\boldsymbol{v}t)$, and Eq.~\eqref{equ:moda} yields $\boldsymbol{J}^\mathrm{d}(\varphi)=\boldsymbol{v}\varphi$ for $\epsilon\rightarrow 0$. Hence, Eq.~\eqref{S_dynamicalPhase_SM} implies 
\begin{align}\label{equ:eprpattern}
 \mathcal{S}= \epsilon^{-1}\vert \boldsymbol{v}\vert^2\sum_i
    \int_V\mathrm{d}\mathbf{r}
 \vert \varphi_i\vert^2 +\mathcal{O}(\epsilon^0).
 \end{align}
Similarly, for an oscillating state of the form $\phi_i^*(\boldsymbol{r},t)=\sin({t}/{\tau^*}+\theta_i)\varphi_i(\boldsymbol{r})$ the time-averaged entropy production is
	\begin{align}\label{equ:eprosc}
		\mathcal{S}=\epsilon^{-1}  \left(\frac{1}{\sqrt{2}\tau^*}\right)^2 \sum_i\int\limits_V\mathrm{d}\mathbf{r}
		\big\vert\nabla^{-1} \varphi_i\big\vert^2 +\mathcal{O}(\epsilon^0)\,,
	\end{align}
 \red{with $\nabla^{-1}\equiv\nabla\nabla^{-2}$~\footnote{\red{Note that, for domains with periodic boundaries, the inverse Laplacian of a scalar field $\varphi(\boldsymbol{r},t)$, with $\int_V\mathrm{d}\boldsymbol{r}\varphi(\boldsymbol{r},t)=0$, is uniquely defined up to \red{a} constant. Thus, the operator $\nabla^{-1}$ is well defined. For a given mass conserving flux $\dot{\phi}$ it precisely returns the current $\boldsymbol{J}$ such that $\dot{\phi}=\nabla\cdot\boldsymbol{J}$, with the gauge choice of $\boldsymbol{J}$ being curl-free and $\int_V\mathrm{d}\boldsymbol{J}=0$}.}.}

 We conclude that, in both cases, the entropy production rate is related to the squared inverse of the characteristic timescale of the deterministic motion.

 Recalling the probabilistic interpretation of $\mathcal{S}$ in Eq.~\eqref{def:Pathprobratio-S}, it measures the uncertainty about the direction of time, given a stochastic trajectory. 
The dynamical phase exhibits a deterministic flux with a direction encoded in the average system configurations, which means that, on the average level, each piece of a trajectory expresses the arrow of time.
The remaining uncertainty comes from the fact that the noise current, $\sqrt{2\epsilon}\boldsymbol{\Lambda}$, can transiently invert the deterministic mass current (which is $\boldsymbol{v}\varphi$ or $\nabla^{-1} \varphi_i/\tau^*$, respectively).
According to Eqs.~\eqref{equ:eprpattern} and \eqref{equ:eprosc}, this uncertainty is then simply given by the ratio of the squared mass and noise currents, integrated over the observed spatial volume.

\subsection{TRSB at phase transitions}\label{sec:eprsing}

Next, we study the general behavior of $\mathcal{S}$ in the vicinity of phase transitions. In particular, we want to address the question, which types of singularities of $\mathcal{S}^*$ can occur and how they are related to the nature of the transition.
	
\subsubsection{The squared deterministic current}	
	As an important first step, we derive an expression for the entropy production rate that allows to draw a connection to the notion of bifurcations (introduced in Sec.~\ref{sec:overview-transitions}). Again, we aim to calculate the averages in Eq.~\eqref{eq:EPRgeneral} within the small noise expansion~\cite{gardiner2009}. First, we consider the second term, which is the contribution from the functional derivative of $\nabla\cdot\boldsymbol{J}^\mathrm{d}$ and find
	\begin{align}\label{equ:eprl}
{\mathcal{S}}=&\int_V\mathrm{d}\mathbf{{\boldsymbol{r}}}\frac{\sum_i\big\langle \vert {{\boldsymbol{J}}}_i^\mathrm{d}({\boldsymbol{r}},t)\vert^2 \big\rangle}{\epsilon}+
\nonumber\\&\int_V\mathrm{d}\mathbf{r}\sum_i\left\langle\frac{\delta}{\delta{\phi^*_i(\boldsymbol{r},t)}}\nabla\cdot {\boldsymbol{J}}_i^\mathrm{d}({\boldsymbol{r}},t) \right\rangle+\mathcal{O}(\epsilon)\,.
	\end{align}
 Inserting the definition of the Jacobian $\mathcal{J}$ of $F$ from Eq.~\eqref{def:Jacobian}, we obtain for the asymptotic form ${\mathcal{S}^*}\equiv\lim_{\epsilon\rightarrow0}\mathcal{S}$  of $\mathcal{S}$,
	\begin{align}\label{equ:ptrate}
		\mathcal{S}^*=
\mathrm{SCNR}+\mathrm{tr}\mathcal{J}\,,
	\end{align}
where 
\begin{align}\label{defSCNR}
     \mathrm{SCNR}\equiv\lim\limits_{\epsilon\rightarrow 0}\int_V\mathrm{d}\mathbf{{\boldsymbol{r}}}
		\frac{\sum_i\big\langle \vert {{\boldsymbol{J}}}_i^\mathrm{d}({\boldsymbol{r}},t)\vert^2 \big\rangle}{\epsilon}		
 \end{align}
is the average of the integrated \textit{squared-deterministic current to noise ratio}. 

To analyze the qualitative behavior of ${\mathcal{S}^*}$ at a transition, we inspect both terms in Eq.~\eqref{equ:ptrate}, separately. 
Denoting the eigenvalues of $\mathcal{J}$ by $\{\lambda_{i}\}$, 
 the trace of the Jacobian can be expressed as
		\begin{align}\label{equ:RealOperatorIdentity}
		\mathrm{tr}\mathcal{J}=
		\sum\limits_i\mathrm{Re}\lambda_i<0\,,
	\end{align}
 where we have used that $\mathcal{J}$ is a real operator and $\phi^*$ is a linearly stable fixed point.
	Therefore, this term will, in any case, remain finite toward the transition. 
 As implied by the inequality~\eqref{equ:RealOperatorIdentity}, the $\mathrm{SCNR}$ \red{defined in Eq.~\eqref{defSCNR}} is, in general, an upper bound to $\mathcal{S}^*$ ($\mathcal{S}^*<\mathrm{SCNR}$), and the decisive contribution when approaching the transitions. 
 Concretely, expanding the average in Eq.~\eqref{defSCNR} to lowest order in $\epsilon$, we can deduce  from Eq.~\eqref{equ:ptrate} the behavior of $\mathcal{S}^*$.

\subsubsection{TRSB and characteristic eigenvalues}\label{sec:connection}

To fully unravel the asymptotic form of $\mathcal{S^*}$ as a transition is approached, we now derive an equation that expresses the $\mathrm{SCNR}$ [defined in Eq.~\eqref{defSCNR}] exclusively in terms of the eigenvalues and eigenvectors of the Jacobian, or linearized dynamical operator, $\mathcal{J}$ \red{from Eq.~\eqref{def:Jacobian}}. 
	\red{As a first step, we define the $d$-dimensional Fourier components of the deterministic current
\begin{align} 
\left[\boldsymbol{J}_i^{\mathrm{d}}(t)\right]^{k}\equiv V^{-1}\int_{V}\mathrm{d}\boldsymbol{r}{{\boldsymbol{J}}}_i^\mathrm{d}({\boldsymbol{r}},t)\,\mathrm{exp}\{-i\boldsymbol{q}^{k}\boldsymbol{r}\}, 
 \end{align}
 }with $\boldsymbol{q}^{k}={(2\pi)^{d}}V^{-1}(q^k_1,\dots,q^k_d)$.
 In Fourier space, the integral over the average squared deterministic current in Eq.~\eqref{defSCNR} can then be expressed as the sum \begin{align}\label{equ:intsdc}
	&	\int_V\mathrm{d}\mathbf{{\boldsymbol{r}}}
		{\sum_i\left\langle \left\vert {{\boldsymbol{J}}}_i^\mathrm{d}({\boldsymbol{r}},t)\right\vert^2 \right\rangle}
  =\sum_{k }V\sum_i\left\langle\big\vert 
  \right[\boldsymbol{J}_i^{\mathrm{d}}(t)\left]^{k}\right\vert^2\big\rangle \,.
\end{align} 
To proceed further, we will first only consider the case where $\mathcal{J}$ is \textit{diagonal} in Fourier space. 
The case of {nondiagonal} $\mathcal{J}$ is considered separately below, in Sec.~\ref{sec:nondiagonal-J}.
\red{ Given this assumption, the Fourier representation of the small noise expansion~\cite{gardiner2009} of the deterministic current in Eq.~\eqref{eq:detcur} reads}
\red{
 \begin{align}\label{currentrepr}
		 (\boldsymbol{J}_i^{\mathrm{d}})^{k}
=-i\sqrt{\epsilon} \boldsymbol{q}^{k}\sum_j\mathcal{J}_{ij}^k\Delta{\phi}_j^k+\mathcal{O}(\epsilon)\,,
	\end{align}
}which involves the \red{Fourier representations} of the $\mathcal{O}(\sqrt{\epsilon})$-coefficient of the expansions in Eq.~\eqref{snexpansion}, 
\red{
\begin{align}
\Delta{\phi}^k_i\equiv V^{-1}\int_{V}\mathrm{d}\boldsymbol{r}\Delta\phi_i\,\mathrm{exp}\{-i\boldsymbol{q}^{k}\boldsymbol{r}\}
\end{align}
and the Jacobian
\begin{align}
 \mathcal{J}_{ij}^k\equiv V^{-1} \int_V\!\mathrm{d}\boldsymbol{r}'\int_V\!\mathrm{d}\boldsymbol{r}\frac{\delta\boldsymbol{J}_i^{\mathrm{d}}(\boldsymbol{r},t)}{\delta{\phi_j}^*(\boldsymbol{r}',t)}\,\mathrm{exp}\{i\boldsymbol{q}^{k}(\boldsymbol{r}'-\boldsymbol{r})\}.
\end{align}
}
Due to the diagonal form of $\mathcal{J}$, the fluctuations in each Fourier wavenumber can be treated independently.
Hence, all wavenumbers affected by an arising instability can be treated separately, too, so that we can assume, without loss of generality, that the considered instability occurs in a single wavenumber $l$, only. 
Importantly, if $\mathcal{J}^{k}$ is real and \textit{non-Hermitian},
$\mathcal{J}^{k}$ is automatically not normal, which implies that the set of its eigenvectors is not orthogonal~\cite{Ashida2020}.
However, we assume that for all considered types of transition scenarios, the eigenvectors still form a basis away from the instability. Thus, $\mathcal{J}^{k}$ has the diagonal  representation 
	\begin{align}\label{equ:digrep}
		\big(\mathcal{J}^{k}\big)'_{ij}=\sum_{nm}\big(T^{k}\big)^{-1}_{in}\mathcal{J}^{k}_{nm}T^{k}_{mj}=\lambda^k_i\delta_{ij}\,.
	\end{align}
Here, we introduced the transformation matrix $T^k$, given by the column matrix of right eigenvectors.
	The transformation to the eigenbasis coordinates reads
	\begin{align}\label{equ:bchange}
		\Delta{\psi}^{k}_i=\sum_j\big(T^{k}\big)^{-1}_{ij}\Delta{\phi}^{k}_j\,.
	\end{align}
Combining Eqs.~\eqref{equ:bchange} and \eqref{equ:digrep} and inserting the identity $T^{-1}T=TT^{-1}=\mathrm{id}$, we find that the average value of the squared deterministic current can be expressed as
	\begin{align}\label{equ:z1}
		\sum_i\bigg\langle\Big\vert\sum_j\mathcal{J}_{ij}^k\Delta{\phi}_j^k\Big\vert^2\bigg\rangle=\sum\limits_{ijn}\bar\lambda^k_{i}\bar T^k_{ni}T^k_{nj}\lambda^k_j\left\langle\Delta{\psi}^{-k}_{i}\Delta{\psi}^{k}_j\right\rangle\,,
	\end{align}
where the overbars indicate {element-wise} complex conjugation. Next, we evaluate the 
cross-correlations appearing in \red{Eq.~}\eqref{equ:z1}, using the dynamical equation for $\Delta\phi$, which is given by the linearization of Eq.~\eqref{equ:moda} and reads
 \begin{align}\label{equ:Fdef}
	\partial_t\Delta\phi^k_i(t)=\sum_j\mathcal{J}_{ij}^k\Delta{\phi}_j^k+i\boldsymbol{q}_k\cdot\sqrt{2}\boldsymbol{\Lambda}_i^k(t)\,,
	\end{align} 
 with $\langle\boldsymbol{\Lambda}_i^k(t)\boldsymbol{\Lambda}_j^{k'}(t')\rangle=V^{-1}\delta_{ij}\delta_{k,k'}\delta(t-t')$.
We apply the It\^o formula to obtain the dynamical equation for $\Delta{{\psi}^{-k}_i} \Delta{\psi}^k_j$ and take the noise average, which reads 
 \begin{align}\label{equ:steadystate}
\partial_t\langle \Delta{{\psi}^{-k}_i} \Delta{\psi}^k_j\rangle=&
(\bar\lambda^k_i+\lambda^k_j)\langle\, \Delta\psi_{i}^{-k}\Delta\psi_{j}^{k}\,\rangle 
\nonumber\\
  &
+
  \frac{2 \vert\boldsymbol{q}^{k}\vert^{2}}{V}\sum\limits_{n} \left(\bar T^k\right)^{-1}_{in}\left(T^k\right)_{jn}^{-1}\, .
 \end{align}
 Imposing the stationarity condition
 $\partial_t\langle \Delta{{\psi}^{-k}_i} \Delta{\psi}^k_j\rangle=0$, we find
 \begin{align}\label{itoaverage}
    \left\langle\, \Delta\psi_{i}^{-k}\Delta\psi_{j}^{k}\,\right\rangle 
=
  \frac{2 \vert\boldsymbol{q}^{k}\vert^{2}}{V}\sum\limits_{n} \frac{\left(\bar T^k\right)^{-1}_{in}\left(T^k\right)_{jn}^{-1}}{(\bar\lambda^k_i+\lambda^k_j)}\, . 
 \end{align}
Combining Eqs.~\eqref{equ:intsdc},\,\eqref{currentrepr},\,\eqref{equ:z1}, and \eqref{itoaverage}, we arrive at the central result of this paper,
\begin{subequations}\label{eq:main}
\begin{eqnarray}\label{equ:eppt2}
	\mathrm{SNCR}
=\sum_k\zeta^k\,,
	\end{eqnarray}
	with
 \begin{align}\label{equ:defzeta}
     \zeta^k&=-
		\sum\limits_{ij}2\frac{\bar\lambda^k_i\lambda^k_j}{\bar\lambda^k_i+\lambda^k_j}\frac{C^k_{ij}}{\vert\det T^k\vert^2}\, ,
  \\
		C^k_{ij}&=\sum\limits_{nm}\bar T^k_{mi}T^k_{mj}\left( (\bar T^k)^{-1}_{in}\right)\left(T^k\right)^{-1}_{jn}\left\vert\det T^k\right\vert^2
		  \nonumber
		\\
  &=
  \sum\limits_{nm}\bar T^k_{mi}T^k_{mj}\left(\mathrm{adj} ( \bar T^k)^T\right)_{ni}\left(\mathrm{adj}(T^k)^T\right)_{nj}
  \nonumber
		\\
  &=\left((\bar T^k)^TT^k\right)_{ij}\left(\mathrm{adj} \bar T^k\mathrm{adj} (T^k)^T\right)_{ij}\,.\label{equ:defC}
	\end{align}  
Recalling Eqs.~\eqref{equ:ptrate} and \eqref{equ:RealOperatorIdentity}, the entropy production rate in the zero-noise limit can hence be expressed as
\begin{align}\label{equ:sdecomfin}
    \mathcal{S}^*= \sum_k\zeta^k+\sum\limits_i\mathrm{Re}\lambda_i \, .
\end{align}
 \end{subequations}
The above formulae~\eqref{eq:main} express the SCNR \red{[defined in Eq.~\eqref{defSCNR}]} and thus $\mathcal{S}^*$
exclusively in terms of properties of the eigensystem of $\mathcal{J}$. We note that the value of $\zeta^k$ is determined by the spectrum and the \textit{geometry} of the eigenvectors  of $\mathcal{J}$. This provides
insight into the fundamental connection between global symmetries, dynamical properties, and the fluctuations at phase transitions of systems of \red{the type of Eq.~\eqref{equ:moda}.}

A first implication is that the contributions $\zeta^{k}$ of all wavenumbers $k\neq l$
to the SCNR and $\mathcal{S}^*$ generally remain finite. The only possible exception is the contribution from 
the unstable mode $l$, for which the denominator of Eq.~\eqref{equ:defzeta} can, in principle, become zero. This wavenumber thus deserves special attention.

As a reference, let us now first explicitly consider the situation for systems for which  $\mathcal{J}^{k}$ is {Hermitian}. This holds for models \red{of the type of Eq.~\eqref{equ:moda}} if ${\mu}$ has a gradient representation, thus, in particular, for any system in thermal equilibrium. Then, $T^k$ is orthogonal, resulting in $C^k_{ij}=\delta_{ij}$. Hence, $\zeta^k=-\sum_i\lambda_i^k$ for all $k$, including $k=l$.
Equations~\eqref{equ:eppt2} and~\eqref{equ:sdecomfin} then imply that, for Hermitian systems of \red{the type of Eq.~\eqref{equ:moda}},
the SCNR remains  {strictly} finite, even at a phase transition, so that, in general, $\mathcal{S}^* \equiv 0$. 
 
In contrast, for systems with a \textit{non-Hermitian} $\mathcal{J}^{k}$, 
the SCNR can, in principle, become singular at the transition, as can be already anticipated from the form of the $\zeta^k$ \red{in Eq.~\eqref{equ:defzeta}}. With regard to this aspect, we now 
examine the three different transition scenarios presented in Sec.~\ref{sec:overview-transitions}.
To study them, we consider paths in the parameter space of  Eq.~\eqref{equ:moda} that end at the respective transition points. Naturally, the latter are parameterized by the real part $w=\mathrm{Re}(\lambda^l_i)$ of the eigenvalue \red{of $\mathcal{J}^{l}$}
that vanishes at the transition. Accordingly, in the following, $T^k(w)$ and $\{\lambda^k_n(w)\}$ are always assumed to depend on this single parameter $w$. For sake of readability, we explicitly state the argument $w$ in the following only when we evaluate a quantity at the \red{transition} point $w=0$.

\subsubsection{TRSB near conventional critical points}\label{sec:typical}

Let us first consider transitions where a \textit{single} mode  $\hat{e}^l_0$ with eigenvalue $\lambda^l_0$ becomes unstable at the transition, while the eigenvectors form a basis, yielding $\vert \det T^l(0) \vert \neq 0$. We refer to these transitions as ``conventional'' critical points. Examples of this type of transition are the second-order transitions that commonly occur in thermal equilibrium, and also their nonequilibrium analogues. We assume that the deterministic solution $\phi^*$ has the full symmetry of the system, thus, away from any transition all eigenvalues \red{of the Jacobian} have strictly negative real parts~\cite{Goldstone62}. 
From the structure of Eq.~\eqref{equ:defzeta}, it is directly apparent that in this case, also the contribution from the unstable mode with wavenumber $l$ to the entropy production must remain finite at the transition. Thus, $\mathcal{S}$ is in general regular across such critical transitions.
We emphasize that this is the case even though the \red{ typical strength of $\phi$ fluctuations in \red{response to} the noise, measured by}
\red{
	\begin{align}\label{eq:covphi}
		\mathrm{tr}\mathrm{Cov}(\phi) \equiv \sum_i\int_V\mathrm{d}\boldsymbol{r}\langle \vert\phi_i-\phi_i^*\vert^2\rangle\,,
	\end{align}
 }\red{surges dramatically} and the \red{static} susceptibility
\begin{align}\label{eq:sus}
    \chi\equiv\lim_{\epsilon\rightarrow 0}\epsilon^{-1} \mathrm{tr}\mathrm{Cov}(\phi)\,,
\end{align}
\red{which measures the \red{response in the low noise regime}, diverges. In analogy to equilibrium phase transitions, the divergence of $\chi$ serves as an indicator of the phase transition.} \red{This entirely different behavior of $\mathcal{S}$ and $\chi$ is \red{exemplified} in Fig.~\ref{fig:pre_2}(a,d) for a conventional critical transition in the paradigmatic nonreciprocal Cahn-Hilliard model~\cite{suchanek2023entropy}.}

\red{We note that }the divergence of the susceptibility can be \red{formally} seen from
	\begin{align}\label{equ:fluc}
		\chi
		&=V\sum_k\sum\limits_i \left\langle\, \Delta \phi_i^{-k}\Delta\phi_i^{k}\,\right\rangle
\nonumber
\\
&=		
  -\sum_k 2\left\vert\boldsymbol{q}^{k}\right\vert^{2}\sum\limits_{ijnm}\frac{\bar T^k_{ni}T^k_{nj}\left(\bar T^k\right)^{-1}_{im}\left(T^k\right)^{-1}_{jm}}{\bar \lambda^k_i+\lambda^k_j}\, ,
	\end{align}
which is dominated by
 \begin{align}
     \chi^{l} \sim& 
		\, -\frac{\big\vert\boldsymbol{q}^{l}\big\vert^{2}}{\lambda^l_0}\sum\limits_{nm}{\bar T^l_{n0}(0)T^l_{n0}(0)(\bar T^l)^{-1}_{0m}(0)(T^l)^{-1}_{0m}(0)}
  \nonumber
\\
&=
  -\frac{\big\vert\boldsymbol{q}^{l}\big\vert^{2}}{\lambda^l_0}\frac{C^l_{00}(0)}{\vert\det T^l(0)\vert^2}\,,~~~\text{as}~
  \mathrm{Re}\lambda^l_0\rightarrow0 \, .
 \end{align}
Since Eq.~\eqref{equ:defC} implies that $C^k_{ii}>0$, the numerator in the above expression is strictly positive, causing the divergence of $\chi$ \red{when the eigenvalue $\lambda_0$ in the denominator vanishes}.

\subsubsection{TRSB near oscillatory instabilities}
\label{sec:EPRosc}
 
Now we address oscillatory instabilities, which are characterized by the occurrence of unstable modes whose eigenvalues retain a nonzero imaginary part across the transition. These can only occur in non-Hermitian dynamics, as the spectrum of Hermitian operators is completely real. Since $\mathcal{J}$ is real, according to our assumption of a $\mathcal{P}$-symmetric dynamical operator, the modes involved in the transition come, even in the non-Hermitian case, as pairs with complex conjugate eigenvalues
\cite{jeffrey2005complex}. Thus, at the oscillatory instability, there is always a conjugate pair of unstable monochromatic modes of wavenumber $l$, whose eigenvalues we denote by 
 \begin{align}
     \lambda^l_{i_\pm}=-\sigma\pm i\omega.
 \end{align} 
 Furthermore, all eigenmodes are assumed to remain linearly independent across the transition, resulting in $\vert \det T(0) \vert  \neq 0$~\footnote{Otherwise, the transition would correspond to an exotic combination of CEP and oscillatory instability, which we do not consider here.}.  We further assume that the instability is a primary one, i.e., the real parts of all eigenvalues are nonzero and negative, away from the transition.
 
Splitting the sum in Eq.~\eqref{equ:defzeta} into a diagonal and an off-diagonal part, we find
	\begin{align}\
	\!\!	\zeta^l&=-
		\sum\limits_{i}\frac{(\mathrm{Re}\lambda_i^l)^2+(\mathrm{Im}\lambda_i^l)^2}{\mathrm{Re}\lambda^l_i}\frac{C^l_{ii}}{\vert\det T^l\vert^2}
  \nonumber
\\
&-
		\sum\limits_{i\neq j}\frac{2\bar \lambda^l_i\lambda^l_j}{\mathrm{Re}\lambda^l_i+\mathrm{Re}\lambda^l_j+i(\mathrm{Im}\lambda^l_j-\mathrm{Im}\lambda^l_i)}\frac{C^l_{ij}}{\vert\det T^l\vert^2}\,.
	\end{align}
	Therefore, according to Eqs.~\eqref{equ:ptrate} and \eqref{equ:eppt2}, close to such a transition, the asymptotic form of $\mathcal{S}^*$ is given by
	\begin{align}\label{equ:eprsosc2}
 \mathcal{S}^*\sim
		\frac{\omega^2}{\sigma}\frac{C^l_{i_+i_+}(0)+C^l_{i_-i_-}(0)}{\vert\det T^l(0)\vert^2}\,,~\text{as}~
  \sigma\rightarrow0\,.
	\end{align}
Again, since Eq.~\eqref{equ:defC} implies $C^k_{ii}>0$, the numerator in the above expression is strictly positive. Hence, $\mathcal{S}^*$ diverges, as the transition at $\sigma=0$ is approached.
	
	The last expression can further be transformed using the susceptibility $\chi$ defined in Eq.~\eqref{eq:sus}.
 The explicit expression for $\chi$ given in Eq.~\eqref{equ:fluc} now has the scaling
 \begin{align}
     \chi\sim &
		\, \frac{\big\vert\boldsymbol{q}^{l}\big\vert^{2}}{\sigma}\sum\limits_{nm}{T^l_{ni_+}(0)T^l_{ni_+}(0)(T^l)^{-1}_{i_+m}(0)(T^l)^{-1}_{i_+m}(0)}
  \nonumber
\\
&+		\frac{\big\vert\boldsymbol{q}^{l}\big\vert^{2}}{\sigma}\sum\limits_{nm}{T^l_{ni_-}(0)T^l_{ni_-}(0)(T^l)^{-1}_{i_-m}(0)(T^l)^{-1}_{i_-m}(0)} \nonumber\\=& \frac{\big\vert\boldsymbol{q}^{l}\big\vert^{2}}{\sigma}\frac{C^l_{i_+i_+}(0)+C^l_{i_-i_-}(0)}{\vert\det T(0)\vert^2}\,,~~~\text{as}~
  \sigma\rightarrow0\,.
 \end{align}
	In combination with Eq.~\eqref{equ:eprsosc2}, we conclude that
	\begin{align}\label{equ:resosc}
		\mathcal{S}^*\sim \frac{\omega^2}{\big\vert\boldsymbol{q}^{l}\big\vert^{2}}\chi\,,~~~\text{as}~
  \sigma\rightarrow0\,.
	\end{align}
Hence, as the transition is approached, the entropy production rate is proportional 
to the susceptibility $\chi$ and therefore scales with the strength of the fluctuations. \red{This is confirmed by the numerical data for an oscillatory instability in the nonreciprocal Cahn-Hilliard model~\cite{suchanek2023entropy}, presented in Fig.~\ref{fig:pre_2}(b,e).}
The scaling $\sim\omega^2$, further reveals the \red{transient cyclic} currents with characteristic frequency $\omega$, found within a static phase in the vicinity of an oscillatory instability~\cite{Cross93,cross_greenside_2009}, as the culprit of surging entropy production. The divergence of $\mathcal{S}^*$ mirrors the fact that the amplitude of these cyclic currents becomes
systematically positive as the transition is approached.

In Ref.~\cite{suchanek2023irreversible}, we show a kymograph of the fluctuations in a paradigmatic model close to an oscillatory instability, providing striking visual evidence for the oscillatory character of the \red{fluctuating} dynamics.

\subsubsection{TRSB near critical exceptional points}\label{sec:irrexp}

	We finally turn to critical exceptional points. Recall that this scenario involves the co-aligning of two eigenvectors. This immediately implies that $\det T^l$ [appearing in Eq.~\eqref{equ:defzeta}] vanishes. Therefore, for the wavenumber of the unstable mode $l$, we need to carefully evaluate the limit of the ratio $C^l_{ij}/\vert\det T^l\vert^2$ in Eq.~\eqref{equ:defzeta}, as the transition is approached. Its denominator generally vanishes, at every exceptional point, regardless of whether it coincides with a critical point, or not. For this reason, we evaluate Eq.~\eqref{equ:defzeta} also for the more general case of an exceptional point that does not coincide with an instability. 

 First, notice that the term $(\bar \lambda^l_i\lambda^l_j)/(\bar \lambda^l_i+\lambda^l_j)$ appearing in Eq.~\eqref{equ:defzeta} in general remains finite, even at the exceptional point where $\lambda^l_i,\lambda^l_j\rightarrow 0$. This can be seen by applying 
 L'H\^{o}pital's rule~\cite{reed2003methods} to the real and imaginary parts of the expression and taking into account that, away from the exceptional point, $\mathrm{Re}\lambda^l_i<0$ for all $i$. Next, we investigate the behavior of $\det T^l$, close to an exceptional point. We choose an ordering of the eigenbasis $\big\{\hat{e}^l_0,\dots,\hat{e}^l_{N-1}\big\}$
such that $\lambda^l_0$ and $\lambda^l_1$ denote the eigenvalues of the merging eigenvectors. For ease of notation, we will  in the following partially suppress the index $l$. We further define
	\begin{align}
		\Delta\lambda &=\lambda_0-\lambda_1\,,\\
  		\lambda^* &= \lim_{\Delta\lambda\rightarrow 0}  \frac{\lambda_0+\lambda_1}{2} \,,
	\end{align}
and denote by
\begin{align}
	\hat{e}^* =\hat{e}_0\big\rvert_{\Delta\lambda=0}=\hat{e}_1\big\rvert_{\Delta\lambda=0}\,
	\end{align}
the eigenmode into which the two eigenmodes $\hat{e}_1$ and $\hat{e}_1$ merge at the exceptional point.
We further parameterize the path that ends at the exceptional point by $w=\mathrm{Re}\Delta\lambda$, such that
	\begin{align}
		T(w=0)=\begin{pmatrix}
			\hat{e}^*,\hat{e}^*,\hat{e}_2(0),\dots,\hat{e}_{N-1}(0)
		\end{pmatrix}\,.
	\end{align}
	Note that the {reduced} set of $N-1$ eigenvectors $\big\{\hat{e}^*,\hat{e}_2(0),\dots,\hat{e}_{N-1}(0)\big\}$ is still linearly independent at the exceptional point.
	
Now, in order to investigate the properties of $\det T$ close to the exceptional point, we aim to expand it in orders of $\Delta\lambda$.
 For this purpose, it is convenient to define
	\begin{align}
		\hat{e}_c&=\hat{e}_0+\hat{e}_1\,,\\
		\Delta\hat{e}&=\hat{e}_0-\hat{e}_1\,,\\
		\mathcal{L}&=\mathcal{J}^{l}\,,
	\end{align}
 and to employ the Laplace expansion for determinants \cite{poole2014linear}, from which we find
	\begin{align}
		\det T&=\det\Big( \hat{e}_c+\frac{\Delta\hat{e}}{2},\hat{e}_c-\frac{\Delta\hat{e}}{2},\hat{e}_2,\dots,\hat{e}_{N-1}\Big)\nonumber \\
		&=\det\big( {\Delta\hat{e}},\hat{e}_c,\hat{e}_2,\dots,\hat{e}_{N-1}\big)\,.
	\end{align}
Accordingly, we obtain for the linear-order coefficient of the expansion of $\det T$ in $\Delta\lambda$,
	\begin{align}\label{equ:detscal}
		 &\partial_{\Delta\lambda}\det T\big\rvert_{\Delta\lambda=0}
   \nonumber
   \\
 &  =\det\left( \partial_{\Delta\lambda}\Delta\hat{e}(0),\hat{e}^*,\hat{e}_2(0),\dots,\hat{e}_{N-1}(0)\right).
	\end{align}
	This coefficient is nonzero, as long as $\partial_{\Delta\lambda}\Delta\hat{e}(0)$ is nonzero and does not lie in the  $({N}-1)$-dimensional subspace spanned by $(\hat{e}^*,\hat{e}_2(0),\dots,\hat{e}_{N-1}(0))$.
	In the following lines, we prove by contradiction that both conditions are satisfied.
 
First, combining the characteristic equations for $\hat{e}_0$ and $\hat{e}_1$, we find
	\begin{align}\label{equ:char}
    \mathcal{L}\Delta\hat{e}=(\Delta\lambda)\hat{e}_c+(\lambda_0+\lambda_1)\Delta\hat{e}\,.
	\end{align}
	Taking the derivative with respect to $\Delta\lambda$ and the limit
	$\Delta\lambda\rightarrow0$, we obtain
	\begin{align}\label{equ:char0}
		[\mathcal{L}(0)-\lambda^*]\partial_{\Delta\lambda}\Delta\hat{e}(0)=\hat{e}^*\,.
	\end{align}
	This already yields $\partial_{\Delta\lambda}\Delta\hat{e}(0)\neq0$. Further, assuming that $\partial_{\Delta\lambda}\Delta\hat{e}(0)$ lies in the subspace spanned by $\big\{\hat{e}^*,\hat{e}_2(0),\dots,\hat{e}_{N-1}(0)\big\}$ implies that there would exist a tupel of complex numbers $(a_0,a_2,\dots,a_{N-1})\neq0$ such that $\partial_{\Delta\lambda}\Delta\hat{e}(0)=a_0\hat{e}^*+\sum_{i=2}^{N-1}a_i\hat{e}_i(0)$. Inserting this into Eq.~\eqref{equ:char0}, yields
	\begin{align}
		\sum\limits_{i=2}^{N-1}(\lambda_i(0)-\lambda^*)a_i\hat{e}_i(0)=\hat{e}^*\,,
	\end{align}
	which implies, as a contradiction, the linear dependence of $\big\{\hat{e}^*,\hat{e}_2(0),\dots,\hat{e}_{N-1}(0)\big\}$.
	Therefore, it must hold that
 \begin{align}
		\vert \partial_{\Delta\lambda}\det T\big\rvert_{\Delta\lambda=0}\vert \neq 0\,.
	\end{align}
Further, we can bring the left-hand side of Eq.~\eqref{equ:detscal} into a more convenient form, by defining
	\begin{align}
		(\hat{e}_{\perp})_i=\frac{(-1)^iM_{0,i}(0)}{\sqrt{\sum\limits_{i}\vert M_{0,i}(0)\vert^2}}\,,
	\end{align}
 where $M$ denotes the set of minors  of $T$~\cite{poole2014linear}.
 This unit vector has the unique property of being perpendicular to the $({N}-1)$-dimensional subspace spanned by the eigenbasis at the CEP. This property is inferred from
 \begin{align}\label{equ:perp}
     \hat{e}_\perp{\cdot}\hat{e}_i=\det\big(\hat{e_i}(0),\hat{e}_0,\hat{e}_2(0),\dots,\hat{e}_{N-1}(0)\big)=0 \,.
 \end{align}
Using this property, we find
\begin{align}
		&{\big\vert\det( \partial_{\Delta\lambda}\Delta\hat{e}(0),\hat{e}^*,\hat{e}_2(0),\dots,\hat{e}_{N-1}(0))\big\vert^2}
  \nonumber \\
  &={\big\vert \hat{e}_\perp {\cdot} \partial_{\Delta\lambda}\Delta\hat{e}(0)\big\vert^2\big\vert\det(\hat{e}_\perp,\hat{e}^*,\hat{e}_2(0),\dots,\hat{e}_{N-1}(0))\big\vert^2} \nonumber \\
  &={\big\vert \hat{e}_\perp {\cdot} \partial_{\Delta\lambda}\Delta\hat{e}(0)\vert^2\sum\limits_{i}\vert M_{0,i}(0)\big\vert^2}\,.
 \end{align}
Taken together, we conclude that the relevant term in the denominator of Eq.~\eqref{equ:defzeta} can be expanded as
 \begin{align}\label{exp1}
     \vert \det T\vert^2=&{\big\vert \hat{e}_\perp {\cdot} \partial_{\Delta\lambda}\Delta\hat{e}(0)\big\vert^2\sum\limits_{i}\big\vert M_{0,i}(0)\big\vert^2}\big\vert \lambda_0-\lambda_1\big\vert ^2
     \nonumber \\
  &
     +\mathcal{O}\big[(\lambda_0-\lambda_1)^3\big] \, .
 \end{align}
Finally, the remaining term in Eq.~\eqref{equ:defzeta}, which is $C$, can also be expanded to lowest order in $\Delta\lambda$, using similar arguments as for the derivation of Eq.~\eqref{exp1}.
As we explicitly show in App.~\ref{app:cexpansion}, this yields an expansion of the form
\begin{align}
		C=&~C(0)+(\lambda_0-\lambda_1)D+(\bar \lambda_0-\bar \lambda_1)(\bar D)^T
  \nonumber \\
  &+\mathcal{O}\big[(\lambda_0-\lambda_1)^2\big], \label{exp2}
\end{align}
with $C(0)$ and the matrix $D$ given in App.~\ref{app:cexpansion}.
Inserting these expansions into Eq.~\eqref{equ:defzeta}, we find that the summands $\zeta^l$ generally obey
\begin{align}\nonumber
		\zeta^l\sim &~\frac{\vert\lambda_0\vert^2\mathrm{Re}\lambda_1+\vert\lambda_1\vert^2\mathrm{Re}\lambda_0}{\mathrm{Re}\lambda_0\mathrm{Re}\lambda_1\vert\lambda_0+\bar \lambda_1\vert}
		\frac{1}{\vert\lambda_0+\bar \lambda_1\vert}\, X
  \\&+\nonumber
		\mathrm{Re}\left(\frac{\lambda_0^2}{\mathrm{Re}\lambda_0(\lambda_0
 +\bar \lambda_1)}Y_0\right)\\
&+\mathrm{Re}\left(\frac{\lambda_1^2}{\mathrm{Re}\lambda_1(\lambda_1+\bar \lambda_0)}Y_1\right)
 \nonumber\\&+
\sum\limits_{i>1}\mathrm{Re}\left(\frac{\lambda_i^2}{(\lambda_i+\bar \lambda_0)(\bar \lambda_i+\lambda_1)}Z_i\right) \,,
\label{equ:zetaex}
	\end{align}
as $\vert\Delta \lambda\vert\rightarrow 0$, where $X,Y_i,Z_i$ are real, finite constants
given in \eqref{def:termX}, \eqref{def:termY}, \eqref{def:termZ}. 

Close inspection of the denominators in Eq.~\eqref{equ:zetaex} reveals that, as long as the co-aligning pair of eigenmodes remains stable, $\mathrm{Re}(\lambda^*)<0$  \red{[see Fig.~\ref{fig:pre_1}(a,d)]}, $\zeta^l$ remains finite at the exceptional point, and, consequently, $\mathcal{S}^*$  remains finite, too [recall Eq.~\eqref{equ:sdecomfin}].
Otherwise, for $\lambda^*=0$, i.e., when the exceptional point is at the same time a \textit{critical} point, a divergence occurs due to the term in the first line of Eq.~\eqref{equ:zetaex}.
	In this case, $\mathcal{S}^*$ diverges as
	\begin{align}
		\mathcal{S}^*&\sim   \frac{K}{\vert\lambda_0+\bar \lambda_1\vert}\,,~~~\text{as}~ \lambda_0,\lambda_1\rightarrow 0,
	\end{align}
	where
	\begin{align}
		K\equiv\frac{1}{\vert \hat{e}_\perp {\cdot}\partial_{\Delta\lambda}\Delta\hat{e}(0)\vert^2}\lim\limits_{\mathrm{Re}(\Delta\lambda)\rightarrow 0}\frac{\vert\lambda_0\vert^2\mathrm{Re}\lambda_1+\vert\lambda_1\vert^2\mathrm{Re}\lambda_0}{\mathrm{Re}\lambda_0\mathrm{Re}\lambda_1\vert\lambda_0+\bar \lambda_1\vert}
	\end{align}
	is a model-specific, path-dependent finite constant~\footnote{This can be checked by applying L'H\^{o}pital's rule~\cite{reed2003methods} and noting that away from the CEP it holds that $\mathrm{Re}\lambda_1,\mathrm{Re}\lambda_0>0$. }.
We now return to the particularly interesting case where the eigenvector $\hat{e}_0$ is  the Goldstone mode, such that the CEP corresponds to a secondary $\mathcal{PT}-$symmetry breaking transition where a dynamical phase emerges.
Then, also in the vicinity of the CEP, it holds that $\lambda_0=0$ and $\lambda_1 \in \mathbb{R}$. Hence, in this case,
	\begin{align}\label{equ:EPRscale}
		\mathcal{S}^{*}\sim \frac{1}{\lambda_1}\frac{1}{\big\vert \hat{e}_\perp {\cdot} \partial_{\lambda_1}\hat{e}_1(0)\big\vert^2}\,,~~~\text{as}~ \lambda_1\rightarrow0\,. 
	\end{align}
The divergent component of $\mathcal{S}$ thus only depends on properties of the mode that co-aligns with the Goldstone mode; specifically, on the value of $\lambda_1$ and the {orientation} of $\hat{e}_1$ with respect to the eigenbasis.
Interestingly, as for the oscillatory instability and the conventional critical point, $\lambda_1$ controls the susceptibility, also around the CEP. Differently from Eq.~\eqref{equ:fluc}, the susceptibility $\chi$ is now defined as~\footnote{
To have a meaningful measure for the susceptibility in a phase, where a continuous symmetry of the system is already broken, fluctuations in the direction of the Goldstone mode $\hat{e}_0$ (which are unbounded) must be excluded.
}
 \begin{align}\label{suscep2}
     \chi \equiv \lim\limits_{\epsilon\rightarrow0}\epsilon^{-1}\mathrm{tr}\,{\mathrm{Cov}(\Pi_0{{\phi}})}\, ,
 \end{align}
 where
 \begin{align}
     \Pi_0 x\equiv{x-(x{\cdot}\hat{e}_0)\hat{e}_0}
 \end{align}
 is the projector onto the subspace that is orthogonal to the Goldstone mode $\hat{e}_0$. The singular part of the susceptibility stems from the unstable mode, which has a well-defined wavenumber $l$. Therefore,
\begin{align}
    \chi\sim\lim\limits_{\epsilon\rightarrow0}\epsilon^{-1}\mathrm{tr}\,{\mathrm{Cov}(\Pi_0{{{\phi}^l}})}\,,~\text{as}~\lambda_1\rightarrow 0 \, .
\end{align}
According to our derivation in App.~\ref{app:sus}, the susceptibility asymptotically scales like
\begin{align}
    \chi\sim\big\vert\boldsymbol{q}^{l}\big\vert^{2}\frac{\big\vert \partial_{\lambda_1}\hat{e}_1(0)\big\vert^2}{\big\vert \hat{e}_\perp{\cdot}\partial_{\lambda_1}\hat{e}_1(0)\big\vert^2} \frac{1}{\lambda_1}\,,~~~\text{as}~ \lambda_1\rightarrow0\,.
\end{align}
	Hence, by comparison with Eq.~\eqref{equ:EPRscale}, we find
	\begin{align}\label{expfinal}
		\mathcal{S}^*&\sim \frac{1}{\big\vert \partial_{\lambda_1}\hat{e}_1(0)\big\vert^2}\frac{\chi}{\big\vert\boldsymbol{q}^{l}\big\vert^{2}}\,,~~~\text{as}~ \lambda_1\rightarrow0\,. 
	\end{align}
	We conclude that, as in the case of the oscillatory instability in Eq.~\eqref{equ:resosc}, $\mathcal{S}^*$ scales 
as the susceptibility, i.e., as the strength of the fluctuations measured relative to the noise intensity.
\red{We find this prediction confirmed by the numerical data for an CEP in the nonreciprocal Cahn-Hilliard model~\cite{suchanek2023entropy}, presented in Fig.~\ref{fig:pre_2}(c,f).}
The origin of this connection will be discussed in detail in Sec.~\ref{sec:con}. 
	
\subsubsection{Nondiagonal Jacobian}\label{sec:nondiagonal-J}
As the final step of this investigation, we address the general case that $\mathcal{J}$ is not diagonal in Fourier space, such that the unstable mode can have components of 
\textit{multiple} Fourier wavenumbers, which cannot be treated separately from each other. The analysis is nevertheless very similar to the case of diagonal $\mathcal{J}$.

However, since we relied on theorems from finite-dimensional linear algebra, we will now have to additionally assume that the operator $F$ (and hence also $\mathcal{J}$) is {effectively finite-dimensional}. 
 This can always be guaranteed by a UV cutoff. Furthermore, independently of such regularization, mechanisms of wavenumber selection often naturally 
 concentrate the dynamics to a characteristic scale~\cite{Saxena_2019,Cross93} and thereby ensure that phase transitions may already be captured by a finite number of Fourier modes~\cite{Crawford91_symmetry}. This is, for example, the case for the nonreciprocal Cahn-Hilliard model, analyzed in the two companion papers~\cite{suchanek2023irreversible,suchanek2023entropy}.

Expanding the squared deterministic current in the same way as in Sec.~\ref{sec:connection}, we find that~\footnote{Note that we have now assumed that the elements of the eigenbasis of $\mathcal{J}$ are normalized to unity, which was not the case for the Fourier basis used before, which was normalized to $V$.}
	\begin{align}\label{jgen}
		&\mathrm{SNCR}
 =
		\sum\limits_{ijn}\bar \lambda_{i}\big(\bar \nabla^{-1}T\big)_{ji}\bar \lambda_u\big(\nabla^{-1}T\big)_{jn}\big\langle\bar\psi_{i}\psi_n\big\rangle\,,
	\end{align}
	and
	\begin{align}
		\big\langle\, \bar \psi_{i}\psi_{j}\,\big\rangle =2\sum\limits_{nmu} \frac{(\bar T^{-1}_{im})(\bar \nabla_{mn})T_{ju}^{-1}\nabla_{un}}{\bar \lambda_i+\lambda_j}\,.
	\end{align}
Further, defining $\tilde{T}=\nabla^{-1}T$, we can rephrase Eq.~\eqref{jgen}  as
	\begin{align}
	\mathrm{SNCR}=
		\sum\limits_{ij}2\frac{\bar \lambda_i\lambda_j}{\bar \lambda_i+\lambda_j}\frac{\tilde{C}_{ij}}{\vert\det \tilde{T}\vert^2},
	\end{align}
	where $\tilde{C}$ is defined according to Eq.~\eqref{equ:defC}, with $T$ substituted by $\tilde{T}$.
From this point onward, the analysis is completely analogous to the above.
For oscillatory instabilities, we then find
	\begin{align}\label{generalone}
		\mathcal{S}^*\sim \omega^2\sum\limits_{k}\frac{\chi^k}{\vert\boldsymbol{q}^{k}\vert^{2}}\,,~~~\text{as}~ \sigma\rightarrow0\,,
	\end{align}
 with
 \begin{align}\label{eq:susj}
		\chi^k \equiv \lim\limits_{\epsilon\rightarrow0}\epsilon^{-1}\mathrm{tr}\,{\mathrm{Cov}({{\phi}^k})}\,,
	\end{align}
which generalizes the result~\eqref{equ:resosc}.
Analogously, for CEPs we find
\begin{align}\label{generatwo}
		\mathcal{S}^*&\sim \frac{1}{\vert \partial_{\lambda_1}\hat{e}_1(0)\vert^2}\sum_k\frac{\chi^k}{\vert\boldsymbol{q}^{k}\vert^{2}}\,,~~~\text{as}~ \lambda_1\rightarrow0\,, 
	\end{align}
which generalizes Eq.~\eqref{expfinal}.

We conclude that, also under the general condition, that the Fourier modes do not decouple, both transition scenarios, oscillatory instabilities and CEPs, exhibit a  divergence of $\mathcal{S}^*$, as the transition is approached. We note that the same approach also applies if
 the conserved noise $\sqrt{2\epsilon}\nabla \cdot \boldsymbol{\Lambda}_i$ is replaced by a more general one. In particular, this  includes the simpler case of non-conserved (i.e., scalar) noise, which corresponds to $\boldsymbol{q}^k\to 1$ in Eqs. \eqref{generalone},\eqref{generatwo}, and the case of a noise source {with distinct temperatures for each field component}.

\section{Exceptional points, Non-Hermitian dynamics, and $\mathcal{PT}-$symmetry breaking}\label{sec:nonHermitian-PT-CEP}
To deepen our understanding of the mechanisms that cause the surging entropy production toward the static-dynamic phase transition at a CEP, we examine in this section
the dynamics caused by a small perturbation in the vicinity of this type of instability.  
Specifically, we discuss the general connection between the coalescence of eigenmodes at CEPs and the coupling of damped modes to Goldstone modes of the broken symmetries. 
The general mechanism of the mode coupling has already been described in Refs.~\cite{Fruchart2021,hanai2020critical} for two-dimensional systems for the classical and the quantum case. Here, we provide a rigorous mathematical description for systems of arbitrary finite dimension.

\subsection{Eigenmode coupling and noise amplification}
	\label{sec:con}
	We consider some $N$-dimensional dynamical  system $\dot{x}=H(x)$, characterized by a nonlinear dynamical operator $H$, which is assumed to admit a CEP. We denote fixed points of $H$ by ${x}^*$, i.e $H(x^*)=0$. \
Recalling our parametrization throughout, the CEP ${x}^*(w)$ is then located at $w=0$.
	Any steady state of this system corresponds to a stable fixed point of $H$. The linearized dynamics of small perturbations $\Delta x=x-x^*$  around such a steady state $x^*$ is then determined by
	\begin{align}\label{equ:ex5}
 \partial_t \Delta{x}_i =\sum_j \mathcal{L}_{ij} \Delta x_j\,,
	\end{align}
	with 
 \begin{align}
{\mathcal{L}}_{ij}=\frac{\partial H_i}{\partial {x_j}}\bigg\vert_{x={x}^*}.
 \end{align}
 We further assume that the system has a continuous symmetry.
This means, there are symmetry operations $x\rightarrow\mathcal{X}_s{x}$, forming a one-parameter Lie-Group, under the action of which the form of Eq.~\eqref{equ:ex5} is preserved.
	For a given fixed point ${x}^*$ that is not preserved under the action of $\mathcal{X}_s$ and thus breaks the symmetry, the Goldstone theorem~\cite{Goldstone62} ensures that the corresponding Lie algebra contains a single tangent vector $\hat{e}_0$, 
which is an eigenvector of $\mathcal{L}$ with  eigenvalue $0$.
{Since, to linear order, a shift in this direction does not evoke any restoring dynamics, this defines  
 the ``direction of the symmetry operation".}
 Conventionally, this eigenvector is called the Goldstone mode.
 The remaining eigenvectors will be denoted by $\{\hat{e}_1,\hat{e}_2,\dots,\hat{e}_{N-1}\}$, where the ordering is chosen such that $\hat{e}_1$ is the direction that becomes unstable at the CEP and coalesces with the Goldstone mode $\hat{e}_0$. Consequently, the corresponding eigenvalue $\lambda_1$ vanishes at the CEP.

The usual approach to uncovering the effects of perturbations near an instability, would be to represent the dynamics~\eqref{equ:ex5} in the eigenbasis of the linearized dynamical operator $\mathcal{L}$.
At the CEP, this approach fails, since the eigenvectors do not form a complete basis set anymore.
As an alternative choice, the set
$\{\hat{e}_0,\hat{e}_\perp,\hat{e}_2', \dots, \hat{e}_{N-1}'\}$ has turned out to be particularly suitable, for this purpose.
	Here, for $i>1$, $\hat{e}_i'$ is the orthogonal projection of $\hat{e}_i$ on the orthogonal complement of $\hat {e}_0$, i.e., 
 \begin{align}
     \hat{e}_i'\equiv\frac{\hat{e}_i-(\hat{e}_i{\cdot}\hat{e}_0)\hat{e}_0}{\sqrt{1-|\hat{e}_i{\cdot}\hat{e}_0|^2}}
 \end{align}
	and $\hat{e}_\perp$ is  the unit vector defined in Eq.~\eqref{equ:perp}. Since $\hat{e}_\perp$ is perpendicular to the $({N}-1)$-dimensional subspace spanned by the set of eigenvectors at the CEP, the above set indeed forms a basis. Further, this basis allows studying how perturbations, which reside in the orthogonal complement of the Goldstone mode, couple to the Goldstone mode. 
    In App.~\ref{equ:othtrafo}, we show that the representation of the linearized dynamical operator $\mathcal{L}$ in this new basis reads

	\begin{align}\label{equ:fix}
	\!\!\!	\mathcal{L}'\sim\begin{pmatrix}
			0 & m_1 & m_2 &m_3 &\dots & m_{N-1}\\
			0 & \lambda_1  &0 &0 &\dots & 0\\
			0 & o_2  & \lambda_2 &0 &\dots & 0\\
			: & :  &\dots & : &:\\
			0 & o_{N-1}  &\dots & 
			0 &\dots &\lambda_{N-1}
		\end{pmatrix},~~\text{as}~ \lambda_1\rightarrow0,
	\end{align}
	with 
 \begin{align}\label{lprime}
     o_i=
			(-1)^{i-1}\frac{\hat{e}_i(0){\cdot}\partial_{\lambda_1}\hat{e}_1(0)}{\hat{e}_\perp(0){\cdot}\partial_{\lambda_1}\hat{e}_1(0)}\hat{e}_i'(0) {\cdot}\hat{e}_i(0)
 \end{align}
	and
  \begin{align}
  \label{midef}
		\!\!\!\! m_1&=\frac{1+\sum\limits_{i>1}(-1)^{i-1}\lambda_i(0)\hat{e}_i(0){\cdot}\partial_{\lambda_1}\hat{e}_1(0)\hat{e}_i(0) {\cdot}\hat{e}_0(0)}{\hat{e}_\perp(0){\cdot}\partial_{\lambda_1}\hat{e}_1(0)} ,\\
		\!\!\!\! m_{i}&=
			\lambda_i \frac{\hat{e}_i(0) {\cdot}\hat{e}_0(0)}{\sqrt{1-|\hat{e}_i(0){\cdot}\hat{e}_0|^2}} \,,~ i>1 \,.
 \end{align}
 Note that in the matrix given in Eq.~\eqref{equ:fix} all zero entries are identically zero up to \textit{all} orders in $\lambda_1$. Moreover, according to the derivation in Sec.~\ref{sec:irrexp}, $\vert\hat{e}_\perp{\cdot}\partial_{\lambda_1}\hat{e}_1(0)\vert \neq 0$, which appears in the off-diagonal entries.
  
 From this representation, it is apparent that there is a unidirectional (nonreciprocal) coupling from the damped modes $\{\hat{e}_\perp,\hat{e}'_2,\dots,\hat{e}'_{N-1}\}$ to the Goldstone mode $\hat{e}_0$ 
 in the vicinity of the CEP~\footnote{To see that corresponding matrix elements in Eq.~\eqref{equ:fix} cannot vanish, assume the contrary, i.e. $m_i=0$ for all $i$. This implies $\hat{e}_i(0) \hat{e}_0(0)=0$ for all $i>1$, which in turns leads to $m_1>0$, This contradicts the initial statement.}.
 Concerning the dynamics of $\Delta x$, this has two major implications:
First, the one-way coupling to the Goldstone mode implies that excitations pointing in directions perpendicular to $\hat{e}_0$ are unidirectionally transmitted to $\hat{e}_0$, and thereby drastically \textit{amplified}. This can be seen as follows.
Consider an  excitation by white noise $\hat{e}_\perp\xi_\perp(t)$ with $\langle\xi_\perp(t)\xi_\perp(t')\rangle=2\epsilon \delta(t-t')$ and, for sake of illustration, assume $\hat{e}_i {\cdot}\hat{e}_0=0$ for $i>1$. This implies $m_i=0$ for $i>1$ in Eq.~\eqref{equ:fix}.
Hence, this noise in the perpendicular direction $\hat{e}_\perp$ evokes a flow $F_0=m_1\Delta x_\perp\hat{e}_0$ in the direction of the Goldstone mode $\hat{e}_0$ with an average intensity of
\begin{align}\label{F_0scaling}
    \langle\vert F_0\vert^2\rangle\propto\epsilon\lambda^{-1}_1\,.
\end{align} 
Close to the transition, where $\lambda_1\to 0$, the intensity of this noise-induced flow is thus heavily amplified. 
As we show in Sec.~\ref{sec:concoup}, the noise amplification in the Goldstone mode also appears for the general case, where $\hat{e}_i {\cdot}\hat{e}_0\not =0$.
Second, the representation Eq.~\eqref{equ:fix} elucidates the connection between CEPs and the continuous emergence of dynamical phases. In case of a continuous transition scenario, the qualitative properties of a solution that arises after the formerly stable fixed point $x^*$ has become linearly unstable can be deduced via the central manifold approach~\cite{Crawford91_symmetry}. 
First, according to Eq.~\eqref{equ:fix}, the solution must have a component $\vert \Delta x_\perp\vert \neq 0$ in direction $\hat{e}_\perp$. Hence, it lies outside the manifold of the formerly stable fixed points $x^*$. 
The exact form of $\Delta x_\perp$ is determined by the nonlinear terms of the full dynamical operator $H$.
For the remaining directions, perpendicular to $\hat{e}_\perp$, 
we use the representation of $\mathcal{L}$ in the basis $\{\hat{e}_0,\hat{e}_\perp,\hat{e}_2,\dots,\hat{e}_{N-1}\}$, given in App.~\ref{app:ortbasis} [Eq.~\eqref{equ:fix2}], to expand the solution around the CEP. This leads to a solution that satisfies  
  \begin{align}
      \dot{y} \approx \frac{\Delta x_\perp}{\hat{e}_\perp{\cdot}\partial_{\lambda_1}\hat{e}_1(0)} \hat{e}_0\,,
  \end{align}
close to the transition. 
Hence, the solution represents a time-dependent state evolving within the manifold of degenerate steady states.
For the case that $x^*$ represents a static pattern emerging from a parity-symmetric dynamics, this scenario corresponds to a transition to a phase of traveling patterns; implying a breaking of $\mathcal{PT}$ symmetry~\cite{suchanek2023irreversible}.

Notably, these features crucially rely on the mode coalescence, $\hat{e}_1(0) =\hat{e}_0(0)$, at the CEP. To see this, we consider, for comparison, again a conventional critical
transition, where a single eigenmode becomes unstable while there is no co-aligning of eigenmodes. We define a transformation $\{\hat{e}_0,\hat{e}_1,\hat{e}_2,\dots,\,\hat{e}_{N-1}\}\rightarrow\{\hat{e}_0,\hat{e}_\perp,\hat{e}_2',\dots,\hat{e}'_{N-1}\}$, where this time, $\hat{e}_\perp$ is the unique unit vector that lies in the plane spanned by $\hat{e}_0$ and $\hat{e}_1$ and is perpendicular to $\hat{e}_0$.
Following the same procedure as for the CEP above, we find that, up to leading order in $\lambda_1$, the transformed linear operator takes the form
	\begin{align}\
		\mathcal{L}'\sim\begin{pmatrix}
			0 & k_0 & m_1 &m_2 &\dots & m_{N-1}\\
			0 & \lambda_1  &0 &0 &\dots & 0\\
			0 & 0  & \lambda_2 &0 &\dots & 0\\
			: & :  &\dots & : &:\\
			0 & 0  &\dots & 
			0 &\dots &\lambda_{N-1}
		\end{pmatrix},~~\text{as}~ \lambda_1\rightarrow0\,,
	\end{align}
	with
	\begin{align}
		k_0=\lambda_1\frac{\hat{e}_1(0){\cdot}\hat{e}_0(0)}{\hat{e}_1(0){\cdot}\hat{e}_\perp(0)}\,.
	\end{align}
 
Hence, also in this case, there is in general a coupling between unstable modes and the Goldstone mode. However, in sharp contrast to the above results \red{for the CEP, the coupling coefficient $k_0$ \textit{vanishes at the transition}, for a conventional critical point. Hence, there is no noise amplification and no $\mathcal{PT}$ symmetry breaking in this case}.

 \subsection{Connection between coupling of modes and irreversibility}
 \label{sec:concoup}
	Finally, we show how the coupling between damped modes and the Goldstone mode at a CEP is closely connected to the TRSB measured by $\mathcal{S}$. 
	To this end, we consider a stochastic version of the linearized dynamics given in Eq.~\eqref{equ:ex5},
 \begin{align}\label{stochor}
		\partial_t{\Delta x}_i = \sum_j\mathcal{L}_{ij}\Delta x_j +\sqrt{2\epsilon}\sum_j\,\eta_{ij}\xi_j\,,
	\end{align}
 with Gaussian noise satisfying $\langle \xi_i(t)\xi_j(t')\rangle=\delta_{ij}\delta(t-t')$. 
 We show that the basis $\{\hat{e}_0,\hat{e}_\perp,\hat{e}_2', \dots, \hat{e}_{N-1}'\}$
 introduced in Sec.~\ref{sec:con} can be used to achieve a decomposition of $\mathcal{S}$ into a regular and a singular contribution.
 The coordinate representation in this basis is given by
 \begin{align}
     \Delta x'_i=\sum_jU^{-1}_{ij}\Delta x 
 \end{align}
 where $U$ denotes the column matrix of basis vectors.
 We apply the transformation to Eq.~\eqref{stochor} and consider
	\begin{align}
		\partial_t{\Delta x}_i' = F_i(\Delta x') +\sqrt{2\epsilon}\sum_j\,(U^{-1}\eta)_{ij}\xi_j\,
	\end{align}
for $i>0$, with $F_i(\Delta x)\equiv\sum_{j}\mathcal{L}'_{ij} \Delta x_j$, which represents the part of the dynamics lying outside the manifold of degenerate fixed points of $\mathcal{L}$.  Separately, we consider the dynamics of $\Delta x_0$ with
 \begin{align}
     \partial_t \Delta x_0(t)=
     F_0(\Delta x')+\sqrt{2\epsilon}\sum\limits_i(U^{-1}\eta)_{0i}\xi_i\,,
 \end{align}
 which represents a fluctuating motion within this manifold. The transformation matrix $U$ is given in App.~\ref{app:ortbasis}.
 Note that for the case of Hermitian $\mathcal{L}$, the coupling between the Goldstone mode and its orthogonal subspace automatically vanishes and thus $F_0\equiv0$. For conventional critical points of non-Hermitian $\mathcal{L}$, a coupling may be present but vanishes at the transition, as demonstrated in Sec.~\ref{sec:con}.
 
 To evaluate $\mathcal{S}$, we can make use of our general expression~\eqref{equ:ptrate} for field theories. Concretely, we can  utilize the discretized version of Eq.~\eqref{equ:ptrate}, which reads 
 \begin{align}\label{eprdec2}
		\mathcal{S}
		=&~ {\epsilon^{-1}}\sum_{ij}\big[U^T(\eta\eta^T)^{-1}U\big]_{ij}\Big\langle\!(F_{i}(\Delta x') F_j(\Delta{x}')\!\Big\rangle
  \nonumber \\
  &+\sum_i\lambda_i\, .
	\end{align}
First, we consider the case of diagonal uniform noise, i.e., $\eta_{ij}=\eta\delta_{ij}$. In this case, using $(U^TU)_{0i}=\delta_{0i}$, we can separate the entropy production into a contribution 
\begin{align}\label{EPRdecomp}
		\mathcal{S}_0
		=&~  \frac{1}{\epsilon}\Big\langle \big\vert\eta^{-1} F_0(\Delta{x}) \big\vert^2\Big\rangle
	\end{align}
 from the dynamics of $\Delta x_0$ and a contribution from the dynamics $\Delta x_{i>0}$ in the perpendicular subspace.
 From Eq.~\eqref{equ:fix} it can be inferred that ${\mathcal{L}}'$, when restricted to the subspace perpendicular to $\hat{e}_0$,  has negative eigenvalues $\{\lambda_i\}_{i>0}$ and that the eigenvectors  are given by $\{\hat{e}_2',\dots,\hat{e}_{N-1}'\}$ and
 \begin{align}
     \hat{e}_1'\propto\hat{e}_\perp+\sum_i\frac{o_i}{\lambda_i-\lambda_1}\hat{e}_i' \,.
 \end{align}
Since the set $\big\{\hat{e}_2',\dots,\hat{e}_{N-1}'\big\}$ is linearly independent and $\hat{e}_\perp{\cdot}\hat{e}_i'=\hat{e}_\perp{\cdot}\hat{e}_i=0$, the eigenbasis has dimension ${N}-1$.
 Therefore, this  dynamics  exhibits a conventional critical phase transition. Hence, as discussed in Sec.~\ref{sec:eprsing}, it contributes a regular, finite amount to the entropy production.

This, in turn, means that the divergence of $\mathcal{S}^*$ at the CEP, proven in Sec.~\ref{sec:irrexp}, must be originating from the remaining contribution, Eq.~\eqref{EPRdecomp}.
The corresponding dynamics is due to flow 
\begin{align}\label{equ:flow}
F_0=\sum_{j}\mathcal{L}_{0j}' \Delta x'_j=\sum_{j}(U^{-1}\mathcal{L})_{0j} \Delta x_j
\end{align}
 along $\hat{e}_0$, which is driven by fluctuations in all perpendicular directions.
 In particular, taking into account the scaling of $\mathcal{S}^*$ in Eq.~\eqref{equ:EPRscale}, we find again, as in Eq.~\eqref{F_0scaling},
 \begin{align}
    \langle\vert F_0\vert^2\rangle\propto\epsilon\lambda^{-1}_1\,,
\end{align}
where $\lambda_1$ denotes the eigenvalue that becomes unstable at the CEP.
Hence, as already described for the special case of an effectively two-dimensional system in Sec.~\ref{sec:con}, we find that the non-vanishing nonreciprocal mode coupling in Eq.~\eqref{equ:fix} leads to a drastic amplification of noise also in the general case. 
Further, we can conclude that this noise amplification mechanism leads to singular entropy production at the CEP.

As a final remark, we aim to explicitly show the connection between the reasoning presented here, which started with a generic dynamical system with discrete phase space [Eq.~\eqref{stochor}], and the non-Hermitian field theories with conserved dynamics, given in Eq.~\eqref{equ:moda}.
Recall the results of Sec.~\ref{sec:eprsing}, where we showed that the linearized dynamics of a single Fourier mode $\phi^l$ is described by a dynamical equation of the form of Eq.~\eqref{equ:ex5}.
In this case, we have to choose $\eta=\vert q_l\vert V^{-1/2}$, $\Delta x=\Delta{\phi}^l$ and $\mathcal{L}=\mathcal{J}^l$. 

Consequentially, close to an CEP, we can identify the current 
	$J^\mathrm{d}_0(\boldsymbol{r})\equiv\hat{P}_0\nabla^{-1}\mathcal{J}^l{\phi}^le^{iq_l\boldsymbol{r}}$, with the projection operator
 \begin{align}
      \hat{P}_0x\equiv (U^{-1}x)_0\hat{e}_0=(x{\cdot}\hat{e}_0)\hat{e}_0, 
 \end{align}
as the origin of the entropy-producing flow $F_0$ and the divergence of $\mathcal{S}^*$,
	such that, according to Eq.~\eqref{EPRdecomp}, the asymptotic form of $\mathcal{S}^*$ can be represented as
	\begin{align}
		\mathcal{S}^*\sim\lim\limits_{\epsilon\rightarrow0}\epsilon^{-1}\int_V\mathrm{d}\boldsymbol{r}\big\langle\big\vert J_0^\mathrm{d}\big\vert^2 \big\rangle\,,~~\text{as}~ \lambda_1\rightarrow0\,.
	\end{align}
In App.~\ref{sec:currentgeneral}, we show that this, in fact, also holds in the more general case, where $\eta$ is nondiagonal and $\mathcal{L}$ and $K$ do not commute. The main difference, however, is that the projector $\hat{P}_0$ then takes the slightly more complicated form given in Eq.~\eqref{eq:ProjectorGeneral}.

\section{Conclusions}
We have studied the connection between 
time-reversal and parity-time symmetry breaking in non-Hermitian field theories with 
space-time white noise, in the limit of {low} noise intensity.
To quantify \red{the mesoscopically accessible} TRSB, we considered the informatic entropy production rate, $\mathcal{S}$, near two types of $\mathcal{PT}$ symmetry-breaking phase transitions, specifically, oscillatory instabilities and CEPs. As a reference point, we also discussed the behavior of  $\mathcal{S}$ at 
conventional critical transitions.

Our first main result concerns 
the scaling of $\mathcal{S}$ with respect to the noise intensity $\epsilon$, in the different phases. We have derived a general expression for $\mathcal{S}$ in terms of the statistical properties of the deterministic current [Eq.~\eqref{eq:EPRgeneral}].
From it, we find that, in a dynamical phase, $\mathcal{S}\sim \epsilon^{-1}$. The diverging entropy production rate for vanishing noise (independently of UV regularization) reflects the presence of systematic \red{dissipative} transport.
Conversely, in a static phase, we showed that, to leading order, $\mathcal{S}\sim\epsilon^{0}$, 
and thus remains nonzero \red{but finite} in the zero-noise limit.
These general predictions are corroborated and illustrated for the static phases as well as the dynamical phase of the nonreciprocal Cahn-Hilliard model in two companion papers~\cite{suchanek2023irreversible,suchanek2023entropy}, \red{as illustrated in Fig.~\ref{fig:pre_2}}.
Our general theory is also in agreement with earlier literature results 
for $\phi^4$ active field theories that admit activity-driven phase transitions.
Specifically, Ref.~\cite{Borthne2020} studied the entropy production in a polar flocking model, and found $\mathcal{S}\sim \epsilon^{-1}$ in the state of collective motion.
For Active Model B, Ref.~\cite{Nardini2017} reported $\mathcal{S}\sim\epsilon^{0}$ for the phase-separated state, 
and $\mathcal{S}\sim\epsilon^{1}$ for the homogeneous phase. 
This is consistent with our prediction, as this model only falls into the category of non-Hermitian field theories in its demixed phase~\footnote{This can be easily seen from Eq.~(24) in Ref.~\cite{Nardini2017}.}.

Our second main result concerns the behavior of $\mathcal{S}$ in the vicinity of phase transitions, which we obtained by examining the connection between TRSB and the eigensystem of the dynamical operator. We have derived a general 
formula for $\mathcal{S}$ in terms of the spectrum and geometry of the eigenvectors of the linearized dynamical operator [presented in Eqs.~\eqref{equ:ptrate}, \eqref{equ:eppt2}, \eqref{equ:defzeta}]. 
For the $\mathcal{PT}$-breaking transitions, we could show that the entropy production rate $\mathcal{S}^*\equiv\lim_{\epsilon\rightarrow0}\mathcal{S}$ in the low-noise limit is proportional to the susceptibility.  This demonstrates that  
entropy-generating fluctuations reside in the unstable wavelength(s) characterizing emerging \red{dynamical} order, while all other contributions remain finite, thereby testifying the active character of the emerging patterns.
This should be contrasted to a conventional critical point, where the susceptibility diverges too, yet $\mathcal{S}^*$ remains regular. 
The analytical expressions we derived give further insights into the underlying mechanisms that cause the singular contribution to $\mathcal{S}$. 
For oscillatory instabilities, Eq.~\eqref{equ:resosc} suggests that the main source of TRSB are \red{(transient)} cyclic currents with characteristic frequency $\omega$ that herald the emerging dynamical phase.
In the vicinity of CEPs, we found that the coupling of modes leads to 
an entropy-generating 
fluctuating current with diverging intensity along the direction of a Goldstone mode, resulting in gigantically amplified fluctuations. \red{This may justify the notion of a ``pumped" or ``active'' Goldstone mode.} 
In contrast, the TRSB fluctuations in the perpendicular space remain regular, throughout.
The same mechanism also ultimately causes the emergence of a $\mathcal{PT}$ symmetry-breaking phase transition featuring a deterministic entropy producing current, associated with the emergent dissipative dynamical structure.
Thereby, we uncovered a common origin and systematic connection between $\mathcal{PT}$ symmetry breaking, dissipative pattern formation, and TRSB fluctuations. 

The nonreciprocal Cahn-Hilliard model provides an illustrative example of a non-Hermitian field theory to make these abstract mechanisms more tangible. As we show in Ref.~\cite{suchanek2023irreversible},
the TRSB excitations of its Goldstone mode give rise to a persistent interface dynamics that is formally identical to the stochastic motion of an active particle. This active interface dynamics is ``driven" by the (unidirectionally coupled) fluctuating phase shift, which itself exhibits time-symmetric dynamics.

We have studied field theories with conserved dynamics, but we aim to point out that our results are also valid for the cases of non-conserved or partially conserved dynamics.
Beyond field-theoretical models, it would be interesting to apply our results to particle-based models. Specifically, after discretizing the here derived formulae, they also hold for systems with a discrete phase space, such as a collection of (nonreciprocally) coupled, overdamped particles in a thermal heat bath~\cite{loos2020irreversibility,zhang2022entropy,pruessner2022field}. In such models, the connection between TRSB and the underlying friction and dissipation are already well-understood within the framework of stochastic thermodynamics~\cite{Seifert2005}.
In this context, it would be particularly interesting to investigate the relationship between the entropy production of a particle-based model and the mesoscopic entropy production at the field level.

Another broad perspective for future research is the general role of TRSB at nonequilibrium phase transitions.
In thermal equilibrium, phase transitions are events of global symmetry breaking that 
are largely independent of the details of the underlying dynamics. 
Consequentially, many of their fundamental features can be universally characterized solely based on single-time (or ``structural") observables, such as the susceptibility or specific heat. 
In nonequilbrium, dynamical features and global symmetries are more tightly intertwined.
Whether and to what extent the entropy production, which is a dynamical (path-dependent) observable, may serve as a suitable tool to classify nonequilibrium phase transitions remains elusive~\cite{caballero2020stealth}. In this regard, it might be interesting to remark that, similar to the here observed proportionality to the susceptibility, previous numerical studies on spin systems have hinted at a connection between the entropy production rate and the specific heat at continuous phase transitions~\cite{martynec2020entropy,tome2012entropy,noa2019entropy,loos2022long}, but the mechanistic origin of this connection is unknown. \red{To address these very general questions, it seems to be worthwhile to revisit non-Hermitian field theories from the perspective of the renormalization group (RG)  as  in Refs.~\cite{Paoluzzi22,caballero2020stealth}. 
This would be particularly interesting, since unlike in
the models studied so far, the nonequilibrium driving in
non-Hermitian field theories (which appears at the linear
level) is not ``irrelevant in the RG sense." }

\begin{acknowledgments}
SL acknowledges funding by the Deutsche Forschungsgemeinschaft (DFG, German Research Foundation) – through the project 498288081.
TS acknowledges financial
support by the pre-doc award program at Leipzig University. 
SL thanks the Physics Institutes of Leipzig University for their hospitality during several research stays.
\end{acknowledgments}

\appendix

\section{Expansion of $C$}
\label{app:cexpansion}
Here, we provide the explicit expansion of the term $C$ in Eq.~\eqref{equ:defzeta} with respect to $\Delta \lambda = \lambda_0-\lambda_1$.
First, we tackle the expansion of $\mathrm{adj}T$. We find
	\begin{align}
		\mathrm{adj}\,T^T(0)&=
		\begin{pmatrix}
			\hat{d}_0(0),-\hat{d}_0(0),0,\dots,0
		\end{pmatrix}\,,
	\end{align}
	with
	\begin{align}
		\hat{d}_{ij}=(-1)^{i+j}M_{i,j}(T)\,.
	\end{align}
	Note that for $i>1$ all minors are zero at the CEP because of the linear dependence of the first two columns of $T$.
	This results in 
 \begin{widetext}
	\begin{align}
		\big((\bar T)^TT\big)(0)=
		&\begin{pmatrix}
			1 & 1 &\bar{\hat{e}}^*{\cdot}\hat{e}_2(0) &\dots & \bar{\hat{e}}^*{\cdot}\hat{e}_{N-1}(0)\\
			1 & 1 &\bar{\hat{e}}^*{\cdot}\hat{e}_2(0) &\dots & \bar{\hat{e}}^*{\cdot}\hat{e}_{N-1}(0)\\
			\bar{\hat{e}}_2(0){\cdot}\hat{e}^* & \bar{\hat{e}}_2(0){\cdot}\hat{e}^* &1 &\dots & \bar{\hat{e}}_2(0){\cdot}\hat{e}_{N-1}(0)\\
			: & : &: &\dots & :\\
			\bar{\hat{e}}_{N-1}(0){\cdot}\hat{e}^* & \bar{\hat{e}}_{N-1}(0){\cdot}\hat{e}^* &\bar{\hat{e}}_{N-1}(0){\cdot}\hat{e}_2(0) &\dots & 1
		\end{pmatrix}\,,
	\end{align}
and therefore,
	\begin{align}
		C(0)&=\big(\mathrm{adj}\bar T\mathrm{adj} T^T\big)(0)=\begin{pmatrix}
			\sum_i \vert M_{0,i}(0)\vert^2 & -\sum_i  \vert M_{0,i}(0)\vert^2 &0 &\dots & 0\\
			-\sum_i \vert M_{0,i}(0)\vert^2 & \sum_i \vert M_{0,i}(0)\vert^2 &0 &\dots & 0\\
			0 & 0 &0 &\dots & 0&\\
			: & : &: &\dots & :\\
			0 & 0 &0 &\dots & 0
		\end{pmatrix}\,.
	\end{align}
	Further, we denote the derivative of the adjugate of $T$ by
	\begin{align}\label{equ:EPRscale2}
		\partial_{\Delta\lambda}\mathrm{adj}\,T^T(0)&=
		\begin{pmatrix}
			p_0,p_1,p_2,\dots,p_{N-1}
		\end{pmatrix}\,,
	\end{align}
	with
	$p_i=\partial_{\Delta\lambda}\hat{d}_i(0)$.
	Therefore,
	\begin{align}
		C=C(0)+(\lambda_0-\lambda_1)D+(\bar \lambda_0-\bar \lambda_1)(\bar D)^T+\mathcal{O}\big[(\lambda_0-\lambda_1)^2\big]
	\end{align}
	with
	\begin{align}
		D=
		&\begin{pmatrix}
			\bar d_0(0){\cdot}p_0 & \bar d_0(0){\cdot}p_1 &\bar d_0(0){\cdot}p_2\bar{\hat{e}}^*{\cdot}\hat{e}_2(0) &\dots & \bar d_1(0){\cdot}p_n\bar{\hat{e}}^*{\cdot}\hat{e}_{N-1}(0)\\
			-\bar d_0(0){\cdot}p_0 & -\bar d_0(0){\cdot}p_1 &-\bar d_0(0){\cdot}p_2\bar{\hat{e}}^*{\cdot}\hat{e}_2(0) &\dots & -\bar d_0(0){\cdot}p_n\bar{\hat{e}}^*{\cdot}\hat{e}_{N-1}(0)\\
			0 & 0 &0 &\dots & 0\\
			: & : &: &\dots & :\\
			0 & 0 &0 &\dots & 0
		\end{pmatrix}\,.
	\end{align}
 Note that, here we have used 
 \begin{align}
     \partial_{\Delta\lambda}\big((\bar T)^TT\big)_{ii}=0
 \end{align}
 and
 \begin{align}
     &\partial_{\Delta\lambda}\big((\bar T)^TT\big)_{01}+\partial_{\Delta\lambda}\big((\bar T)^TT\big)_{10}
=\partial_{\Delta\lambda}\vert\hat{e}_0\vert^2\big\rvert_{\Delta\lambda=0}+\partial_{\Delta\lambda}\vert\hat{e}_1\vert^2\big\rvert_{\Delta\lambda=0}=0\,.
 \end{align}

\subsection{Expansion of summands in Eq.~\eqref{equ:defzeta}}
	Inserting the expansions given in Eqs.~\eqref{exp1} and \eqref{exp2} into Eq.~\eqref{equ:eppt2}, we find
 \begin{align}
		\zeta^l\sim& ~\frac{1}{\vert \hat{e}_\perp {\cdot} \partial_{\Delta\lambda}\Delta\hat{e}(0)\vert^{2}}\frac{1}{\vert\lambda_0-\lambda_1\vert^2}\left(\frac{\vert \lambda_0\vert^2}{\mathrm{Re}\lambda_0}+\frac{\vert \lambda_1\vert^2}{\mathrm{Re}\lambda_1}-2\frac{\bar \lambda_0\lambda_1}{\bar \lambda_0+\lambda_1}-2\frac{\lambda_0\bar \lambda_1}{\lambda_0+\bar \lambda_1}\right)\nonumber \\
		+&\frac{2}{\vert \hat{e}_\perp {\cdot} \partial_{\Delta\lambda}\Delta\hat{e}(0)\vert^{2}\sum_i \vert M_{0,i}(0)\vert^{2}}\times
  \nonumber
  \\
  &\frac{1}{\vert\lambda_0-\lambda_1\vert^{2}}
		\mathrm{Re}\left[\bar d_0{\cdot}p_0(\lambda_0-\lambda_1)\bigg(\frac{\vert \lambda_0\vert^{2}}{\mathrm{Re}\lambda_0}-2\frac{\bar \lambda_0\lambda_1}{\bar \lambda_0+\lambda_1}\bigg)-\bar d_0{\cdot}p_1(\lambda_0-\lambda_1)\bigg(\frac{\vert \lambda_1\vert^2}{\mathrm{Re}\lambda_1}-2\frac{\lambda_0\bar \lambda_1}{\lambda_0+\bar \lambda_1}\bigg)\right]\\
  +&\frac{4}{\vert \hat{e}_\perp {\cdot} \partial_{\Delta\lambda}\Delta\hat{e}(0)\vert^{2}\sum_i \vert M_{0,i}(0)\vert^{2}}\times
  \nonumber
  \\
  &\frac{1}{\vert\lambda_0-\lambda_1\vert^2}
			\mathrm{Re}\left[\sum_{i>1}\bar d_0{\cdot}p_i\bar{\hat{e}}^*{\cdot}\hat{e}_i(\lambda_0-\lambda_1)\bigg(\frac{\lambda_i\bar \lambda_0}{\lambda_i+\lambda_0}-\frac{\lambda_i\bar \lambda_1}{\lambda_l+\bar \lambda_1}\bigg)\right]
   \,,~~~\text{as}~\vert\lambda_1-\lambda_0\vert\rightarrow 0\, .
	\end{align}
 \end{widetext}
  This cumbersome expression can be further simplified to 
\begin{align}\nonumber
		\zeta^l\sim &~\frac{\vert\lambda_0\vert^2\mathrm{Re}\lambda_1+\vert\lambda_1\vert^2\mathrm{Re}\lambda_0}{\mathrm{Re}\lambda_0\mathrm{Re}\lambda_1\vert\lambda_0+\bar \lambda_1\vert}
		\frac{1}{\vert\lambda_0+\bar \lambda_1\vert}\, X
  \\&+\nonumber
		\mathrm{Re}\left(\frac{\lambda_0^2}{\mathrm{Re}\lambda_0(\lambda_0
 +\bar \lambda_1)}Y_0\right)\\
&+\mathrm{Re}\left(\frac{\lambda_1^2}{\mathrm{Re}\lambda_1(\lambda_1+\bar \lambda_0)}Y_1\right)
 \nonumber\\&+
\sum\limits_{i>1}\mathrm{Re}\left(\frac{\lambda_i^2}{(\lambda_i+\bar \lambda_0)(\bar \lambda_i+\lambda_1)}Z_i\right) \,,
	\end{align}
as $\vert\lambda_1-\lambda_0\vert\rightarrow 0 $, with the constants
\begin{align}
    X&=\frac{1}{\vert \hat{e}_\perp {\cdot} \partial_{\Delta\lambda}\Delta\hat{e}(0)\vert^{2}}\,,
  \label{def:termX}
    \\\nonumber
    Y_0&=\frac{2p_0\bar d_0}{\vert \hat{e}_\perp {\cdot} \partial_{\Delta\lambda}\Delta\hat{e}(0)\vert^{2}\sum_i \vert M_{0,i}(0)\vert^{2}}
    \\\label{def:termY}
    Y_1&=\frac{2p_1\bar d_0}{\vert \hat{e}_\perp {\cdot} \partial_{\Delta\lambda}\Delta\hat{e}(0)\vert^{2}\sum_i \vert M_{0,i}(0)\vert^{2}}
    \\
    Z_i&= \frac{4d_0p_i\bar{\hat{e}}^*\hat{e}_i}{\vert \hat{e}_\perp {\cdot} \partial_{\Delta\lambda}\Delta\hat{e}(0)\vert^{2}\sum_i \vert M_{0,i}(0)\vert^{2}}\,.
     \label{def:termZ}
\end{align}
 
 \section{Susceptibility close to a critical exceptional point}\label{app:sus}
\newpage
  The singular part
of the susceptibility is obtained as follows: expanding the definition of the susceptibility in Eq.~\eqref{suscep2} as \begin{align}\label{equ:chiscaling}
		&~
		\lim\limits_{\epsilon\rightarrow0}\epsilon^{-1}\big\vert\boldsymbol{q}^{l}\big\vert^{-2}\mathrm{tr}\,{\mathrm{Cov}(\Pi_0{\phi}^l)} \nonumber 
  \\
	=&~	\big\vert\boldsymbol{q}^{l}\big\vert^{-2}V\sum\limits_{i} \big\langle\, (\Pi_0\Delta\bar \phi^l)_i(\Pi_0\Delta{\phi}^l)_i\,\big\rangle \nonumber 
 \\
 =&~
		\big\vert\boldsymbol{q}^{l}\big\vert^{-2}V\sum\limits_{ijnmu}\big\langle(\Pi_0)_{ij}T_{jn}\Delta\bar \psi^l_n(\Pi_0)_{im}T_{mu}\Delta\psi^l_u\big\rangle
\nonumber 		
  \\
  =&~\vert\boldsymbol{q}^{l}\vert^{-2}V\sum\limits_{ijn}(\Pi_0\bar{\hat{e}}_i)_j(\Pi_0\hat{e}_n)_j\big\langle \Delta\psi^l_i\Delta\psi^l_n\big\rangle \nonumber 
  \\
  =&~2\sum\limits_{i,j>0 }\frac{\Pi_0\bar{\hat{e}}_i\cdot\Pi_0\hat{e}_j}{\vert\det T\vert^2}\frac{\big(\mathrm{adj}\bar {T}^T\mathrm{adj}T^T\big)_{ij}}{\bar\lambda_i+\lambda_j}
		\, ,
  \end{align}
  and expanding further expanding the sum in the last line, we find
  \newpage
  \begin{widetext}
      \begin{align}&~
		\lim\limits_{\epsilon\rightarrow0}\epsilon^{-1}\big\vert\boldsymbol{q}^{l}\big\vert^{-2}\mathrm{tr}\,{\mathrm{Cov}(\Pi_0{\phi}^l)} \nonumber \\
	=&~
          \frac{\vert\Pi_0\hat{e}_1\vert^2}{\vert\det T\vert^2}\frac{\big(\mathrm{adj}\bar {T}^T\mathrm{adj}T^T\big)_{11}}{\lambda_1}+
  2\epsilon\!\left[
		\sum\limits_{i>1}
		\frac{\Pi_0\bar{\hat{e}}_1\cdot\Pi_0\hat{e}_i}{\vert\det T\vert^2}\frac{\big(\mathrm{adj}\bar {T}^t\mathrm{adj}T^T\big)_{1i}}{\bar \lambda_1+\lambda_i}
		+\!\sum\limits_{i,j>1}
		\frac{\Pi_0\bar{\hat{e}}_i\cdot\Pi_0\hat{e}_j}{\vert\det T\vert^2}\frac{(\mathrm{adj}\bar {T}^T\mathrm{adj}T^T)_{ij}}{\bar \lambda_i+\lambda_j}+\mathrm{c.c.}\right].
      \end{align}
  \end{widetext}
	Using that \begin{align}
\vert\Pi_0\hat{e}_1\vert\sim\vert\partial_{\lambda_1}\hat{e}_1(0)\vert\vert\lambda_1\vert\,,~~ ~\text{as}~ \lambda_1\rightarrow0\,,
	\end{align}
 and taking also into account the precise asymptotic scaling of the  entries of $ \mathrm{adj}\bar {T}^T\mathrm{adj}T^T$ and $\det T$ in $\lambda_1$, given in Eq.~\eqref{equ:EPRscale2} and Eq.~\eqref{exp1},
 which results in 
 \begin{align}
  \!\!\!\frac{\big(\mathrm{adj}\bar {T}^T\mathrm{adj}T^T\big)_{11}}{\vert\det T\vert^2}\sim \frac{1}{\big\vert \hat{e}_\perp{\cdot}\partial_{\lambda_1}\hat{e}_1(0)\big\vert^2} \frac{1}{\lambda_1}\,,~\text{as}~ \lambda_1\rightarrow0,
 \end{align}
we find that only the first term in the last line of Eq.~\eqref{equ:chiscaling} generates the divergence, and the asymptotic form of the susceptibility is given by 
\begin{align}
    \chi\sim\big\vert\boldsymbol{q}^{l}\big\vert^{2}\frac{\big\vert \partial_{\lambda_1}\hat{e}_1(0)\big\vert^2}{\big\vert \hat{e}_\perp{\cdot}\partial_{\lambda_1}\hat{e}_1(0)\big\vert^2} \frac{1}{\lambda_1}\,,~~~\text{as}~ \lambda_1\rightarrow0\,.
\end{align}

\section{Orthogonal basis at an exceptional point}\label{app:ortbasis}
Here, we show how to obtain the representation of the operator $\mathcal{L}$ in the basis introduced in Sec.~\ref{sec:con}.
We denote by $\mathbb{D}$ the diagonal matrix of eigenvalues of ${\mathcal{L}}$, i.e. $\mathbb{D}_{ij}=\lambda_i\delta_{ij}$,  and by $O$ the matrix representation of the orthogonal projection. Again, $T$ is  the matrix of eigenvectors of ${\mathcal{L}}$ and $\tilde{T}=(\hat{e}_0,\hat{e}_\perp,\hat{e}_2,\dots,\hat{e}_{N-1})$ the matrix after $\hat{e}_1$ was substituted by $\hat{e}_\perp$. Accordingly, the matrix representation for the full transformation is given by 
\begin{align}\label{defU}
    {U}^{-1}=(\tilde{T}O)^{-1}\, .
\end{align}\label{appD2}
The representation of the linearized dynamical operator $\mathcal{L}$ in this new basis is obtained from the expression
	\begin{align}\label{equ:othtrafo}
		\mathcal{L}'={U}^{-1}\mathcal{L}{U}=O^{-1}\tilde{T}^{-1}T\,\mathbb{D}\,T^{-1}\tilde{T}O\,,
	\end{align}
 which we evaluate in two steps. First, the explicit expressions for the transformation matrices matrix products in Eq.~\eqref{appD2}  are given by
 \begin{widetext}
	\begin{align}
		T^{-1}\tilde{T}=
		\frac{1}{\det T}
		\begin{pmatrix}
			\det T & \det(\hat{e}_\perp,\hat{e}_1,\dots,\hat{e}_{N-1})  & 0 &0 &\dots & 0\\
			0 & \det \tilde{T}  &0 &0 &\dots & 0\\
			0 & \det(\hat{e}_0,\hat{e}_1,\hat{e}_\perp,\hat{e}_3,\dots,\hat{e}_{N-1})  & \det \tilde{T} &0 &\dots & 0\\
			: & :  &\dots & : &:\\
			0 & \det(\hat{e}_0,\hat{e}_1,\dots,\hat{e}_{{N}-2},\hat{e}_\perp)  &\dots & 
			0 &\dots &\det \tilde{T}
		\end{pmatrix}\,,
	\end{align}
 and
 \begin{align}
		\tilde{T}^{-1}T=
		\frac{1}{\det \tilde{T}}
		\begin{pmatrix}
			\det \tilde{T} & \det(\hat{e}_1,\hat{e}_\perp,\dots,\hat{e}_{N-1})  & 0 &0 &\dots & 0\\
			0 & \det T  &0 &0 &\dots & 0\\
			0 & \det(\hat{e}_0,\hat{e}_\perp,\hat{e}_1,\hat{e}_3,\dots,\hat{e}_{N-1})  & \det T &0 &\dots & 0\\
			: & :  &\dots & : &:\\
			0 & \det(\hat{e}_0,\hat{e}_\perp,\dots,\hat{e}_{{N}-2},\hat{e}_1)  &\dots & 
			0 &\dots &\det T
		\end{pmatrix},
	\end{align}
To obtain these, the adjugate representation of the Matrix inverse \cite{poole2014linear} and the  the Laplace expansion rule were used. Now, using the explicit expressions for the scaling of determinants in the limit $\lambda_1\rightarrow 0$, derived in Sec. \ref{sec:eprsing}, we find up to lowest order in $\lambda_1$
	\begin{align}\label{equ:fix2}
		\tilde{T}^{-1}T\,\mathbb{D}\,T^{-1}\tilde{T}\sim\begin{pmatrix}
			0 & l_0 & 0 &0 &\dots & 0\\
			0 & \lambda_1  &0 &0 &\dots & 0\\
			0 & l_2  & \lambda_2 &0 &\dots & 0\\
			: & :  &\dots & : &:\\
			0 & l_{N-1}  &\dots & 
			0 &\dots &\lambda_{N-1}
		\end{pmatrix}\,,
	\end{align} as $\lambda_1\rightarrow0$,
	where the off-diagonal elements are given
	by
	\begin{align}\label{ldef2}
		l_i=
		\begin{cases}
			\frac{1}{\hat{e}_\perp{\cdot}\partial_{\lambda_1}\hat{e}_1(0)} & i= 0 \\
			\lambda_i (0)\frac{\det(\hat{e}_0,\partial_{\lambda_1}\hat{e}_1(0),\hat{e}_2(0),\dots,\hat{e}_{i-1}(0),\hat{e}_\perp,\hat{e}_{i+1}(0),\dots,\hat{e}_{N-1}(0))}{\det(\hat{e}_0,\partial_{\lambda_1}\hat{e}_1(0),\hat{e}_2(0),\dots,\hat{e}_{N-1}(0))}=(-1)^{i-1}\lambda_i(0)\frac{\hat{e}_i(0){\cdot}\partial_{\lambda_1}\hat{e}_1(0)}{\hat{e}_\perp {\cdot}\partial_{\lambda_1}\hat{e}_1(0)} & i>1
		\end{cases}\,.
	\end{align}
 \end{widetext}
	Note that in Eq.~\eqref{equ:fix2} all zero matrix entries are identically zero up to \textit{all} orders in $\lambda_1$. Moreover, according to the derivation in Sec.~\ref{sec:irrexp}, $\vert\hat{e}_\perp\partial_{\lambda_1}\hat{e}_1(0)\vert \neq 0$.
 This already shows that, at the CEP, the unstable mode $\hat{e}_\perp$ couples to the Goldstone mode $\hat{e}_0$.
	Now, applying the orthogonal projection 
	\begin{align}\label{equ:ortpr}
		\!O=
		\begin{pmatrix}
			1 & 0  & -\frac{\hat{e}_2{\cdot}\hat{e}_0}{\sqrt{1-|\hat{e}_2{\cdot}\hat{e}_0|^2}}&\dots&
			-\frac{\hat{e}_{N-1}{\cdot}\hat{e}_0}{\sqrt{1-|\hat{e}_{N-1}{\cdot}\hat{e}_0|^2}}&\\
			0 & 1   & 0&\dots&0\\
			0 & 0   & \frac{1}{\sqrt{1-|\hat{e}_2{\cdot}\hat{e}_0|^2}}&\dots&0\\
			: & :   & : &: &:\\
			0 & 0   & 
			0 &\dots &\frac{1}{\sqrt{1-|\hat{e}_{N-1}{\cdot}\hat{e}_0)
   |^2}}
		\end{pmatrix},
	\end{align}
  we arrive at the expressions \eqref{equ:fix}, \eqref{lprime} and \eqref{midef}.

\section{Entropy production at CEPs for nondiagonal noise}\label{sec:currentgeneral}
Here, we generalize the result Eq.~\eqref{equ:flow} for the flow that causes the singular contribution to $\mathcal{S}^*$ at a CEP to the case where $\eta$ is a general matrix operator and does not commute with  $\mathcal{L}$. In this case, it does not automatically hold that the noise in the Goldstone mode along $\hat{e}_0$ is statistically independent of the noise in the perpendicular directions, after performing the transformation in Sec.~\ref{sec:con}. For nontrivial (nondiagonal) $\mathcal{L}$, this problem already arises, e.g., if the noise intensity for diverse degrees of freedom does not coincide~\footnote{Note that differing noise temperatures alone indeed not need to be a sign of nonequilibrium in this context, but may also result from a chosen basis representation.}.
 
	To obtain a decomposition of $\mathcal{S}$ equivalent to the one in Sec.~\ref{sec:concoup}, we first have to define the transformation
	\begin{align}
		G
		=\begin{pmatrix}
			g_0 & g_1  & g_2&\dots&
			g_{N-1}\\
			0 & 1   & 0&\dots&0\\
			: & :   & : &: &:\\
			0 & 0   & 
			0 & 1
		\end{pmatrix}^{-1}
		\,,
	\end{align}
	where 
the $w_i$ are implicitly determined through the orthogonality condition
	\begin{align}
		\delta_{i0}=\big \langle \big(G^{-1}U^{-1}\eta\xi\big)_i\big( G^{-1} U^{-1}\eta\xi\big)_0\big \rangle\,,
	\end{align}
so that the noise in the $\hat{e}_0$-direction becomes disentangled from noise in the perpendicular directions.
	Note that the transformation also preserves the direction of $\hat{e}_0$, as well as the coordinate representations of vectors and operators in its orthogonal complement. Now the same arguments as used in the main text apply. Again, we can identify a flow generating the divergent part of $\mathcal{S}$, which now takes the slightly more complicated form
 \begin{align}
F_0\equiv(G^{-1}U^{-1}\mathcal{L}\Delta\boldsymbol{x})_0\,.
 \end{align} 
 Again considering the case $\eta=\nabla$, the projector for the current that generates $F_0$ is then given by 
 \begin{align}\label{eq:ProjectorGeneral}
     \hat{P}_0x\equiv (G^{-1}U^{-1}\nabla x)_0\hat{j}_0\,,
 \end{align}
 with $\hat{j}_0\equiv\nabla^{-1}\hat{e}_0$.

As a concrete example, we refer to the equations of motion for the center-of-mass phase $\theta_c$ and the phase shift $\Delta \theta^{\pi}$ given in the companion paper~\cite{suchanek2023irreversible}, where a noise term with differing intensities arises as a result of the chosen non-orthonormal basis  transformation.

 \bibliography{bib.bib}

\providecommand{\noopsort}[1]{}\providecommand{\singleletter}[1]{#1}%
\begin{thebibliography}{93}%
\makeatletter
\providecommand \@ifxundefined [1]{%
 \@ifx{#1\undefined}
}%
\providecommand \@ifnum [1]{%
 \ifnum #1\expandafter \@firstoftwo
 \else \expandafter \@secondoftwo
 \fi
}%
\providecommand \@ifx [1]{%
 \ifx #1\expandafter \@firstoftwo
 \else \expandafter \@secondoftwo
 \fi
}%
\providecommand \natexlab [1]{#1}%
\providecommand \enquote  [1]{``#1''}%
\providecommand \bibnamefont  [1]{#1}%
\providecommand \bibfnamefont [1]{#1}%
\providecommand \citenamefont [1]{#1}%
\providecommand \href@noop [0]{\@secondoftwo}%
\providecommand \href [0]{\begingroup \@sanitize@url \@href}%
\providecommand \@href[1]{\@@startlink{#1}\@@href}%
\providecommand \@@href[1]{\endgroup#1\@@endlink}%
\providecommand \@sanitize@url [0]{\catcode `\\12\catcode `\$12\catcode
  `\&12\catcode `\#12\catcode `\^12\catcode `\_12\catcode `\%12\relax}%
\providecommand \@@startlink[1]{}%
\providecommand \@@endlink[0]{}%
\providecommand \url  [0]{\begingroup\@sanitize@url \@url }%
\providecommand \@url [1]{\endgroup\@href {#1}{\urlprefix }}%
\providecommand \urlprefix  [0]{URL }%
\providecommand \Eprint [0]{\href }%
\providecommand \doibase [0]{https://doi.org/}%
\providecommand \selectlanguage [0]{\@gobble}%
\providecommand \bibinfo  [0]{\@secondoftwo}%
\providecommand \bibfield  [0]{\@secondoftwo}%
\providecommand \translation [1]{[#1]}%
\providecommand \BibitemOpen [0]{}%
\providecommand \bibitemStop [0]{}%
\providecommand \bibitemNoStop [0]{.\EOS\space}%
\providecommand \EOS [0]{\spacefactor3000\relax}%
\providecommand \BibitemShut  [1]{\csname bibitem#1\endcsname}%
\let\auto@bib@innerbib\@empty
\bibitem [{\citenamefont {Cates}(2019)}]{cates2019active}%
  \BibitemOpen
  \bibfield  {author} {\bibinfo {author} {\bibfnamefont {M.~E.}\ \bibnamefont
  {Cates}},\ }\href@noop {} {\bibinfo {title} {Active field theories}}
  (\bibinfo {year} {2019}),\ \Eprint {https://arxiv.org/abs/1904.01330}
  {arXiv:1904.01330 [cond-mat.stat-mech]} \BibitemShut {NoStop}%
\bibitem [{\citenamefont {Menzel}\ and\ \citenamefont
  {L\"owen}(2013)}]{Menzel13}%
  \BibitemOpen
  \bibfield  {author} {\bibinfo {author} {\bibfnamefont {A.~M.}\ \bibnamefont
  {Menzel}}\ and\ \bibinfo {author} {\bibfnamefont {H.}~\bibnamefont
  {L\"owen}},\ }\href {https://doi.org/10.1103/PhysRevLett.110.055702}
  {\bibfield  {journal} {\bibinfo  {journal} {Phys. Rev. Lett.}\ }\textbf
  {\bibinfo {volume} {110}},\ \bibinfo {pages} {055702} (\bibinfo {year}
  {2013})}\BibitemShut {NoStop}%
\bibitem [{\citenamefont {Alaimo}\ \emph {et~al.}(2016)\citenamefont {Alaimo},
  \citenamefont {Praetorius},\ and\ \citenamefont {Voigt}}]{Alaimo_2016}%
  \BibitemOpen
  \bibfield  {author} {\bibinfo {author} {\bibfnamefont {F.}~\bibnamefont
  {Alaimo}}, \bibinfo {author} {\bibfnamefont {S.}~\bibnamefont {Praetorius}},\
  and\ \bibinfo {author} {\bibfnamefont {A.}~\bibnamefont {Voigt}},\ }\href
  {https://doi.org/10.1088/1367-2630/18/8/083008} {\bibfield  {journal}
  {\bibinfo  {journal} {New J. Phys.}\ }\textbf {\bibinfo {volume} {18}},\
  \bibinfo {pages} {083008} (\bibinfo {year} {2016})}\BibitemShut {NoStop}%
\bibitem [{\citenamefont {Peshkov}\ \emph {et~al.}(2012)\citenamefont
  {Peshkov}, \citenamefont {Ngo}, \citenamefont {Bertin}, \citenamefont
  {Chat{\'{e} }},\ and\ \citenamefont {Ginelli}}]{Peshkov_2012}%
  \BibitemOpen
  \bibfield  {author} {\bibinfo {author} {\bibfnamefont {A.}~\bibnamefont
  {Peshkov}}, \bibinfo {author} {\bibfnamefont {S.}~\bibnamefont {Ngo}},
  \bibinfo {author} {\bibfnamefont {E.}~\bibnamefont {Bertin}}, \bibinfo
  {author} {\bibfnamefont {H.}~\bibnamefont {Chat{\'{e} }}},\ and\ \bibinfo
  {author} {\bibfnamefont {F.}~\bibnamefont {Ginelli}},\ }\bibfield  {journal}
  {\bibinfo  {journal} {Phys. Rev. Lett.}\ }\textbf {\bibinfo {volume} {109}},\
  \href {https://doi.org/10.1103/physrevlett.109.098101}
  {10.1103/physrevlett.109.098101} (\bibinfo {year} {2012})\BibitemShut
  {NoStop}%
\bibitem [{\citenamefont {Wittkowski}\ \emph {et~al.}(2017)\citenamefont
  {Wittkowski}, \citenamefont {Stenhammar},\ and\ \citenamefont
  {Cates}}]{Wittkowski_2017}%
  \BibitemOpen
  \bibfield  {author} {\bibinfo {author} {\bibfnamefont {R.}~\bibnamefont
  {Wittkowski}}, \bibinfo {author} {\bibfnamefont {J.}~\bibnamefont
  {Stenhammar}},\ and\ \bibinfo {author} {\bibfnamefont {M.~E.}\ \bibnamefont
  {Cates}},\ }\href {https://doi.org/10.1088/1367-2630/aa8195} {\bibfield
  {journal} {\bibinfo  {journal} {New J. Phys.}\ }\textbf {\bibinfo {volume}
  {19}},\ \bibinfo {pages} {105003} (\bibinfo {year} {2017})}\BibitemShut
  {NoStop}%
\bibitem [{\citenamefont {Jülicher}\ \emph {et~al.}(2018)\citenamefont
  {Jülicher}, \citenamefont {Grill},\ and\ \citenamefont
  {Salbreux}}]{Jülicher_2018}%
  \BibitemOpen
  \bibfield  {author} {\bibinfo {author} {\bibfnamefont {F.}~\bibnamefont
  {Jülicher}}, \bibinfo {author} {\bibfnamefont {S.~W.}\ \bibnamefont
  {Grill}},\ and\ \bibinfo {author} {\bibfnamefont {G.}~\bibnamefont
  {Salbreux}},\ }\href {https://doi.org/10.1088/1361-6633/aab6bb} {\bibfield
  {journal} {\bibinfo  {journal} {Rep. Prog. Phys.}\ }\textbf {\bibinfo
  {volume} {81}},\ \bibinfo {pages} {076601} (\bibinfo {year}
  {2018})}\BibitemShut {NoStop}%
\bibitem [{\citenamefont {Tiribocchi}\ \emph {et~al.}(2015)\citenamefont
  {Tiribocchi}, \citenamefont {Wittkowski}, \citenamefont {Marenduzzo},\ and\
  \citenamefont {Cates}}]{Tiribocchi_15_model_H}%
  \BibitemOpen
  \bibfield  {author} {\bibinfo {author} {\bibfnamefont {A.}~\bibnamefont
  {Tiribocchi}}, \bibinfo {author} {\bibfnamefont {R.}~\bibnamefont
  {Wittkowski}}, \bibinfo {author} {\bibfnamefont {D.}~\bibnamefont
  {Marenduzzo}},\ and\ \bibinfo {author} {\bibfnamefont {M.~E.}\ \bibnamefont
  {Cates}},\ }\href {https://doi.org/10.1103/PhysRevLett.115.188302} {\bibfield
   {journal} {\bibinfo  {journal} {Phys. Rev. Lett.}\ }\textbf {\bibinfo
  {volume} {115}},\ \bibinfo {pages} {188302} (\bibinfo {year}
  {2015})}\BibitemShut {NoStop}%
\bibitem [{\citenamefont {Liebchen}\ and\ \citenamefont
  {Löwen}(2018)}]{liebchen2018synthetic}%
  \BibitemOpen
  \bibfield  {author} {\bibinfo {author} {\bibfnamefont {B.}~\bibnamefont
  {Liebchen}}\ and\ \bibinfo {author} {\bibfnamefont {H.}~\bibnamefont
  {Löwen}},\ }\href@noop {} {\bibfield  {journal} {\bibinfo  {journal} {Acc.
  Chem. Res.}\ }\textbf {\bibinfo {volume} {51}},\ \bibinfo {pages} {2982}
  (\bibinfo {year} {2018})}\BibitemShut {NoStop}%
\bibitem [{\citenamefont {Weber}\ \emph {et~al.}(2019)\citenamefont {Weber},
  \citenamefont {Zwicker}, \citenamefont {Jülicher},\ and\ \citenamefont
  {Lee}}]{Weber_2019}%
  \BibitemOpen
  \bibfield  {author} {\bibinfo {author} {\bibfnamefont {C.~A.}\ \bibnamefont
  {Weber}}, \bibinfo {author} {\bibfnamefont {D.}~\bibnamefont {Zwicker}},
  \bibinfo {author} {\bibfnamefont {F.}~\bibnamefont {Jülicher}},\ and\
  \bibinfo {author} {\bibfnamefont {C.~F.}\ \bibnamefont {Lee}},\ }\href
  {https://doi.org/10.1088/1361-6633/ab052b} {\bibfield  {journal} {\bibinfo
  {journal} {Rep. Prog. Phys.}\ }\textbf {\bibinfo {volume} {82}},\ \bibinfo
  {pages} {064601} (\bibinfo {year} {2019})}\BibitemShut {NoStop}%
\bibitem [{\citenamefont {Hickey}\ \emph {et~al.}(2023)\citenamefont {Hickey},
  \citenamefont {Golestanian},\ and\ \citenamefont
  {Vilfan}}]{hickey2023nonreciprocal}%
  \BibitemOpen
  \bibfield  {author} {\bibinfo {author} {\bibfnamefont {D.~J.}\ \bibnamefont
  {Hickey}}, \bibinfo {author} {\bibfnamefont {R.}~\bibnamefont
  {Golestanian}},\ and\ \bibinfo {author} {\bibfnamefont {A.}~\bibnamefont
  {Vilfan}},\ }\href {https://doi.org/doi:10.1073/pnas.2307279120} {\bibfield
  {journal} {\bibinfo  {journal} {Proc. Natl. Acad. Sci. U.S.A.}\ }\textbf
  {\bibinfo {volume} {120}},\ \bibinfo {pages} {e2307279120} (\bibinfo {year}
  {2023})}\BibitemShut {NoStop}%
\bibitem [{\citenamefont {Ramaswamy}\ \emph {et~al.}(2000)\citenamefont
  {Ramaswamy}, \citenamefont {Toner},\ and\ \citenamefont
  {Prost}}]{Ramaswamy2000}%
  \BibitemOpen
  \bibfield  {author} {\bibinfo {author} {\bibfnamefont {S.}~\bibnamefont
  {Ramaswamy}}, \bibinfo {author} {\bibfnamefont {J.}~\bibnamefont {Toner}},\
  and\ \bibinfo {author} {\bibfnamefont {J.}~\bibnamefont {Prost}},\ }\href
  {https://link.aps.org/doi/10.1103/PhysRevLett.84.3494} {\bibfield  {journal}
  {\bibinfo  {journal} {Phys. Rev. Lett.}\ }\textbf {\bibinfo {volume} {84}},\
  \bibinfo {pages} {3494} (\bibinfo {year} {2000})}\BibitemShut {NoStop}%
\bibitem [{\citenamefont {Kozyreff}\ and\ \citenamefont
  {Tlidi}(2007)}]{Kozyreff07_Swift_H}%
  \BibitemOpen
  \bibfield  {author} {\bibinfo {author} {\bibfnamefont {G.}~\bibnamefont
  {Kozyreff}}\ and\ \bibinfo {author} {\bibfnamefont {M.}~\bibnamefont
  {Tlidi}},\ }\href {https://doi.org/10.1063/1.2759436} {\bibfield  {journal}
  {\bibinfo  {journal} {Chaos (Woodbury, N.Y.)}\ }\textbf {\bibinfo {volume}
  {17}},\ \bibinfo {pages} {037103} (\bibinfo {year} {2007})}\BibitemShut
  {NoStop}%
\bibitem [{\citenamefont {Bhattacharya}\ \emph {et~al.}(2020)\citenamefont
  {Bhattacharya}, \citenamefont {Banerjee}, \citenamefont {Miao}, \citenamefont
  {Zhan}, \citenamefont {Devreotes},\ and\ \citenamefont
  {Iglesias}}]{Bhattacharya20}%
  \BibitemOpen
  \bibfield  {author} {\bibinfo {author} {\bibfnamefont {S.}~\bibnamefont
  {Bhattacharya}}, \bibinfo {author} {\bibfnamefont {T.}~\bibnamefont
  {Banerjee}}, \bibinfo {author} {\bibfnamefont {Y.}~\bibnamefont {Miao}},
  \bibinfo {author} {\bibfnamefont {H.}~\bibnamefont {Zhan}}, \bibinfo {author}
  {\bibfnamefont {P.}~\bibnamefont {Devreotes}},\ and\ \bibinfo {author}
  {\bibfnamefont {P.}~\bibnamefont {Iglesias}},\ }\href
  {https://www.science.org/doi/10.1126/sciadv.aay7682} {\bibfield  {journal}
  {\bibinfo  {journal} {Sci. Adv.}\ }\textbf {\bibinfo {volume} {6}},\ \bibinfo
  {pages} {eaay7682} (\bibinfo {year} {2020})}\BibitemShut {NoStop}%
\bibitem [{\citenamefont {Kohyama}\ \emph {et~al.}(2019)\citenamefont
  {Kohyama}, \citenamefont {Yoshinaga}, \citenamefont {Yanagisawa},
  \citenamefont {Fujiwara},\ and\ \citenamefont {Doi}}]{Kohyama19}%
  \BibitemOpen
  \bibfield  {author} {\bibinfo {author} {\bibfnamefont {S.}~\bibnamefont
  {Kohyama}}, \bibinfo {author} {\bibfnamefont {N.}~\bibnamefont {Yoshinaga}},
  \bibinfo {author} {\bibfnamefont {M.}~\bibnamefont {Yanagisawa}}, \bibinfo
  {author} {\bibfnamefont {K.}~\bibnamefont {Fujiwara}},\ and\ \bibinfo
  {author} {\bibfnamefont {N.}~\bibnamefont {Doi}},\ }\href
  {https://doi.org/10.7554/eLife.44591} {\bibfield  {journal} {\bibinfo
  {journal} {eLife}\ }\textbf {\bibinfo {volume} {8}},\ \bibinfo {pages}
  {e44591} (\bibinfo {year} {2019})}\BibitemShut {NoStop}%
\bibitem [{\citenamefont {John}\ and\ \citenamefont
  {B\"ar}(2005)}]{John2005_membranes}%
  \BibitemOpen
  \bibfield  {author} {\bibinfo {author} {\bibfnamefont {K.}~\bibnamefont
  {John}}\ and\ \bibinfo {author} {\bibfnamefont {M.}~\bibnamefont {B\"ar}},\
  }\href {https://doi.org/10.1103/PhysRevLett.95.198101} {\bibfield  {journal}
  {\bibinfo  {journal} {Phys. Rev. Lett.}\ }\textbf {\bibinfo {volume} {95}},\
  \bibinfo {pages} {198101} (\bibinfo {year} {2005})}\BibitemShut {NoStop}%
\bibitem [{\citenamefont {Zheng}\ and\ \citenamefont {Shen}(2015)}]{ZHENG2015}%
  \BibitemOpen
  \bibfield  {author} {\bibinfo {author} {\bibfnamefont {Q.}~\bibnamefont
  {Zheng}}\ and\ \bibinfo {author} {\bibfnamefont {J.}~\bibnamefont {Shen}},\
  }\href {https://doi.org/https://doi.org/10.1016/j.camwa.2015.06.031}
  {\bibfield  {journal} {\bibinfo  {journal} {Comput. Math. with Appl.}\
  }\textbf {\bibinfo {volume} {70}},\ \bibinfo {pages} {1082} (\bibinfo {year}
  {2015})}\BibitemShut {NoStop}%
\bibitem [{\citenamefont {Peña}\ \emph {et~al.}(2004)\citenamefont {Peña},
  \citenamefont {Pérez-García},\ and\ \citenamefont
  {Bestehorn}}]{Pena04_wave_insta}%
  \BibitemOpen
  \bibfield  {author} {\bibinfo {author} {\bibfnamefont {B.}~\bibnamefont
  {Peña}}, \bibinfo {author} {\bibfnamefont {C.}~\bibnamefont
  {Pérez-García}},\ and\ \bibinfo {author} {\bibfnamefont {M.}~\bibnamefont
  {Bestehorn}},\ }\href {https://doi.org/10.1142/S021812740401165X} {\bibfield
  {journal} {\bibinfo  {journal} {I. J. Bifurcation and Chaos}\ }\textbf
  {\bibinfo {volume} {14}},\ \bibinfo {pages} {3899} (\bibinfo {year}
  {2004})}\BibitemShut {NoStop}%
\bibitem [{\citenamefont {Okuzono}\ and\ \citenamefont
  {Ohta}(2003)}]{Okuzono2003}%
  \BibitemOpen
  \bibfield  {author} {\bibinfo {author} {\bibfnamefont {T.}~\bibnamefont
  {Okuzono}}\ and\ \bibinfo {author} {\bibfnamefont {T.}~\bibnamefont {Ohta}},\
  }\href {https://doi.org/10.1103/PhysRevE.67.056211} {\bibfield  {journal}
  {\bibinfo  {journal} {Phys. Rev. E}\ }\textbf {\bibinfo {volume} {67}},\
  \bibinfo {pages} {056211} (\bibinfo {year} {2003})}\BibitemShut {NoStop}%
\bibitem [{\citenamefont {Tong}\ and\ \citenamefont
  {Yang}(2002)}]{tong2002phase}%
  \BibitemOpen
  \bibfield  {author} {\bibinfo {author} {\bibfnamefont {C.}~\bibnamefont
  {Tong}}\ and\ \bibinfo {author} {\bibfnamefont {Y.}~\bibnamefont {Yang}},\
  }\href@noop {} {\bibfield  {journal} {\bibinfo  {journal} {J. Chem. Phys.}\
  }\textbf {\bibinfo {volume} {116}},\ \bibinfo {pages} {1519} (\bibinfo {year}
  {2002})}\BibitemShut {NoStop}%
\bibitem [{\citenamefont {You}\ \emph {et~al.}(2020)\citenamefont {You},
  \citenamefont {Baskaran},\ and\ \citenamefont {Marchetti}}]{You_2020}%
  \BibitemOpen
  \bibfield  {author} {\bibinfo {author} {\bibfnamefont {Z.}~\bibnamefont
  {You}}, \bibinfo {author} {\bibfnamefont {A.}~\bibnamefont {Baskaran}},\ and\
  \bibinfo {author} {\bibfnamefont {M.~C.}\ \bibnamefont {Marchetti}},\ }\href
  {http://dx.doi.org/10.1073/pnas.2010318117} {\bibfield  {journal} {\bibinfo
  {journal} {Proc. Natl. Acad. Sci. U. S. A.}\ }\textbf {\bibinfo {volume}
  {117}},\ \bibinfo {pages} {19767–19772} (\bibinfo {year}
  {2020})}\BibitemShut {NoStop}%
\bibitem [{\citenamefont {Saha}\ \emph {et~al.}(2020)\citenamefont {Saha},
  \citenamefont {Agudo-Canalejo},\ and\ \citenamefont
  {Golestanian}}]{Saha_2020}%
  \BibitemOpen
  \bibfield  {author} {\bibinfo {author} {\bibfnamefont {S.}~\bibnamefont
  {Saha}}, \bibinfo {author} {\bibfnamefont {J.}~\bibnamefont
  {Agudo-Canalejo}},\ and\ \bibinfo {author} {\bibfnamefont {R.}~\bibnamefont
  {Golestanian}},\ }\href {http://dx.doi.org/10.1103/PhysRevX.10.041009}
  {\bibfield  {journal} {\bibinfo  {journal} {Phys. Rev. X}\ }\textbf {\bibinfo
  {volume} {10}},\ \bibinfo {pages} {041009} (\bibinfo {year}
  {2020})}\BibitemShut {NoStop}%
\bibitem [{\citenamefont {Frohoff-H{\"u}lsmann}\ and\ \citenamefont
  {Thiele}(2023)}]{frohoff2023nonreciprocal}%
  \BibitemOpen
  \bibfield  {author} {\bibinfo {author} {\bibfnamefont {T.}~\bibnamefont
  {Frohoff-H{\"u}lsmann}}\ and\ \bibinfo {author} {\bibfnamefont
  {U.}~\bibnamefont {Thiele}},\ }\href@noop {} {\bibinfo {title} {Nonreciprocal
  {C}ahn-{H}illiard equations emerging as one of eight universal amplitude
  equations}} (\bibinfo {year} {2023}),\ \Eprint
  {https://arxiv.org/abs/2301.05568} {arXiv:2301.05568 [cond-mat.soft]}
  \BibitemShut {NoStop}%
\bibitem [{\citenamefont {Frohoff-H{\"u}lsmann}\ \emph
  {et~al.}(2023)\citenamefont {Frohoff-H{\"u}lsmann}, \citenamefont {Thiele},\
  and\ \citenamefont {Pismen}}]{frohoff2023non}%
  \BibitemOpen
  \bibfield  {author} {\bibinfo {author} {\bibfnamefont {T.}~\bibnamefont
  {Frohoff-H{\"u}lsmann}}, \bibinfo {author} {\bibfnamefont {U.}~\bibnamefont
  {Thiele}},\ and\ \bibinfo {author} {\bibfnamefont {L.~M.}\ \bibnamefont
  {Pismen}},\ }\href@noop {} {\bibfield  {journal} {\bibinfo  {journal}
  {Philosophical Transactions of the Royal Society A}\ }\textbf {\bibinfo
  {volume} {381}},\ \bibinfo {pages} {20220087} (\bibinfo {year}
  {2023})}\BibitemShut {NoStop}%
\bibitem [{\citenamefont {Malomed}\ and\ \citenamefont
  {Tribelsky}(1984)}]{MALOMED1984}%
  \BibitemOpen
  \bibfield  {author} {\bibinfo {author} {\bibfnamefont {B.}~\bibnamefont
  {Malomed}}\ and\ \bibinfo {author} {\bibfnamefont {M.}~\bibnamefont
  {Tribelsky}},\ }\href
  {https://www.sciencedirect.com/science/article/pii/0167278984900058}
  {\bibfield  {journal} {\bibinfo  {journal} {Phys. D: Nonlinear Phenom.}\
  }\textbf {\bibinfo {volume} {14}},\ \bibinfo {pages} {67} (\bibinfo {year}
  {1984})}\BibitemShut {NoStop}%
\bibitem [{\citenamefont {Cummins}\ \emph {et~al.}(1993)\citenamefont
  {Cummins}, \citenamefont {Fourtune},\ and\ \citenamefont
  {Rabaud}}]{Cummins93}%
  \BibitemOpen
  \bibfield  {author} {\bibinfo {author} {\bibfnamefont {H.~Z.}\ \bibnamefont
  {Cummins}}, \bibinfo {author} {\bibfnamefont {L.}~\bibnamefont {Fourtune}},\
  and\ \bibinfo {author} {\bibfnamefont {M.}~\bibnamefont {Rabaud}},\ }\href
  {https://link.aps.org/doi/10.1103/PhysRevE.47.1727} {\bibfield  {journal}
  {\bibinfo  {journal} {Phys. Rev. E}\ }\textbf {\bibinfo {volume} {47}},\
  \bibinfo {pages} {1727} (\bibinfo {year} {1993})}\BibitemShut {NoStop}%
\bibitem [{\citenamefont {Dangelmayr}\ \emph {et~al.}(1997)\citenamefont
  {Dangelmayr}, \citenamefont {Hettel},\ and\ \citenamefont
  {Knobloch}}]{Dangelmayr_1997}%
  \BibitemOpen
  \bibfield  {author} {\bibinfo {author} {\bibfnamefont {G.}~\bibnamefont
  {Dangelmayr}}, \bibinfo {author} {\bibfnamefont {J.}~\bibnamefont {Hettel}},\
  and\ \bibinfo {author} {\bibfnamefont {E.}~\bibnamefont {Knobloch}},\ }\href
  {https://doi.org/10.1088/0951-7715/10/5/006} {\bibfield  {journal} {\bibinfo
  {journal} {Nonlinearity}\ }\textbf {\bibinfo {volume} {10}},\ \bibinfo
  {pages} {1093} (\bibinfo {year} {1997})}\BibitemShut {NoStop}%
\bibitem [{\citenamefont {Zhabotinsky}\ \emph {et~al.}(1995)\citenamefont
  {Zhabotinsky}, \citenamefont {Dolnik},\ and\ \citenamefont
  {Epstein}}]{Zhabotinsky95}%
  \BibitemOpen
  \bibfield  {author} {\bibinfo {author} {\bibfnamefont {A.~M.}\ \bibnamefont
  {Zhabotinsky}}, \bibinfo {author} {\bibfnamefont {M.}~\bibnamefont
  {Dolnik}},\ and\ \bibinfo {author} {\bibfnamefont {I.~R.}\ \bibnamefont
  {Epstein}},\ }\href {https://doi.org/10.1063/1.469932} {\bibfield  {journal}
  {\bibinfo  {journal} {The Journal of Chemical Physics}\ }\textbf {\bibinfo
  {volume} {103}},\ \bibinfo {pages} {10306} (\bibinfo {year} {1995})},\
  \Eprint
  {https://arxiv.org/abs/https://pubs.aip.org/aip/jcp/article-pdf/103/23/10306/9436240/10306\_1\_online.pdf}
  {https://pubs.aip.org/aip/jcp/article-pdf/103/23/10306/9436240/10306\_1\_online.pdf}
  \BibitemShut {NoStop}%
\bibitem [{\citenamefont {Pan}\ and\ \citenamefont {de~Bruyn}(1994)}]{Pan94}%
  \BibitemOpen
  \bibfield  {author} {\bibinfo {author} {\bibfnamefont {L.}~\bibnamefont
  {Pan}}\ and\ \bibinfo {author} {\bibfnamefont {J.~R.}\ \bibnamefont
  {de~Bruyn}},\ }\href {https://link.aps.org/doi/10.1103/PhysRevE.49.483}
  {\bibfield  {journal} {\bibinfo  {journal} {Phys. Rev. E}\ }\textbf {\bibinfo
  {volume} {49}},\ \bibinfo {pages} {483} (\bibinfo {year} {1994})}\BibitemShut
  {NoStop}%
\bibitem [{\citenamefont {Bestehorn}\ \emph {et~al.}(1989)\citenamefont
  {Bestehorn}, \citenamefont {Friedrich},\ and\ \citenamefont
  {Haken}}]{BESTEHORN1989}%
  \BibitemOpen
  \bibfield  {author} {\bibinfo {author} {\bibfnamefont {M.}~\bibnamefont
  {Bestehorn}}, \bibinfo {author} {\bibfnamefont {R.}~\bibnamefont
  {Friedrich}},\ and\ \bibinfo {author} {\bibfnamefont {H.}~\bibnamefont
  {Haken}},\ }\href
  {https://www.sciencedirect.com/science/article/pii/0167278989901371}
  {\bibfield  {journal} {\bibinfo  {journal} {Phys. D: Nonlinear Phenom.}\
  }\textbf {\bibinfo {volume} {37}},\ \bibinfo {pages} {295} (\bibinfo {year}
  {1989})}\BibitemShut {NoStop}%
\bibitem [{\citenamefont {Coullet}\ \emph {et~al.}(1989)\citenamefont
  {Coullet}, \citenamefont {Goldstein},\ and\ \citenamefont
  {Gunaratne}}]{Coullet89}%
  \BibitemOpen
  \bibfield  {author} {\bibinfo {author} {\bibfnamefont {P.}~\bibnamefont
  {Coullet}}, \bibinfo {author} {\bibfnamefont {R.~E.}\ \bibnamefont
  {Goldstein}},\ and\ \bibinfo {author} {\bibfnamefont {G.~H.}\ \bibnamefont
  {Gunaratne}},\ }\href {https://link.aps.org/doi/10.1103/PhysRevLett.63.1954}
  {\bibfield  {journal} {\bibinfo  {journal} {Phys. Rev. Lett.}\ }\textbf
  {\bibinfo {volume} {63}},\ \bibinfo {pages} {1954} (\bibinfo {year}
  {1989})}\BibitemShut {NoStop}%
\bibitem [{\citenamefont {Rabaud}\ \emph {et~al.}(1990)\citenamefont {Rabaud},
  \citenamefont {Michalland},\ and\ \citenamefont {Couder}}]{Rabaud90_viscous}%
  \BibitemOpen
  \bibfield  {author} {\bibinfo {author} {\bibfnamefont {M.}~\bibnamefont
  {Rabaud}}, \bibinfo {author} {\bibfnamefont {S.}~\bibnamefont {Michalland}},\
  and\ \bibinfo {author} {\bibfnamefont {Y.}~\bibnamefont {Couder}},\ }\href
  {https://doi.org/10.1103/PhysRevLett.64.184} {\bibfield  {journal} {\bibinfo
  {journal} {Phys. Rev. Lett.}\ }\textbf {\bibinfo {volume} {64}},\ \bibinfo
  {pages} {184} (\bibinfo {year} {1990})}\BibitemShut {NoStop}%
\bibitem [{\citenamefont {Goldstein}\ \emph {et~al.}(1991)\citenamefont
  {Goldstein}, \citenamefont {Gunaratne}, \citenamefont {Gil},\ and\
  \citenamefont {Coullet}}]{Goldstein91PT}%
  \BibitemOpen
  \bibfield  {author} {\bibinfo {author} {\bibfnamefont {R.~E.}\ \bibnamefont
  {Goldstein}}, \bibinfo {author} {\bibfnamefont {G.~H.}\ \bibnamefont
  {Gunaratne}}, \bibinfo {author} {\bibfnamefont {L.}~\bibnamefont {Gil}},\
  and\ \bibinfo {author} {\bibfnamefont {P.}~\bibnamefont {Coullet}},\ }\href
  {https://doi.org/10.1103/PhysRevA.43.6700} {\bibfield  {journal} {\bibinfo
  {journal} {Phys. Rev. A}\ }\textbf {\bibinfo {volume} {43}},\ \bibinfo
  {pages} {6700} (\bibinfo {year} {1991})}\BibitemShut {NoStop}%
\bibitem [{\citenamefont {Ophaus}\ \emph {et~al.}(2021)\citenamefont {Ophaus},
  \citenamefont {Knobloch}, \citenamefont {Gurevich},\ and\ \citenamefont
  {Thiele}}]{Ophaus21}%
  \BibitemOpen
  \bibfield  {author} {\bibinfo {author} {\bibfnamefont {L.}~\bibnamefont
  {Ophaus}}, \bibinfo {author} {\bibfnamefont {E.}~\bibnamefont {Knobloch}},
  \bibinfo {author} {\bibfnamefont {S.~V.}\ \bibnamefont {Gurevich}},\ and\
  \bibinfo {author} {\bibfnamefont {U.}~\bibnamefont {Thiele}},\ }\href
  {https://doi.org/10.1103/PhysRevE.103.032601} {\bibfield  {journal} {\bibinfo
   {journal} {Phys. Rev. E}\ }\textbf {\bibinfo {volume} {103}},\ \bibinfo
  {pages} {032601} (\bibinfo {year} {2021})}\BibitemShut {NoStop}%
\bibitem [{\citenamefont {Frohoff-H{\"u}lsmann}\ and\ \citenamefont
  {Thiele}(2021)}]{frohoff2021localized}%
  \BibitemOpen
  \bibfield  {author} {\bibinfo {author} {\bibfnamefont {T.}~\bibnamefont
  {Frohoff-H{\"u}lsmann}}\ and\ \bibinfo {author} {\bibfnamefont
  {U.}~\bibnamefont {Thiele}},\ }\href@noop {} {\bibfield  {journal} {\bibinfo
  {journal} {IMA J. Appl. Math.}\ }\textbf {\bibinfo {volume} {86}},\ \bibinfo
  {pages} {924} (\bibinfo {year} {2021})}\BibitemShut {NoStop}%
\bibitem [{\citenamefont {Stegemerten}\ \emph {et~al.}(2022)\citenamefont
  {Stegemerten}, \citenamefont {John},\ and\ \citenamefont
  {Thiele}}]{Stegemerten22}%
  \BibitemOpen
  \bibfield  {author} {\bibinfo {author} {\bibfnamefont {F.}~\bibnamefont
  {Stegemerten}}, \bibinfo {author} {\bibfnamefont {K.}~\bibnamefont {John}},\
  and\ \bibinfo {author} {\bibfnamefont {U.}~\bibnamefont {Thiele}},\ }\href
  {https://doi.org/10.1039/D2SM00648K} {\bibfield  {journal} {\bibinfo
  {journal} {Soft Matter}\ }\textbf {\bibinfo {volume} {18}},\ \bibinfo {pages}
  {5823} (\bibinfo {year} {2022})}\BibitemShut {NoStop}%
\bibitem [{\citenamefont {Fruchart}\ \emph {et~al.}(2021)\citenamefont
  {Fruchart}, \citenamefont {Hanai}, \citenamefont {Littlewood},\ and\
  \citenamefont {Vitelli}}]{Fruchart2021}%
  \BibitemOpen
  \bibfield  {author} {\bibinfo {author} {\bibfnamefont {M.}~\bibnamefont
  {Fruchart}}, \bibinfo {author} {\bibfnamefont {R.}~\bibnamefont {Hanai}},
  \bibinfo {author} {\bibfnamefont {P.~B.}\ \bibnamefont {Littlewood}},\ and\
  \bibinfo {author} {\bibfnamefont {V.}~\bibnamefont {Vitelli}},\ }\href
  {http://dx.doi.org/10.1038/s41586-021-03375-9} {\bibfield  {journal}
  {\bibinfo  {journal} {Nature}\ }\textbf {\bibinfo {volume} {592}},\ \bibinfo
  {pages} {363–369} (\bibinfo {year} {2021})}\BibitemShut {NoStop}%
\bibitem [{\citenamefont {Zhang}\ \emph {et~al.}(2020)\citenamefont {Zhang},
  \citenamefont {Sokolov},\ and\ \citenamefont
  {Snezhko}}]{zhang2020reconfigurable}%
  \BibitemOpen
  \bibfield  {author} {\bibinfo {author} {\bibfnamefont {B.}~\bibnamefont
  {Zhang}}, \bibinfo {author} {\bibfnamefont {A.}~\bibnamefont {Sokolov}},\
  and\ \bibinfo {author} {\bibfnamefont {A.}~\bibnamefont {Snezhko}},\
  }\href@noop {} {\bibfield  {journal} {\bibinfo  {journal} {Nature
  communications}\ }\textbf {\bibinfo {volume} {11}},\ \bibinfo {pages} {4401}
  (\bibinfo {year} {2020})}\BibitemShut {NoStop}%
\bibitem [{\citenamefont {Liao}\ and\ \citenamefont
  {Klapp}(2021)}]{Liao21_chiral}%
  \BibitemOpen
  \bibfield  {author} {\bibinfo {author} {\bibfnamefont {G.-J.}\ \bibnamefont
  {Liao}}\ and\ \bibinfo {author} {\bibfnamefont {S.~H.~L.}\ \bibnamefont
  {Klapp}},\ }\href {https://doi.org/10.1039/D1SM00545F} {\bibfield  {journal}
  {\bibinfo  {journal} {Soft Matter}\ }\textbf {\bibinfo {volume} {17}},\
  \bibinfo {pages} {6833} (\bibinfo {year} {2021})}\BibitemShut {NoStop}%
\bibitem [{\citenamefont {Cross}\ and\ \citenamefont
  {Hohenberg}(1993)}]{Cross93}%
  \BibitemOpen
  \bibfield  {author} {\bibinfo {author} {\bibfnamefont {M.~C.}\ \bibnamefont
  {Cross}}\ and\ \bibinfo {author} {\bibfnamefont {P.~C.}\ \bibnamefont
  {Hohenberg}},\ }\href {https://link.aps.org/doi/10.1103/RevModPhys.65.851}
  {\bibfield  {journal} {\bibinfo  {journal} {Rev. Mod. Phys.}\ }\textbf
  {\bibinfo {volume} {65}},\ \bibinfo {pages} {851} (\bibinfo {year}
  {1993})}\BibitemShut {NoStop}%
\bibitem [{\citenamefont {Cross}\ and\ \citenamefont
  {Greenside}(2009)}]{cross_greenside_2009}%
  \BibitemOpen
  \bibfield  {author} {\bibinfo {author} {\bibfnamefont {M.}~\bibnamefont
  {Cross}}\ and\ \bibinfo {author} {\bibfnamefont {H.}~\bibnamefont
  {Greenside}},\ }\bibinfo {title} {Oscillatory patterns},\ in\ \href@noop {}
  {\emph {\bibinfo {booktitle} {Pattern Formation and Dynamics in
  Nonequilibrium Systems}}}\ (\bibinfo  {publisher} {Cambridge University
  Press},\ \bibinfo {year} {2009})\ p.\ \bibinfo {pages}
  {358–400}\BibitemShut {NoStop}%
\bibitem [{\citenamefont {Hanai}\ and\ \citenamefont
  {Littlewood}(2020)}]{hanai2020critical}%
  \BibitemOpen
  \bibfield  {author} {\bibinfo {author} {\bibfnamefont {R.}~\bibnamefont
  {Hanai}}\ and\ \bibinfo {author} {\bibfnamefont {P.~B.}\ \bibnamefont
  {Littlewood}},\ }\href
  {https://journals.aps.org/prresearch/abstract/10.1103/PhysRevResearch.2.033018}
  {\bibfield  {journal} {\bibinfo  {journal} {Phys. {R}ev. Research}\ }\textbf
  {\bibinfo {volume} {2}},\ \bibinfo {pages} {033018} (\bibinfo {year}
  {2020})}\BibitemShut {NoStop}%
\bibitem [{\citenamefont {Zhang}\ \emph {et~al.}(2019)\citenamefont {Zhang},
  \citenamefont {Sweeney}, \citenamefont {Hsu}, \citenamefont {Yang},
  \citenamefont {Stone},\ and\ \citenamefont {Jiang}}]{Zhang19}%
  \BibitemOpen
  \bibfield  {author} {\bibinfo {author} {\bibfnamefont {M.}~\bibnamefont
  {Zhang}}, \bibinfo {author} {\bibfnamefont {W.}~\bibnamefont {Sweeney}},
  \bibinfo {author} {\bibfnamefont {C.~W.}\ \bibnamefont {Hsu}}, \bibinfo
  {author} {\bibfnamefont {L.}~\bibnamefont {Yang}}, \bibinfo {author}
  {\bibfnamefont {A.~D.}\ \bibnamefont {Stone}},\ and\ \bibinfo {author}
  {\bibfnamefont {L.}~\bibnamefont {Jiang}},\ }\href
  {https://doi.org/10.1103/PhysRevLett.123.180501} {\bibfield  {journal}
  {\bibinfo  {journal} {Phys. Rev. Lett.}\ }\textbf {\bibinfo {volume} {123}},\
  \bibinfo {pages} {180501} (\bibinfo {year} {2019})}\BibitemShut {NoStop}%
\bibitem [{\citenamefont {Farrell}\ and\ \citenamefont
  {Ioannou}(1994)}]{Farrell94}%
  \BibitemOpen
  \bibfield  {author} {\bibinfo {author} {\bibfnamefont {B.~F.}\ \bibnamefont
  {Farrell}}\ and\ \bibinfo {author} {\bibfnamefont {P.~J.}\ \bibnamefont
  {Ioannou}},\ }\href {https://doi.org/10.1103/PhysRevLett.72.1188} {\bibfield
  {journal} {\bibinfo  {journal} {Phys. Rev. Lett.}\ }\textbf {\bibinfo
  {volume} {72}},\ \bibinfo {pages} {1188} (\bibinfo {year}
  {1994})}\BibitemShut {NoStop}%
\bibitem [{\citenamefont {Biancalani}\ \emph {et~al.}(2017)\citenamefont
  {Biancalani}, \citenamefont {Jafarpour},\ and\ \citenamefont
  {Goldenfeld}}]{Biancalani17}%
  \BibitemOpen
  \bibfield  {author} {\bibinfo {author} {\bibfnamefont {T.}~\bibnamefont
  {Biancalani}}, \bibinfo {author} {\bibfnamefont {F.}~\bibnamefont
  {Jafarpour}},\ and\ \bibinfo {author} {\bibfnamefont {N.}~\bibnamefont
  {Goldenfeld}},\ }\href {https://doi.org/10.1103/PhysRevLett.118.018101}
  {\bibfield  {journal} {\bibinfo  {journal} {Phys. Rev. Lett.}\ }\textbf
  {\bibinfo {volume} {118}},\ \bibinfo {pages} {018101} (\bibinfo {year}
  {2017})}\BibitemShut {NoStop}%
\bibitem [{\citenamefont {Zhong}\ \emph {et~al.}(2020)\citenamefont {Zhong},
  \citenamefont {Ozdemir}, \citenamefont {Eisfeld}, \citenamefont {Metelmann},\
  and\ \citenamefont {El-Ganainy}}]{Zhong20}%
  \BibitemOpen
  \bibfield  {author} {\bibinfo {author} {\bibfnamefont {Q.}~\bibnamefont
  {Zhong}}, \bibinfo {author} {\bibfnamefont {S.}~\bibnamefont {Ozdemir}},
  \bibinfo {author} {\bibfnamefont {A.}~\bibnamefont {Eisfeld}}, \bibinfo
  {author} {\bibfnamefont {A.}~\bibnamefont {Metelmann}},\ and\ \bibinfo
  {author} {\bibfnamefont {R.}~\bibnamefont {El-Ganainy}},\ }\href
  {https://doi.org/10.1103/PhysRevApplied.13.014070} {\bibfield  {journal}
  {\bibinfo  {journal} {Phys. Rev. Appl.}\ }\textbf {\bibinfo {volume} {13}},\
  \bibinfo {pages} {014070} (\bibinfo {year} {2020})}\BibitemShut {NoStop}%
\bibitem [{\citenamefont {Nardini}\ \emph {et~al.}(2017)\citenamefont
  {Nardini}, \citenamefont {Fodor}, \citenamefont {Tjhung}, \citenamefont {van
  Wijland}, \citenamefont {Tailleur},\ and\ \citenamefont
  {Cates}}]{Nardini2017}%
  \BibitemOpen
  \bibfield  {author} {\bibinfo {author} {\bibfnamefont {C.}~\bibnamefont
  {Nardini}}, \bibinfo {author} {\bibfnamefont {{\'E}.}~\bibnamefont {Fodor}},
  \bibinfo {author} {\bibfnamefont {E.}~\bibnamefont {Tjhung}}, \bibinfo
  {author} {\bibfnamefont {F.}~\bibnamefont {van Wijland}}, \bibinfo {author}
  {\bibfnamefont {J.}~\bibnamefont {Tailleur}},\ and\ \bibinfo {author}
  {\bibfnamefont {M.~E.}\ \bibnamefont {Cates}},\ }\href
  {https://link.aps.org/doi/10.1103/PhysRevX.7.021007} {\bibfield  {journal}
  {\bibinfo  {journal} {Phys. Rev. X}\ }\textbf {\bibinfo {volume} {7}},\
  \bibinfo {pages} {021007} (\bibinfo {year} {2017})}\BibitemShut {NoStop}%
\bibitem [{\citenamefont {Seara}\ \emph {et~al.}(2021)\citenamefont {Seara},
  \citenamefont {Machta},\ and\ \citenamefont {Murrell}}]{seara2021}%
  \BibitemOpen
  \bibfield  {author} {\bibinfo {author} {\bibfnamefont {D.~S.}\ \bibnamefont
  {Seara}}, \bibinfo {author} {\bibfnamefont {B.~B.}\ \bibnamefont {Machta}},\
  and\ \bibinfo {author} {\bibfnamefont {M.~P.}\ \bibnamefont {Murrell}},\
  }\href {https://doi.org/10.1038/s41467-020-20281-2} {\bibfield  {journal}
  {\bibinfo  {journal} {Nat. Commun.}\ }\textbf {\bibinfo {volume} {12}},\
  \bibinfo {pages} {1} (\bibinfo {year} {2021})}\BibitemShut {NoStop}%
\bibitem [{\citenamefont {Caballero}\ and\ \citenamefont
  {Cates}(2020)}]{caballero2020stealth}%
  \BibitemOpen
  \bibfield  {author} {\bibinfo {author} {\bibfnamefont {F.}~\bibnamefont
  {Caballero}}\ and\ \bibinfo {author} {\bibfnamefont {M.~E.}\ \bibnamefont
  {Cates}},\ }\href {https://doi.org/10.1103/PhysRevLett.124.240604} {\bibfield
   {journal} {\bibinfo  {journal} {Phys. Rev. Lett.}\ }\textbf {\bibinfo
  {volume} {124}},\ \bibinfo {pages} {240604} (\bibinfo {year}
  {2020})}\BibitemShut {NoStop}%
\bibitem [{\citenamefont {Meibohm}\ and\ \citenamefont
  {Esposito}(2023)}]{meibohm2023landau}%
  \BibitemOpen
  \bibfield  {author} {\bibinfo {author} {\bibfnamefont {J.}~\bibnamefont
  {Meibohm}}\ and\ \bibinfo {author} {\bibfnamefont {M.}~\bibnamefont
  {Esposito}},\ }\href
  {https://iopscience.iop.org/article/10.1088/1367-2630/acbc41/meta} {\bibfield
   {journal} {\bibinfo  {journal} {New J. Phys.}\ }\textbf {\bibinfo {volume}
  {25}},\ \bibinfo {pages} {023034} (\bibinfo {year} {2023})}\BibitemShut
  {NoStop}%
\bibitem [{\citenamefont {Fiore}\ \emph {et~al.}(2021)\citenamefont {Fiore},
  \citenamefont {Harunari}, \citenamefont {Noa},\ and\ \citenamefont
  {Landi}}]{fiore2021current}%
  \BibitemOpen
  \bibfield  {author} {\bibinfo {author} {\bibfnamefont {C.~E.}\ \bibnamefont
  {Fiore}}, \bibinfo {author} {\bibfnamefont {P.~E.}\ \bibnamefont {Harunari}},
  \bibinfo {author} {\bibfnamefont {C.~F.}\ \bibnamefont {Noa}},\ and\ \bibinfo
  {author} {\bibfnamefont {G.~T.}\ \bibnamefont {Landi}},\ }\href@noop {}
  {\bibfield  {journal} {\bibinfo  {journal} {Phys. {R}ev. E}\ }\textbf
  {\bibinfo {volume} {104}},\ \bibinfo {pages} {064123} (\bibinfo {year}
  {2021})}\BibitemShut {NoStop}%
\bibitem [{\citenamefont {Barato}\ \emph {et~al.}(2010)\citenamefont {Barato},
  \citenamefont {Chetrite}, \citenamefont {Hinrichsen},\ and\ \citenamefont
  {Mukamel}}]{barato2010entropy}%
  \BibitemOpen
  \bibfield  {author} {\bibinfo {author} {\bibfnamefont {A.}~\bibnamefont
  {Barato}}, \bibinfo {author} {\bibfnamefont {R.}~\bibnamefont {Chetrite}},
  \bibinfo {author} {\bibfnamefont {H.}~\bibnamefont {Hinrichsen}},\ and\
  \bibinfo {author} {\bibfnamefont {D.}~\bibnamefont {Mukamel}},\ }\href@noop
  {} {\bibfield  {journal} {\bibinfo  {journal} {Journal of Statistical
  Mechanics: Theory and Experiment}\ }\textbf {\bibinfo {volume} {2010}},\
  \bibinfo {pages} {P10008} (\bibinfo {year} {2010})}\BibitemShut {NoStop}%
\bibitem [{\citenamefont {Tom{\'e}}\ and\ \citenamefont
  {de~Oliveira}(2012)}]{tome2012entropy}%
  \BibitemOpen
  \bibfield  {author} {\bibinfo {author} {\bibfnamefont {T.}~\bibnamefont
  {Tom{\'e}}}\ and\ \bibinfo {author} {\bibfnamefont {M.~J.}\ \bibnamefont
  {de~Oliveira}},\ }\href
  {https://journals.aps.org/prl/abstract/10.1103/PhysRevLett.108.020601}
  {\bibfield  {journal} {\bibinfo  {journal} {Phys. Rev. Lett.}\ }\textbf
  {\bibinfo {volume} {108}},\ \bibinfo {pages} {020601} (\bibinfo {year}
  {2012})}\BibitemShut {NoStop}%
\bibitem [{\citenamefont {Paoluzzi}(2022)}]{Paoluzzi22}%
  \BibitemOpen
  \bibfield  {author} {\bibinfo {author} {\bibfnamefont {M.}~\bibnamefont
  {Paoluzzi}},\ }\href {https://doi.org/10.1103/PhysRevE.105.044139} {\bibfield
   {journal} {\bibinfo  {journal} {Phys. Rev. E}\ }\textbf {\bibinfo {volume}
  {105}},\ \bibinfo {pages} {044139} (\bibinfo {year} {2022})}\BibitemShut
  {NoStop}%
\bibitem [{\citenamefont {GrandPre}\ \emph {et~al.}(2021)\citenamefont
  {GrandPre}, \citenamefont {Klymko}, \citenamefont {Mandadapu},\ and\
  \citenamefont {Limmer}}]{GrandPre21}%
  \BibitemOpen
  \bibfield  {author} {\bibinfo {author} {\bibfnamefont {T.}~\bibnamefont
  {GrandPre}}, \bibinfo {author} {\bibfnamefont {K.}~\bibnamefont {Klymko}},
  \bibinfo {author} {\bibfnamefont {K.~K.}\ \bibnamefont {Mandadapu}},\ and\
  \bibinfo {author} {\bibfnamefont {D.~T.}\ \bibnamefont {Limmer}},\ }\href
  {https://doi.org/10.1103/PhysRevE.103.012613} {\bibfield  {journal} {\bibinfo
   {journal} {Phys. Rev. E}\ }\textbf {\bibinfo {volume} {103}},\ \bibinfo
  {pages} {012613} (\bibinfo {year} {2021})}\BibitemShut {NoStop}%
\bibitem [{\citenamefont {Bowick}\ \emph {et~al.}(2022)\citenamefont {Bowick},
  \citenamefont {Fakhri}, \citenamefont {Marchetti},\ and\ \citenamefont
  {Ramaswamy}}]{Bowick22}%
  \BibitemOpen
  \bibfield  {author} {\bibinfo {author} {\bibfnamefont {M.~J.}\ \bibnamefont
  {Bowick}}, \bibinfo {author} {\bibfnamefont {N.}~\bibnamefont {Fakhri}},
  \bibinfo {author} {\bibfnamefont {M.~C.}\ \bibnamefont {Marchetti}},\ and\
  \bibinfo {author} {\bibfnamefont {S.}~\bibnamefont {Ramaswamy}},\ }\href
  {https://doi.org/10.1103/PhysRevX.12.010501} {\bibfield  {journal} {\bibinfo
  {journal} {Phys. Rev. X}\ }\textbf {\bibinfo {volume} {12}},\ \bibinfo
  {pages} {010501} (\bibinfo {year} {2022})}\BibitemShut {NoStop}%
\bibitem [{\citenamefont {Suchanek}\ \emph
  {et~al.}(2023{\natexlab{a}})\citenamefont {Suchanek}, \citenamefont {Kroy},\
  and\ \citenamefont {Loos}}]{suchanek2023irreversible}%
  \BibitemOpen
  \bibfield  {author} {\bibinfo {author} {\bibfnamefont {T.}~\bibnamefont
  {Suchanek}}, \bibinfo {author} {\bibfnamefont {K.}~\bibnamefont {Kroy}},\
  and\ \bibinfo {author} {\bibfnamefont {S.~A.~M.}\ \bibnamefont {Loos}},\
  }\href@noop {} {\bibinfo {title} {Irreversible mesoscale fluctuations herald
  the emergence of dynamical phases}} (\bibinfo {year} {2023}{\natexlab{a}}),\
  \Eprint {https://arxiv.org/abs/2303.16701} {arXiv:2303.16701
  [cond-mat.stat-mech]} \BibitemShut {NoStop}%
\bibitem [{\citenamefont {Suchanek}\ \emph
  {et~al.}(2023{\natexlab{b}})\citenamefont {Suchanek}, \citenamefont {Kroy},\
  and\ \citenamefont {Loos}}]{suchanek2023entropy}%
  \BibitemOpen
  \bibfield  {author} {\bibinfo {author} {\bibfnamefont {T.}~\bibnamefont
  {Suchanek}}, \bibinfo {author} {\bibfnamefont {K.}~\bibnamefont {Kroy}},\
  and\ \bibinfo {author} {\bibfnamefont {S.~A.~M.}\ \bibnamefont {Loos}},\
  }\href@noop {} {\bibinfo {title} {Entropy production rate in the
  nonreciprocal {C}ahn-{H}illiard model}} (\bibinfo {year}
  {2023}{\natexlab{b}}),\ \Eprint {https://arxiv.org/abs/arXiv:2305.00744}
  {arXiv:arXiv:2305.00744 [cond-mat.soft]} \BibitemShut {NoStop}%
\bibitem [{\citenamefont {Hohenberg}\ and\ \citenamefont
  {Krekhov}(2015)}]{Hohenberg_2015}%
  \BibitemOpen
  \bibfield  {author} {\bibinfo {author} {\bibfnamefont {P.}~\bibnamefont
  {Hohenberg}}\ and\ \bibinfo {author} {\bibfnamefont {A.}~\bibnamefont
  {Krekhov}},\ }\href {https://doi.org/10.1016%2Fj.physrep.2015.01.001}
  {\bibfield  {journal} {\bibinfo  {journal} {Physics Reports}\ }\textbf
  {\bibinfo {volume} {572}},\ \bibinfo {pages} {1} (\bibinfo {year}
  {2015})}\BibitemShut {NoStop}%
\bibitem [{\citenamefont {Bose}\ and\ \citenamefont {Ghosh}(2019)}]{Bose_2019}%
  \BibitemOpen
  \bibfield  {author} {\bibinfo {author} {\bibfnamefont {I.}~\bibnamefont
  {Bose}}\ and\ \bibinfo {author} {\bibfnamefont {S.}~\bibnamefont {Ghosh}},\
  }\href {https://dx.doi.org/10.1088/1742-5468/ab11d8} {\bibfield  {journal}
  {\bibinfo  {journal} {J. Stat. Mech. Theory Exp.}\ }\textbf {\bibinfo
  {volume} {2019}},\ \bibinfo {pages} {043403} (\bibinfo {year}
  {2019})}\BibitemShut {NoStop}%
\bibitem [{\citenamefont {Goldstone}\ \emph {et~al.}(1962)\citenamefont
  {Goldstone}, \citenamefont {Salam},\ and\ \citenamefont
  {Weinberg}}]{Goldstone62}%
  \BibitemOpen
  \bibfield  {author} {\bibinfo {author} {\bibfnamefont {J.}~\bibnamefont
  {Goldstone}}, \bibinfo {author} {\bibfnamefont {A.}~\bibnamefont {Salam}},\
  and\ \bibinfo {author} {\bibfnamefont {S.}~\bibnamefont {Weinberg}},\ }\href
  {https://doi.org/10.1103/PhysRev.127.965} {\bibfield  {journal} {\bibinfo
  {journal} {Phys. Rev.}\ }\textbf {\bibinfo {volume} {127}},\ \bibinfo {pages}
  {965} (\bibinfo {year} {1962})}\BibitemShut {NoStop}%
\bibitem [{Note1()}]{Note1}%
  \BibitemOpen
  \bibinfo {note} {This implication does not hold for hydrodynamic theories of
  flocking, where the dynamical operator of the polarization field commonly
  includes advection terms~\cite {Marchetti2013}, and therefore spatial
  derivatives of uneven order, yet is still parity invariant}\BibitemShut
  {NoStop}%
\bibitem [{\citenamefont {Dangelmayr}\ \emph {et~al.}(1987)\citenamefont
  {Dangelmayr}, \citenamefont {Knobloch},\ and\ \citenamefont
  {Zeeman}}]{Dangelmayr87}%
  \BibitemOpen
  \bibfield  {author} {\bibinfo {author} {\bibfnamefont {G.}~\bibnamefont
  {Dangelmayr}}, \bibinfo {author} {\bibfnamefont {E.}~\bibnamefont
  {Knobloch}},\ and\ \bibinfo {author} {\bibfnamefont {E.~C.}\ \bibnamefont
  {Zeeman}},\ }\href {https://doi.org/10.1098/rsta.1987.0050} {\bibfield
  {journal} {\bibinfo  {journal} {Philosophical Transactions of the Royal
  Society of London. Series A, Mathematical and Physical Sciences}\ }\textbf
  {\bibinfo {volume} {322}},\ \bibinfo {pages} {243} (\bibinfo {year}
  {1987})}\BibitemShut {NoStop}%
\bibitem [{Note2()}]{Note2}%
  \BibitemOpen
  \bibinfo {note} {Assuming that two eigenvectors with complex eigenvalues
  $\lambda _0$ and $\lambda _1$ merge at a certain fixed point $\phi ^*$, the
  complex conjugate root theorem~\cite {jeffrey2005complex} implies that the
  complex conjugates of these are also eigenvectors with eigenvalues $\protect
  \bar {\lambda }_0$ and $\protect \bar {\lambda }_1$, which accordingly also
  merge.}\BibitemShut {Stop}%
\bibitem [{\citenamefont {Li}\ and\ \citenamefont {Cates}(2021)}]{Li_2021}%
  \BibitemOpen
  \bibfield  {author} {\bibinfo {author} {\bibfnamefont {Y.~I.}\ \bibnamefont
  {Li}}\ and\ \bibinfo {author} {\bibfnamefont {M.~E.}\ \bibnamefont {Cates}},\
  }\href {https://doi.org/10.1088/1742-5468/abd311} {\bibfield  {journal}
  {\bibinfo  {journal} {J. Stat. Mech. Theory Exp.}\ }\textbf {\bibinfo
  {volume} {2021}},\ \bibinfo {pages} {013211} (\bibinfo {year}
  {2021})}\BibitemShut {NoStop}%
\bibitem [{\citenamefont {Markovich}\ \emph {et~al.}(2021)\citenamefont
  {Markovich}, \citenamefont {Fodor}, \citenamefont {Tjhung},\ and\
  \citenamefont {Cates}}]{Cates2021}%
  \BibitemOpen
  \bibfield  {author} {\bibinfo {author} {\bibfnamefont {T.}~\bibnamefont
  {Markovich}}, \bibinfo {author} {\bibfnamefont {{\'E}.}~\bibnamefont
  {Fodor}}, \bibinfo {author} {\bibfnamefont {E.}~\bibnamefont {Tjhung}},\ and\
  \bibinfo {author} {\bibfnamefont {M.~E.}\ \bibnamefont {Cates}},\ }\href
  {https://link.aps.org/doi/10.1103/PhysRevX.11.021057} {\bibfield  {journal}
  {\bibinfo  {journal} {Phys. Rev. X}\ }\textbf {\bibinfo {volume} {11}},\
  \bibinfo {pages} {021057} (\bibinfo {year} {2021})}\BibitemShut {NoStop}%
\bibitem [{\citenamefont {Seifert}(2005)}]{Seifert2005}%
  \BibitemOpen
  \bibfield  {author} {\bibinfo {author} {\bibfnamefont {U.}~\bibnamefont
  {Seifert}},\ }\href {https://link.aps.org/doi/10.1103/PhysRevLett.95.040602}
  {\bibfield  {journal} {\bibinfo  {journal} {Phys. Rev. Lett.}\ }\textbf
  {\bibinfo {volume} {95}},\ \bibinfo {pages} {040602} (\bibinfo {year}
  {2005})}\BibitemShut {NoStop}%
\bibitem [{\citenamefont {Onsager}\ and\ \citenamefont
  {Machlup}(1953)}]{Onsager53}%
  \BibitemOpen
  \bibfield  {author} {\bibinfo {author} {\bibfnamefont {L.}~\bibnamefont
  {Onsager}}\ and\ \bibinfo {author} {\bibfnamefont {S.}~\bibnamefont
  {Machlup}},\ }\href {https://doi.org/10.1103/PhysRev.91.1505} {\bibfield
  {journal} {\bibinfo  {journal} {Phys. Rev.}\ }\textbf {\bibinfo {volume}
  {91}},\ \bibinfo {pages} {1505} (\bibinfo {year} {1953})}\BibitemShut
  {NoStop}%
\bibitem [{\citenamefont {Gradenigo}\ \emph {et~al.}(2012)\citenamefont
  {Gradenigo}, \citenamefont {Puglisi},\ and\ \citenamefont
  {Sarracino}}]{Gradenigo2012}%
  \BibitemOpen
  \bibfield  {author} {\bibinfo {author} {\bibfnamefont {G.}~\bibnamefont
  {Gradenigo}}, \bibinfo {author} {\bibfnamefont {A.}~\bibnamefont {Puglisi}},\
  and\ \bibinfo {author} {\bibfnamefont {A.}~\bibnamefont {Sarracino}},\ }\href
  {https://doi.org/10.1063/1.4731633} {\bibfield  {journal} {\bibinfo
  {journal} {J. Phys. Chem.}\ }\textbf {\bibinfo {volume} {137}},\ \bibinfo
  {pages} {014509} (\bibinfo {year} {2012})}\BibitemShut {NoStop}%
\bibitem [{\citenamefont {Cont}\ and\ \citenamefont {Fournie}(2010)}]{Itofunc}%
  \BibitemOpen
  \bibfield  {author} {\bibinfo {author} {\bibfnamefont {R.}~\bibnamefont
  {Cont}}\ and\ \bibinfo {author} {\bibfnamefont {D.}~\bibnamefont {Fournie}},\
  }\href {https://www.sciencedirect.com/science/article/pii/S1631073X09003951}
  {\bibfield  {journal} {\bibinfo  {journal} {Comptes Rendus Math.}\ }\textbf
  {\bibinfo {volume} {348}},\ \bibinfo {pages} {57} (\bibinfo {year}
  {2010})}\BibitemShut {NoStop}%
\bibitem [{\citenamefont {Gardiner}(2009)}]{gardiner2009}%
  \BibitemOpen
  \bibfield  {author} {\bibinfo {author} {\bibfnamefont {C.}~\bibnamefont
  {Gardiner}},\ }\href@noop {} {\emph {\bibinfo {title} {Stochastic
  methods}}},\ Vol.~\bibinfo {volume} {4}\ (\bibinfo  {publisher} {Springer
  Berlin},\ \bibinfo {year} {2009})\BibitemShut {NoStop}%
\bibitem [{\citenamefont {Wittkowski}\ \emph {et~al.}(2014)\citenamefont
  {Wittkowski}, \citenamefont {Tiribocchi}, \citenamefont {Stenhammar},
  \citenamefont {Allen}, \citenamefont {Marenduzzo},\ and\ \citenamefont
  {Cates}}]{Wittkowski_2014}%
  \BibitemOpen
  \bibfield  {author} {\bibinfo {author} {\bibfnamefont {R.}~\bibnamefont
  {Wittkowski}}, \bibinfo {author} {\bibfnamefont {A.}~\bibnamefont
  {Tiribocchi}}, \bibinfo {author} {\bibfnamefont {J.}~\bibnamefont
  {Stenhammar}}, \bibinfo {author} {\bibfnamefont {R.~J.}\ \bibnamefont
  {Allen}}, \bibinfo {author} {\bibfnamefont {D.}~\bibnamefont {Marenduzzo}},\
  and\ \bibinfo {author} {\bibfnamefont {M.~E.}\ \bibnamefont {Cates}},\
  }\href@noop {} {\bibfield  {journal} {\bibinfo  {journal} {Nat. Commun.}\
  }\textbf {\bibinfo {volume} {5}} (\bibinfo {year} {2014})}\BibitemShut
  {NoStop}%
\bibitem [{\citenamefont {Borthne}\ \emph {et~al.}(2020)\citenamefont
  {Borthne}, \citenamefont {Fodor},\ and\ \citenamefont {Cates}}]{Borthne2020}%
  \BibitemOpen
  \bibfield  {author} {\bibinfo {author} {\bibfnamefont {{\O}.}~\bibnamefont
  {Borthne}}, \bibinfo {author} {\bibfnamefont {{\'E}.}~\bibnamefont {Fodor}},\
  and\ \bibinfo {author} {\bibfnamefont {M.}~\bibnamefont {Cates}},\ }\href
  {https://iopscience.iop.org/article/10.1088/1367-2630/abcd66} {\bibfield
  {journal} {\bibinfo  {journal} {New J. Phys.}\ }\textbf {\bibinfo {volume}
  {22}},\ \bibinfo {pages} {123012} (\bibinfo {year} {2020})}\BibitemShut
  {NoStop}%
\bibitem [{Note3()}]{Note3}%
  \BibitemOpen
  \bibinfo {note} {\protect \leavevmode {\protect \color {black}Note that, for
  domains with periodic boundaries, the inverse Laplacian of a scalar field
  $\varphi (\protect \bm {r},t)$, with $\DOTSI \intop \ilimits@ _V\protect
  \mathrm {d}\protect \bm {r}\varphi (\protect \bm {r},t)=0$, is uniquely
  defined up to \protect \leavevmode {\protect \color {black}a} constant. Thus,
  the operator $\nabla ^{-1}$ is well defined. For a given mass conserving flux
  $\protect \dot {\phi }$ it precisely returns the current $\protect \bm {J}$
  such that $\protect \dot {\phi }=\nabla \cdot \protect \bm {J}$, with the
  gauge choice of $\protect \bm {J}$ being curl-free and $\DOTSI \intop
  \ilimits@ _V\protect \mathrm {d}\protect \bm {J}=0$}.}\BibitemShut {Stop}%
\bibitem [{\citenamefont {Ashida}\ \emph {et~al.}(2020)\citenamefont {Ashida},
  \citenamefont {Gong},\ and\ \citenamefont {Ueda}}]{Ashida2020}%
  \BibitemOpen
  \bibfield  {author} {\bibinfo {author} {\bibfnamefont {Y.}~\bibnamefont
  {Ashida}}, \bibinfo {author} {\bibfnamefont {Z.}~\bibnamefont {Gong}},\ and\
  \bibinfo {author} {\bibfnamefont {M.}~\bibnamefont {Ueda}},\ }\href
  {http://dx.doi.org/10.1080/00018732.2021.1876991} {\bibfield  {journal}
  {\bibinfo  {journal} {Adv. Phys.}\ }\textbf {\bibinfo {volume} {69}},\
  \bibinfo {pages} {249–435} (\bibinfo {year} {2020})}\BibitemShut {NoStop}%
\bibitem [{\citenamefont {Jeffrey}(2005)}]{jeffrey2005complex}%
  \BibitemOpen
  \bibfield  {author} {\bibinfo {author} {\bibfnamefont {A.}~\bibnamefont
  {Jeffrey}},\ }\href@noop {} {\emph {\bibinfo {title} {Complex analysis and
  applications}}}\ (\bibinfo  {publisher} {Chapman and Hall/CRC},\ \bibinfo
  {year} {2005})\BibitemShut {NoStop}%
\bibitem [{Note4()}]{Note4}%
  \BibitemOpen
  \bibinfo {note} {Otherwise, the transition would correspond to an exotic
  combination of CEP and oscillatory instability, which we do not consider
  here.}\BibitemShut {Stop}%
\bibitem [{\citenamefont {Reed}\ and\ \citenamefont
  {Simon}(1975)}]{reed2003methods}%
  \BibitemOpen
  \bibfield  {author} {\bibinfo {author} {\bibfnamefont {M.}~\bibnamefont
  {Reed}}\ and\ \bibinfo {author} {\bibfnamefont {B.}~\bibnamefont {Simon}},\
  }\href {https://books.google.de/books?id=Kz7s7bgVe8gC} {\emph {\bibinfo
  {title} {II: Fourier Analysis, Self-Adjointness}}},\ Methods of Modern
  Mathematical Physics\ (\bibinfo  {publisher} {Elsevier Science},\ \bibinfo
  {year} {1975})\BibitemShut {NoStop}%
\bibitem [{\citenamefont {Poole}(2014)}]{poole2014linear}%
  \BibitemOpen
  \bibfield  {author} {\bibinfo {author} {\bibfnamefont {D.}~\bibnamefont
  {Poole}},\ }\href@noop {} {\emph {\bibinfo {title} {Linear algebra: A modern
  introduction}}}\ (\bibinfo  {publisher} {Cengage Learning},\ \bibinfo {year}
  {2014})\BibitemShut {NoStop}%
\bibitem [{Note5()}]{Note5}%
  \BibitemOpen
  \bibinfo {note} {This can be checked by applying L'H\^{o}pital's rule~\cite
  {reed2003methods} and noting that away from the CEP it holds that $\protect
  \mathrm {Re}\lambda _1,\protect \mathrm {Re}\lambda _0>0$.}\BibitemShut
  {Stop}%
\bibitem [{Note6()}]{Note6}%
  \BibitemOpen
  \bibinfo {note} {To have a meaningful measure for the susceptibility in a
  phase, where a continuous symmetry of the system is already broken,
  fluctuations in the direction of the Goldstone mode $\protect \hat {e}_0$
  (which are unbounded) must be excluded.}\BibitemShut {Stop}%
\bibitem [{\citenamefont {Saxena}\ and\ \citenamefont
  {Kosterlitz}(2019)}]{Saxena_2019}%
  \BibitemOpen
  \bibfield  {author} {\bibinfo {author} {\bibfnamefont {S.}~\bibnamefont
  {Saxena}}\ and\ \bibinfo {author} {\bibfnamefont {J.~M.}\ \bibnamefont
  {Kosterlitz}},\ }\href {https://doi.org/10.1103%2Fphysreve.100.022223}
  {\bibfield  {journal} {\bibinfo  {journal} {Phys. Rev. E}\ }\textbf {\bibinfo
  {volume} {100}} (\bibinfo {year} {2019})}\BibitemShut {NoStop}%
\bibitem [{\citenamefont {Crawford}\ and\ \citenamefont
  {Knobloch}(1991)}]{Crawford91_symmetry}%
  \BibitemOpen
  \bibfield  {author} {\bibinfo {author} {\bibfnamefont {J.~D.}\ \bibnamefont
  {Crawford}}\ and\ \bibinfo {author} {\bibfnamefont {E.}~\bibnamefont
  {Knobloch}},\ }\href {https://doi.org/10.1146/annurev.fl.23.010191.002013}
  {\bibfield  {journal} {\bibinfo  {journal} {Annual Review of Fluid
  Mechanics}\ }\textbf {\bibinfo {volume} {23}},\ \bibinfo {pages} {341}
  (\bibinfo {year} {1991})},\ \Eprint
  {https://arxiv.org/abs/https://doi.org/10.1146/annurev.fl.23.010191.002013}
  {https://doi.org/10.1146/annurev.fl.23.010191.002013} \BibitemShut {NoStop}%
\bibitem [{Note7()}]{Note7}%
  \BibitemOpen
  \bibinfo {note} {Note that we have now assumed that the elements of the
  eigenbasis of $\protect \mathcal {J}$ are normalized to unity, which was not
  the case for the Fourier basis used before, which was normalized to
  $V$.}\BibitemShut {Stop}%
\bibitem [{Note8()}]{Note8}%
  \BibitemOpen
  \bibinfo {note} {To see that corresponding matrix elements in Eq.~\protect
  \textup {\hbox {\mathsurround \z@ \protect \normalfont (\ignorespaces \ref
  {equ:fix}\unskip \@@italiccorr )}} cannot vanish, assume the contrary, i.e.
  $m_i=0$ for all $i$. This implies $\protect \hat {e}_i(0) \protect \hat
  {e}_0(0)=0$ for all $i>1$, which in turns leads to $m_1>0$, This contradicts
  the initial statement.}\BibitemShut {Stop}%
\bibitem [{Note9()}]{Note9}%
  \BibitemOpen
  \bibinfo {note} {This can be easily seen from Eq.~(24) in Ref.~\cite
  {Nardini2017}.}\BibitemShut {Stop}%
\bibitem [{\citenamefont {Loos}\ and\ \citenamefont
  {Klapp}(2020)}]{loos2020irreversibility}%
  \BibitemOpen
  \bibfield  {author} {\bibinfo {author} {\bibfnamefont {S.~A.~M.}\
  \bibnamefont {Loos}}\ and\ \bibinfo {author} {\bibfnamefont {S.~H.~L.}\
  \bibnamefont {Klapp}},\ }\href
  {https://iopscience.iop.org/article/10.1088/1367-2630/abcc1e} {\bibfield
  {journal} {\bibinfo  {journal} {New J. Phys.}\ }\textbf {\bibinfo {volume}
  {22}},\ \bibinfo {pages} {123051} (\bibinfo {year} {2020})}\BibitemShut
  {NoStop}%
\bibitem [{\citenamefont {Zhang}\ and\ \citenamefont
  {Garcia-Millan}(2022)}]{zhang2022entropy}%
  \BibitemOpen
  \bibfield  {author} {\bibinfo {author} {\bibfnamefont {Z.}~\bibnamefont
  {Zhang}}\ and\ \bibinfo {author} {\bibfnamefont {R.}~\bibnamefont
  {Garcia-Millan}},\ }\href@noop {} {\bibinfo {title} {Entropy production of
  non-reciprocal interactions}} (\bibinfo {year} {2022}),\ \Eprint
  {https://arxiv.org/abs/2209.09721} {arXiv:2209.09721 [cond-mat.stat-mech]}
  \BibitemShut {NoStop}%
\bibitem [{\citenamefont {Pruessner}\ and\ \citenamefont
  {Garcia-Millan}(2022)}]{pruessner2022field}%
  \BibitemOpen
  \bibfield  {author} {\bibinfo {author} {\bibfnamefont {G.}~\bibnamefont
  {Pruessner}}\ and\ \bibinfo {author} {\bibfnamefont {R.}~\bibnamefont
  {Garcia-Millan}},\ }\href@noop {} {\bibinfo {title} {Field theories of active
  particle systems and their entropy production}} (\bibinfo {year} {2022}),\
  \Eprint {https://arxiv.org/abs/2211.11906} {arXiv:2211.11906
  [cond-mat.stat-mech]} \BibitemShut {NoStop}%
\bibitem [{\citenamefont {Martynec}\ \emph {et~al.}(2020)\citenamefont
  {Martynec}, \citenamefont {Klapp},\ and\ \citenamefont
  {Loos}}]{martynec2020entropy}%
  \BibitemOpen
  \bibfield  {author} {\bibinfo {author} {\bibfnamefont {T.}~\bibnamefont
  {Martynec}}, \bibinfo {author} {\bibfnamefont {S.~H.~L.}\ \bibnamefont
  {Klapp}},\ and\ \bibinfo {author} {\bibfnamefont {S.~A.~M.}\ \bibnamefont
  {Loos}},\ }\href@noop {} {\bibfield  {journal} {\bibinfo  {journal} {New J.
  Phys.}\ }\textbf {\bibinfo {volume} {22}},\ \bibinfo {pages} {093069}
  (\bibinfo {year} {2020})}\BibitemShut {NoStop}%
\bibitem [{\citenamefont {Noa}\ \emph {et~al.}(2019)\citenamefont {Noa},
  \citenamefont {Harunari}, \citenamefont {de~Oliveira},\ and\ \citenamefont
  {Fiore}}]{noa2019entropy}%
  \BibitemOpen
  \bibfield  {author} {\bibinfo {author} {\bibfnamefont {C.~E.~F.}\
  \bibnamefont {Noa}}, \bibinfo {author} {\bibfnamefont {P.~E.}\ \bibnamefont
  {Harunari}}, \bibinfo {author} {\bibfnamefont {M.~J.}\ \bibnamefont
  {de~Oliveira}},\ and\ \bibinfo {author} {\bibfnamefont {C.~E.}\ \bibnamefont
  {Fiore}},\ }\href@noop {} {\bibfield  {journal} {\bibinfo  {journal} {Phys.
  Rev. E}\ }\textbf {\bibinfo {volume} {100}},\ \bibinfo {pages} {012104}
  (\bibinfo {year} {2019})}\BibitemShut {NoStop}%
\bibitem [{\citenamefont {Loos}\ \emph {et~al.}(2023)\citenamefont {Loos},
  \citenamefont {Klapp},\ and\ \citenamefont {Martynec}}]{loos2022long}%
  \BibitemOpen
  \bibfield  {author} {\bibinfo {author} {\bibfnamefont {S.~A.~M.}\
  \bibnamefont {Loos}}, \bibinfo {author} {\bibfnamefont {S.~H.~L.}\
  \bibnamefont {Klapp}},\ and\ \bibinfo {author} {\bibfnamefont
  {T.}~\bibnamefont {Martynec}},\ }\href
  {https://link.aps.org/doi/10.1103/PhysRevLett.130.198301} {\bibfield
  {journal} {\bibinfo  {journal} {Phys. Rev. Lett.}\ }\textbf {\bibinfo
  {volume} {130}},\ \bibinfo {pages} {198301} (\bibinfo {year}
  {2023})}\BibitemShut {NoStop}%
\bibitem [{Note10()}]{Note10}%
  \BibitemOpen
  \bibinfo {note} {Note that differing noise temperatures alone indeed not need
  to be a sign of nonequilibrium in this context, but may also result from a
  chosen basis representation.}\BibitemShut {Stop}%
\bibitem [{\citenamefont {Marchetti}\ \emph {et~al.}(2013)\citenamefont
  {Marchetti}, \citenamefont {Joanny}, \citenamefont {Ramaswamy}, \citenamefont
  {Liverpool}, \citenamefont {Prost}, \citenamefont {Rao},\ and\ \citenamefont
  {Simha}}]{Marchetti2013}%
  \BibitemOpen
  \bibfield  {author} {\bibinfo {author} {\bibfnamefont {M.~C.}\ \bibnamefont
  {Marchetti}}, \bibinfo {author} {\bibfnamefont {J.~F.}\ \bibnamefont
  {Joanny}}, \bibinfo {author} {\bibfnamefont {S.}~\bibnamefont {Ramaswamy}},
  \bibinfo {author} {\bibfnamefont {T.~B.}\ \bibnamefont {Liverpool}}, \bibinfo
  {author} {\bibfnamefont {J.}~\bibnamefont {Prost}}, \bibinfo {author}
  {\bibfnamefont {M.}~\bibnamefont {Rao}},\ and\ \bibinfo {author}
  {\bibfnamefont {R.~A.}\ \bibnamefont {Simha}},\ }\href
  {https://link.aps.org/doi/10.1103/RevModPhys.85.1143} {\bibfield  {journal}
  {\bibinfo  {journal} {Rev. Mod. Phys.}\ }\textbf {\bibinfo {volume} {85}},\
  \bibinfo {pages} {1143} (\bibinfo {year} {2013})}\BibitemShut {NoStop}%
\end{thebibliography}%
\end{document}